\documentclass[prd,tightenlines,twocolumn]{revtex4}
\usepackage{amsmath}
\usepackage{amssymb}
\usepackage{graphicx}
\usepackage{bm}

\usepackage{enumerate}
\usepackage{color}
\newcommand{\be}{\begin{equation}}
\newcommand{\ee}{\end{equation}}
\newcommand{\bea}{\begin{eqnarray}}
\newcommand{\eea}{\end{eqnarray}}
\newcommand{\ba}{\begin{eqnarray}}
\newcommand{\ea}{\end{eqnarray}}
\newcommand{\Dslash}{D\hspace{-1.6ex}/\hspace{0.6ex} }

\newcommand{\Tr}{\mbox{Tr}\;}
\newcommand{\tr}{\mbox{tr}\;}
\newcommand{\ket}[1]{\left|#1\right\rangle}

\newcommand{\avg}[1]{\left\langle #1\right\rangle}

\newcommand{\pair}{\raisebox{-7pt}{\includegraphics[height=20pt]{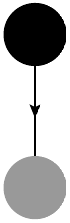}}}
\newcommand{\paircrs}{\raisebox{-7pt}{\includegraphics[height=20pt]{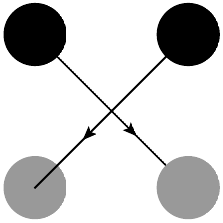}}}
\newcommand{\paircc}{\raisebox{-7pt}{\includegraphics[height=20pt]{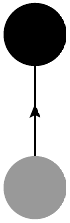}}}
\newcommand{\paircrscc}{\raisebox{-7pt}{\includegraphics[height=20pt]{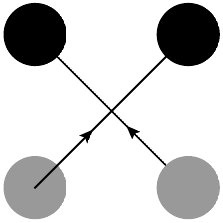}}}
\newcommand{\pairloop}{\raisebox{-7pt}{\includegraphics[height=20pt]{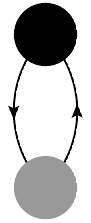}}}
\newcommand{\pairloopf}{\raisebox{-7pt}{\includegraphics[height=20pt]{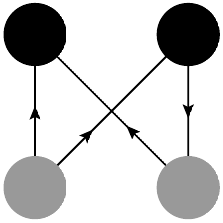}}}
\newcommand{\pairlooph}{\raisebox{-7pt}{\includegraphics[height=20pt]{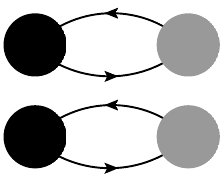}}}

\begin{document}

\title{The   Chiral Symmetry Breaking/Restoration \\ in Dyonic Vacuum }
\author{Edward Shuryak$^{1,2}$ and Tin Sulejmanpasic$^{1,3}$}
\affiliation{$^1$Department of Physics and Astronomy, \\ Stony Brook University,\\
Stony Brook, NY 11794, USA\vspace{0.3cm}\\ 
$^2$Kavli Institute of Theoretical Physics,\\ University of California at Santa Barbara\vspace{0.3cm} \\
$^3$Institut f\"ur Theoretische Physik\\
Universi\"at Regensburg\\
D-93040 Regensburg, Germany\vspace{0.3cm}}

\date{\today}

\begin{abstract}
We discuss the topological phenomena in the QCD-like theories with a variable number of fundamental fermions $N_f$, 
focusing on the temperatures  
at or above the critical value $T_c$ of chiral symmetry restoration. The nonzero average
of the Polyakov line, or holonomy, splits instantons into (anti)self-dual dyons, and we study both the bosonic and fermionic interactions between them. The high temperature phase is a dilute gas of  ``molecules" made of $2N_c$ dyons, neutral in topological, electric, and magnetic charges. At intermediate temperatures the diluteness of the ``molecular gas" reaches some critical
level at which chiral symmetry gets restored: we comment on why it is  different
for the fundamental and adjoint fermions.  At high density the ensemble is a strongly coupled liquid with
crystal-like short range order: we speculate about its possible structure at small and large $N_f$. We finally show that certain lattice observations are in agreement with the proposed model, and  suggest a number of further lattice tests.
\end{abstract}

\maketitle
\section{Introduction}

\subsection{Qualitative overview}

Here we outline the qualitative findings that emerged during the course of this study, and provide
the answers to some of the questions that followed from it.
   A history of the works and ideas that lead us to these answers will be provided in the next subsection.

It is perhaps necessary to  explain what we mean by the ``dyonic vacuum." Dyons in general are objects possessing both electric and magnetic 
charges.  However, the term is used in two very different contexts.  Historically, the first one  is the ``particle dyons" (Julia-Zee), the  excitations 
 of 't Hooft-Polyakov  monopoles,  well known in the Georgi-Glashow model and many supersymmetric theories: we will $not$ discuss those in this work. The second type is the ``self-dual dyons," which appear as constituents of the instantons.  Rather than being particlelike \emph{excitations} of the vacuum, as the monopoles and the first type of dyons are, 
they are part of the vacuum itself, describing a certain topologically nontrivial  configurations
of the gauge fields.  Not being particles, they do not have momenta or kinetic energies: they appear in the QCD partition function integrated
over their collective variables. While instantons had among such variables color orientations, the dyons have their positions and Abelian (diagonal) color charges. In the case of
the SU(2) color group, there are two dyons per instanton,  commonly called the $M$-type and the $L$ (or twisted) type, (see Table \ref{tab_su2dyons}). For a review of instanton dyons see \cite{Bruckmann:2003yq}, \cite{Diakonov:2009ln} and references therein.


Here are some physical questions we will discuss:
\begin{enumerate}
\item What are the interactions between dyons, especially between the self-dual and anti--self-dual ones?

\item How do fermions contribute to  the interaction between dyons?

\item  What is the qualitative picture of the dyon statistical ensemble, 
 as a function of three key parameters, the temperature $T$ and the number of
fundamental $N_f$ or adjoint $N_a$ quarks in the theory? In particular, why does the 
chiral transition moves to a stronger coupling? 

\item
In the high-T limit,  gauge field topology was described as a dilute gas of instanton--anti-instanton molecules \cite{Ilgenfritz:1988dh}.
 How are these objects modified for the case of the nonzero holonomy, in the language of dyons?

\item What is the Dirac eigenvalue spectra for different dyonic ensembles? At which
condition chiral symmetry breaking takes place? 

\item Can one explain
the dependence of the chiral phase transition on $N_f$ and/or  $N_a$?

\item Can one evaluate the ``gaps" in the Dirac eigenspectra
which are developed at $T>T_c$ using the dyonic ensemble?

\item At $T>T_c$,  using the quenched ensemble of gauge fields, it has been found on the lattice that
the chiral properties crucially depend on the particular periodicity conditions for the fermions. 
In particular, the  \emph{periodic} ones
do not show a chiral restoration transition, unlike the (physical)
 \emph{anti-periodic} fermions. How can one understand these observations? 

\item 
Why does the chiral transition strongly depend on the color representation of the fermions, such as the fundamental or adjoint ones?

\end{enumerate}

  The reader perhaps noticed that this list of questions  includes
neither a discussion of the holonomy potential nor other questions
   related to confinement (such as e.g. in   \cite{Diakonov:2004jn}.)
 We think that any  assessment of the  
   back reaction of the dyons on the holonomy can only be done after a more quantitative
   theory of their ensemble emerges. The purpose of this paper is to
    take a step 
   toward developing such a theory. Thus here we will
   consider the holonomy $\avg{P(T)}$ as given, e.g. by the lattice data.
   
   Now we outline the picture. It is convenient to discuss it 
   by defining
three regimes,  from high to low $T$. We will call them: 
\begin{enumerate}[(i)]
 \item High $T\gg T_c$ case, the regime of a dilute molecular gas
 \item Intermediate regime, $T=(1..2) T_c$, interacting molecular gas
 \item Dense regime, $T<T_c$, dyons form a strongly coupled  plasma, in their liquid phase
\end{enumerate}

Here are  further explanatory comments on each of those:

(i) High temperature implies weak coupling and thus the semiclassical
treatment of instantons/dyons is applicable. Since these objects have nonzero electric fields, subject to the perturbative Debye screening, their density at high $T$ is strongly suppressed. In a resulting
dilute regime, the ensemble forms a ``molecular gas" of objects that have 
zero topological, electric, and magnetic charges.
 The average Polyakov line in this regime is close to 1, i.e. the ``Higgs Vacuum Expectation Value (VEV)"  $v\approx 0$,  so all $M$-type dyons are light, while ``heavy" $L,\bar{L}$ dyons have nearly all the action of instantons/antiinstantons. Instanton-antiinstanton molecules were described in \cite{Ilgenfritz:1988dh}
 and subsequent works: and in the high-$T$ limit we expect to be close to those results.
 
 \begin{figure}[htb]
 \includegraphics[width=\columnwidth]{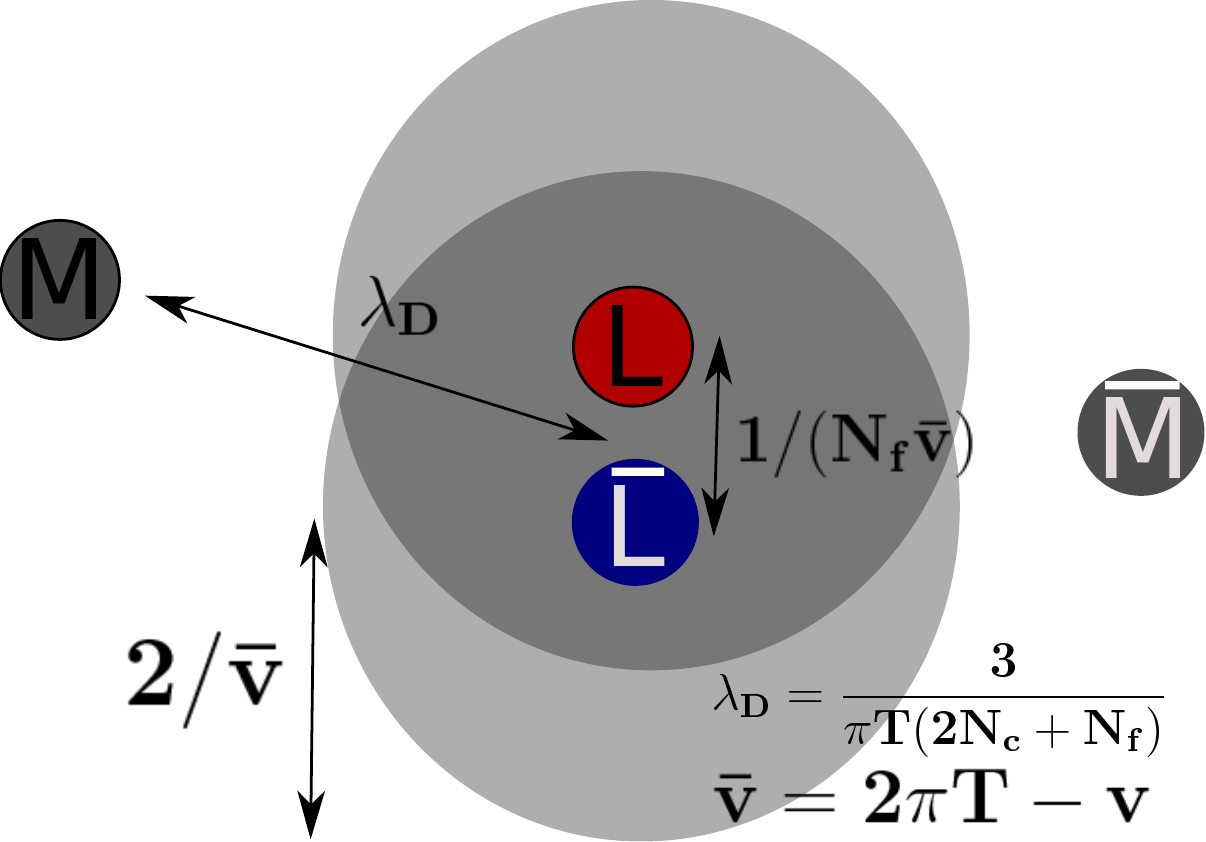}
 \caption{ (Color online)The schematic picture of the dyonic molecule, for 2 colors and large $N_f$. }\label{fig:mol}
\end{figure}
 
 Fermionic zero modes of the instantons are shared by their constituents in a way
 that depends on the type of fermion of the theory.   Physical antiperiodic fermions have zero mode of the (twisted)  $L,\bar{L}$ dyons. 
 As the number of fundamental fermions $N_f$ in the theory increases, 
 they bind them into tight $L\bar{L}$
``clusters,''  which play a role of the nucleus of these molecules.   Consequently, the chiral symmetry is unbroken and the lowest Dirac eigenstates ``at the gap" correspond to  independent  $L\bar{L}$
 clusters (see Fig. \ref{fig:mol}).
 
 The standard Abelian electric charges of both  $L$ and $\bar{L}$ are equal to -1, so the clusters  has the  charge -2.  
 (The molecule thus looks like anti-He, with $M,\bar{M}$ as ``positrons" around it).
A particular sign of a charge does not violate C parity, of course, because the Abelian fields are color-projected to the color direction of the
 holonomy field $\sim \Tr(F_{\mu\nu} A_4)$, or more precisely $\sim \Tr F_{\mu\nu} L$, where $L$ is the Polyakov loop. Since both non-Abelian fields in it are C odd, the product is C even. These signs
 are just matter of the definitions  used in the field. 
 
 Lattice practitioners  sometimes use the so called ``valence" or non-dynamical quarks  (not included in the
 partition function) as a tool for the analysis of the gauge configurations.
  Those may have arbitrary properties and
periodicity conditions.  
 The ``valence quarks"  $periodic$ over the Matsubara time have completely different zero
modes and interact with the lighter  $M$- type dyons. Those are also more weakly correlated than the $L$-type ones. The difference in their Dirac eigenspectra will be important tool in testing the structure of
 the dyonic vacuum.

 (ii) As $T$ is lowered,  the effective electric coupling $\alpha_s(T)=g^2(T)/4\pi$ grows and  eventually becomes large $\alpha_s=O(1)$. 
A quite specific point  introduced in \cite{Liao:2006ry}
is the so-called ``E/M equilibrium point," at which $\alpha_s(T)=1$. 
Because of the Dirac condition for electric and magnetic couplings
 \be \alpha_s \alpha_{\text{magnetic}}=\text{integer} \ee
 at this point, with integer being 1, the magnetic  alpha is also 1. It has been argued and confirmed on the lattice
 that at this point magnetic excitations -- monopoles -- become about as numerous as the usual
 electric excitations, quarks and gluons. In gluodynamics (no fermions, $N_f=0$) this happens at
 $T\approx 1.4 \,T_c$: how this depends on the presence of fermions remains to be studied. 
 
 Below $T_c$ the 
 confinement  forces the density of the electric objects (quarks and gluons) to be zero, while the magnetic (monopoles) retain the finite density. Only at $T\rightarrow 0$ does it disappear
 as well, with only the magnetic condensate remaining.
  Consequently, the electric Debye mass disappear at $T<T_c$, while the magnetic screening mass remains finite. 
  This implies that the electric screening of instantons at high $T$ is substituted
  by a  magnetic screening at $T< T_c$.
 As argued in \cite{Shuryak:1999fe},  the latter creates 
 a  factor in the density of instantons $\sim \exp(-\rho^2\times const.)$, where $\rho$ is the instanton radius and the constant, due to \emph{magnetic} screening, has a nonzero value
  even at $T\rightarrow 0$ and is related to
 {\em Bose-condensed} magnetic monopoles. In certain dual models
 this constant was further related to the QCD string tension $const=2\pi\sigma$  \cite{Shuryak:1999fe}. The expression
 describes well the lattice data on instanton size distribution and also explains why
in the QCD instanton ensemble remains relatively dilute even at $T=0$.  It  would be very interesting to see if any of that
remains to be true at large $N_f$. 

Near $T_c$  magnetic screening should be induced mostly by the scattering of \emph{non}-Bose-condensed monopoles.
To our knowledge no study of the effect has ever been done, and 
we also defer it to future studies.
   
   The interrelation between the ``particle-monopoles" (inducing confinement by their BEC) and the
   instanton-dyons (inducing chiral symmetry breaking as we discuss below) is of course an intriguing
   open problem. On the level of gauge configurations themselves or their zero modes one finds
   no direct relation between them. However, at the level  of the effective 't Hooft Lagrangian
   an intriguing relation has been found by Poppitz and Unsal \cite{Poppitz:2011wy} in {\cal N}=2
   compactified Super Yang Mills (SYM) case. It appears at the level of the partition functions, one being the sum of the particle-dyon excitations and another the sum over the periodic instanton-dyon semiclassical solutions. Such relation
   clearly deserves further study and generalizations.  

 While any perturbative expressions/intuition  is obviously not to be trusted in the regime with $\alpha_s(T)\approx 1$, the lattice simulations  treat this region consistently. 
 Furthermore, as we will detail below, in QCD-like
  theories with many fermions the plasma phase extends to
  even stronger coupling of $\alpha_s\sim 3$ or so.  Perhaps  the dual-magnetic-formulation of such theories can be used in this case, as the magnetic coupling is getting weak $\alpha_{magnetic}\sim 1/3$.
 
 Our paper, as many others,
 rely on the robustness of the topological effects under  deformations, even if
 the amplitude of  those is not small.
  Furthermore,
topology is related (by index theorems)   to fermionic zero modes.
A ``collectivized" set of such fermionic states contributes significantly to the quark condensate, pions and 
strongly influence the structure of the lowest hadronic states and correlation functions  \cite{Schafer:1996wv}. 
In contrast to the papers discussed in that review, we now approach this problem ``from above," starting from
the hot symmetric phase at $T>T_c$ and follow the evolution of the topological quark states, from
localized to delocalized ones as the transition temperature is approached.

Lattice data  tell us that in the temperature interval $(2..1)T_c$ the average Polyakov line
$ \avg{P(T)}$ changes from $\approx 1$ to $\approx 0$. The holonomy
changes  from $\nu=0$ to $ \nu=1/2$, at which point the masses (i.e. actions) of the L,M-type dyons become comparable. 
We also know that at the latter point confinement phenomenon takes place.

 (iii) We can only qualitatively discuss the dense regime near and below $T_c$,
 as the interaction between the dyons gets very strong. 
 We  try to approach the problem  from the  perspective of  the strongly coupled classical plasma. 
 
 For large $N_f$ the basic objects include the $L\bar{L}$ clusters which, we will argue, strongly repel each other.
 Therefore, the optimal correlations in such a medium would be similar to other systems which experience
 strong repulsive forces,
 such as closely packed liquids. While the global order is absent, locally those are strongly correlated,
 with the type of correlations being similar to those in certain best-packed crystals.
 
 For zero $N_f$ the dominant forces are Coulomb-like and corrections to them are in a form of the determinant 
proposed by Diakonov and collaborators, as well as the screening ones. If so, we suggest dyonic crystals resembling 
salt: cubic with alternating $L,M$ dyons. 

(Needless to say, we do not think that the solid phase is
actually reached; it is well known that strongly correlated liquid have short-range correlations that are the same as their fully ordered, crystalline form. While the ``dyonic crystals" discussed provide examples of configurations in which the
interaction is minimized, thermal fluctuations do kill the long range order, making it a liquid. Perhaps it is worth mentioning that
  the main parameter in the Boltzmann exponents,  the mean ratio of the interaction potential per particle to $T$, also called $\Gamma$, needs to be 
$\Gamma > \Gamma_c\sim O(100)$ for solidification. In the dyon problem discussed this is not so, as
  $\Gamma\sim O(10)\ll \Gamma_c$.)

   Let us now focus on the main observable to be discussed, the Dirac eigenvalue spectrum and possible chiral symmetry breaking.
With the increasing  number of fundamental fermions $N_f$ in the theory, they induce stronger correlations 
and reduce the size of the $L\bar L$ ``clusters.''  If one wants to follow the lines of constant quark condensate, e.g. the 
chiral restoration line,  one has to increase the density of the clusters accordingly. This can only be achieved 
by a shift to stronger  coupling. As the dyon masses and interactions are $\sim 1/g^2$,  they  become lighter and 
 less interacting.  (Needless to say,  their fermionic zero modes and related interaction
must still be there, for topological reasons: they do not depend on the coupling.) The (dimensionless) density of dyons continue to grow
to the situation in which the inter-dyon distances become comparable to the $L\bar{L}$  molecule 
size.  
 
The adjoint fermions are very different from the fundamental ones. Some of them remain ``massless'' (in the sense of the ``holonomy mass'') after adjoint Higgsing, and this drastically changes
the dependence of the ``hopping amplitudes" on the distance, from exponential to powerlike.
The chiral symmetry is
unbroken above such T when not only heavy $L$ but also light $M$ dyons have zeromodes.    
This puts the chiral phase transition at much weaker coupling (higher T)

\subsection{From instantons to dyons}
 
  The discovery of the instanton solution \cite{Belavin:1975fg} has created a great deal of literature,
 including electroweak physics of baryon charge nonconservation as well as the famous exact results for various supersymmetric theories. Obviously we cannot review this amount of literature here.
 
 In the context of the QCD-like theories, the predecessor of this paper is the so-called instanton liquid model, for a review see
 \cite{Schafer:1996wv}. Its main point was to account for the so-called 't Hooft 
interactions to arbitrary order, by including the fermionic determinant in certain approximation
in numerically simulated statistical ensembles. The calculated
point-to-point correlation functions have reproduced many lattice results related
to chiral $SU(N_f)$ and $U(1)$ symmetries.  Chiral restoration can be viewed  as the disappearance
of the nontrivial solution to the so called gap equation. Alternatively,  it was explained  \cite{Ilgenfritz:1988dh} as a consequence of
a structural phase transition in the instanton ensemble,  
from a random plasma at low $T$ into a  gas of strongly correlated  $\bar{I} I$  instanton--anti-instanton pairs. 
The pairing mechanism is due to the fermion exchange, thus it gets stronger as $N_f$ grows.

  Let us recall its basic ideas which will be used below. In the basis spanned by the zero modes of individual instantons/anti-instantons, one can write the 
Dirac operator as
\ba 
\label{D_zmz}
i D\!\!\!\!/ &=& \left (
\begin{array}{cc}  0 & T_{IA}\\
                   T_{AI} & 0 
\end{array} \right ),
\label{eqn_tij1}
\ea
where we have introduced the overlap sub-matrix  $T_{IA}$ 
\newcommand{\dslash}{D\!\!\!\!/\,}
\ba
\label{def_TIA}
T_{IA} &=& \int d^4 x\, \psi_{0,I}^\dagger (x-z_I)i\dslash
\psi_{0,A}(x-z_A)
\ea
where $I,A$ are indices
which run over all instantons and antiinstantons in the configuration.
Here, $\psi_{0,I}$ is the fermionic zero mode. The individual matrix 
elements have the meaning of a hopping amplitude for a quark from 
one pseudoparticle to another, and the determinant of this matrix is
nothing else but the sum over the loop diagrams in which quarks visit
each instanton once.
 Note that two $\psi$s have opposite chirality, so if i=instanton then j=antiinstanton or v.v.
 The fermionic determinant is approximated by $|det(T_{ij})|^2$.
The low-T ensemble is a dense liquid that breaks chiral symmetry, but at high $T$ (small size in $\tau$ direction) 
it breaks into ``$\bar{I} I$ molecules"
and chiral symmetry gets restored. The actual calculation was a simulation of the ensemble with the weight containing 
$|det(T_{ij})|^2$, which was then used for
the evaluation of  the Dirac spectra and hadronic correlation functions.
At high $T$ the approximate factorization of the Dirac matrix into independent $2\times 2$  boxes (for separate clusters) 
explains the deformation of the Dirac eigenvalue spectra and disappearance
of near-zero eigenvalues and the existence and the magnitude of the spectral  
gap $G$. 

We will now extend these ideas to the case of the nonzero holonomy,
 the 
gauge-invariant closed loop integral over the $x^4=\tau$ circle $ \int_0^{1/T}  d\tau A_4$. Its exponent, the so called Polyakov line, averaged over the statistical ensemble of fields, has  a nonzero
 value
\be  \avg{P}=\avg{\Tr\exp\left(i \int d\tau A_4\right)} \ne 0 \ee
This calls for classical solutions that do not
approach zero fields at spatial infinity but rather some constant value $v$ of the $A_4^3$ (in SU(2)).
We will also use dimensionless notations \be \nu= {v \over 2\pi T},   \,\,\,\,\, \bar{\nu}=1-\nu \ee
 Explicit solutions of such type 
\cite{Lee:1998bb,Kraan:1998sn}
 demonstrate that  an instanton gets split into
the $N_c$ constituent dyons. 
The names and quantum numbers (for the simplest $SU(2)$ gauge group
we will discuss in this work) cover all four possibilities for the electric and magnetic charges,
see Table \ref{tab_su2dyons}. For $SU(N_c)$ 
in general there are $M_1,M_2... M_{Nc-1}$ static dyons with all diagonal charges and one ``twisted" $L$-dyon. 

\begin{table}[h!]
\begin{tabular}{| c | c|c| c| c|} \hline
name & E & M & mass \\ \hline
$M$         & + &+ &$v$ \\
$\bar{M}$ & + &- &$v$ \\
$L$          & - & - &$2\pi T -v$ \\
$\bar{L}$ & - & + &$2\pi T -v $ \\ \hline
\end{tabular} 
\caption{ The charges and the mass (in units of $8\pi^2/g^2 T$) for 4 SU(2) dyons.} \label{tab_su2dyons}
\end{table}

  Let us indicate here the qualitative difference that the nonzero holonomy brings into this problem.
The fundamental  fermions  in the  ``Higgsed" vacuum with a VEV of $A_4$   are ``massive,'' \footnote{In the text we often refer to the ``holonomy mass'' or simply ``mass'', occasionally without the quotation marks, to indicate the mass given by the Higgs effect of $A_4$. This ``mass'' does not break chiral symmetry and should in no way be confused with the fermion mass. In our study fermions are taken as massless.}, with masses (in SU(2)) $m_f=\pm g v/2$. Therefore,
the zero modes at large distances $r\rightarrow \infty$ decrease exponentially with the distance, unlike the power behavior typical for the zero holonomy case.
These rapidly decreasing fermionic amplitudes are  of course further enhanced by the
  number $N_m$ of fermionic zero modes 
\be e^{-V}\sim \det{T} \sim e^{-N_m m_f r} \ee
which creates strong linear confining potential for the corresponding dyons 
and thus produces small-size ``clusters" of the size
\be  \avg{r}\sim (N_m m_f)^{-1} \ee 

The number of the modes is dependent on the fermion's color charge and the number of 
its copies. For the usual fundamental quarks $N_m=2 N_f$, as there is a zero mode
for a quark and for an antiquark. 

For the adjoint fermion $N_m=2N_c N_a$.
Furthermore, adjoint fermions that are diagonal with respect to the Polyakov line VEV remain ``massless.''
 The interaction between the dyons due to an exchange of the adjoint fermions
 has been discussed by Unsal \cite{Unsal:2007jx}. The center of his proposal is a ``bion,''
 a cluster of $L\bar{M}$ dyons with the magnetic charge 2.  Such bions can induce Polyakov-like
  confinement in the spatially compactified QCD \cite{Unsal:2007jx}.
 
 \subsection{Overview of lattice data on chiral symmetry restoration and deconfinement } \label{sec_lattice}

\begin{figure*}[t] 
   \centering
   \includegraphics[width=7cm]{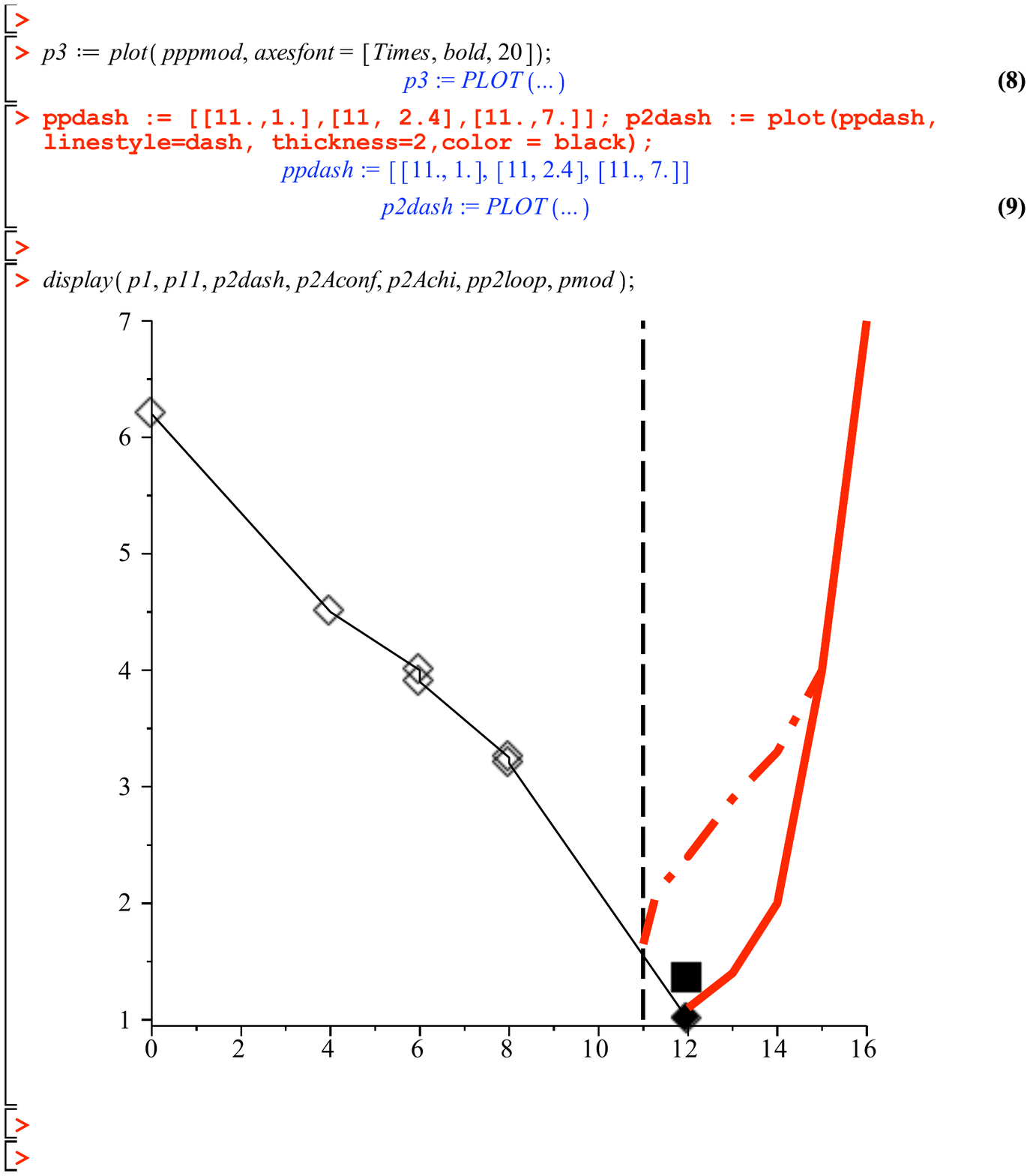}   \includegraphics[width=7cm]{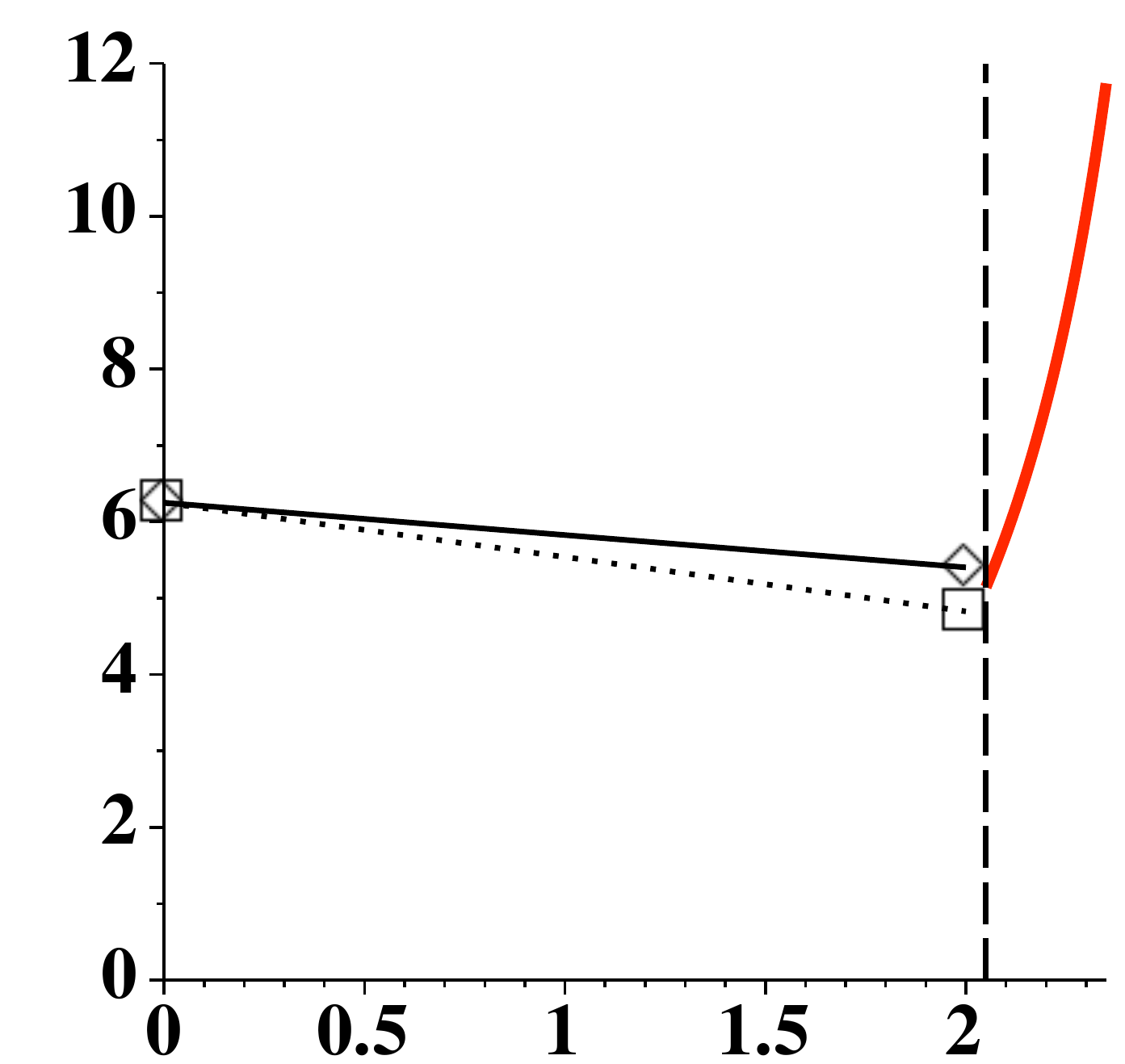} 
   \caption{ (Color online)   The critical lines for  chiral restoration (solid line and diamonds) and deconfinement (dotted line and boxes)  of the $N_c=3$ gauge theory.
 We plot the critical lattice coupling $\beta_c(T_c)=6/g^2_c(T_c)$ 
   versus  the number of fundamental quarks $N_f$ in (a) or a number of adjoint quarks $N_a$ in (b).
     Both paths of the infrared fixed point, calculated in the two-loop approximation, are shown by the thick (red) lines. The vertical  dashed lines
separate the ``conformal window domain": its location is a guess. In Fig.(a) we also
show, by the   dash-dotted (red) line, our guess for the actual location of the fixed point.
For the meaning of the data points see the text.
    }
   \label{fig:romanplot}
\end{figure*}

\subsubsection{ The critical lines versus the number of fermions $N_f,N_a$}
Let us start by reviewing some recent  lattice results. Our  version of the  phase  diagram uses  the ``critical lattice coupling"
\be \beta_c(T_c)={2N_c \over g^2(T_c)} \ee
as a function of $N_f$ or $N_a$. The ``bare" coupling values in lattice works
are defined at the lattice UV scale $a$.
In order to make it lattice independent, we have
evolved the scale from $a$ by a factor of $N_t$ (the number of points in the time direction)
 to the physical scale $N_t a=1/T_c$
using the two-loop beta function.  The near overlapping points in Fig. \ref{fig:romanplot}(a) are from different $N_t$ simulations: their spread is a measure of the inaccuracy
of the two-loop beta function used.

The open diamonds in  Fig.\ref{fig:romanplot}(a) correspond to lattice data from Ref.  \cite{Miura:2011mc} : they show the critical line of the chiral restoration
(thin solid line). Above the line one finds  symmetric [quark-gluon plasma (QGP)] phase, while
below is the chirally broken (and confining) one. 
Moving downward on this figure means increasing the gauge coupling or decreasing the temperature.
Thus, at $N_f\sim 10$ one may reach  ``the most strongly coupled QGP,''
which is by itself a very interesting phenomenon. Why is it happening?

   (There are many other simulations reported in the lattice literature, of course: we decided not to put those on this plot because the rather different actions used
produce rather random spreading of the couplings, confusing the picture.)    

   The situation at $N_f=12$ is special. 
   Reference \cite{Deuzeman:2010gb}  argued that the chiral symmetry 
   remains unbroken. 
 A more recent paper \cite{Cheng:2011ic}
   have studied the region of even stronger coupling $6/g^2\sim 1$,
   and  perhaps clarified the picture. Two distinct phase transitions are observed, with the chiral condensate disappearing at
   stronger coupling  (the closed diamond in  Fig.\ref{fig:romanplot}(a))
   than the confinement (the closed box). 
    Thus, a novel  intermediate phase in between is confining but
   chirally symmetric. An understanding of its precise nature remains a challenge,
   although such examples are known in supersymmetric theories.
   
   The same authors \cite{Cheng:2011ic} also concluded
    that for $N_f=12$ those transitions are also
   separated by a bulk transition from a weaker coupling domain, in which
  there    seem to 
   exist a conformal (infrared fixed point) behavior. 
       If so, the two phase transitions must be below (on the other side of) the true trajectory of the fixed point
schematically  shown by  the (red) dash-dotted line.  

[This line may deviate from the thick solid (red) line, showing a fixed point line using the
two-loop beta function, because of the very strong coupling involved.
The issue of where exactly the conformal window starts  remains unresolved. We  tentatively put
     the vertical dashed line separating it at $N_f=11$ in the figure.
 This is not a prediction but just a  guess, and should not be used in any way.]

Changing the fundamental quarks  into either (i) adjoint  (triplet in SU(2), octet in SU(3))  or (ii) symmetric 
rang-2-tensor  (triplet in SU(2), sextet in SU(3)) shifts the $T_c$ of the chiral transition upwards. 
The theory with two adjoint fermions, $N_a=2$, has been studied in detail, see Fig. \ref{fig:romanplot}(b) based on 
 \cite{Cossu:2008wh}, but (to our knowledge) not for other values, notably $N_a=1$. (Of course, introducing
 variable masses may allow to follow the lines continuously.)
Two distinct transitions were found,
but in the opposite order to the previous case of the fundamental fermions.
 So, between the  solid and the dotted lines there exists the {\em deconfined but chirally broken}
phase (a plasma of constituent quarks). Furthermore, while the difference between these two points
may appear small on this plot, the actual $T_c$ scales
of the two transitions are different by about factor 8. 

(It is also noteworthy that these points are close to the conformal window perturbatively, or 
perhaps  already
inside it nonperturbatively. We tentatively  put its boundary -- the vertical dashed line-- at $N_a>2$: it is not a prediction. Its exact location needs to be found numerically in future works. )

\subsubsection{The magnitude of the chiral splittings versus $N_f$.}
   For $N_f$ up to at least 8, the low-$T$ theories retain both confinement 
   and chiral symmetry breaking, but the quantitative relation between them  changes. Let us
   characterize it  by  the relative splittings between the chiral partners,
   such as vector-axial $\rho-A_1$ mesons or the nucleon and the lowest $1/2^-$ $N^*$ resonance at $T=0$
   \be  \Delta_{\rho A_1}=2 {m_{A_1}- m_\rho \over m_{A_1}+m_\rho}  \ee
    \be  \Delta_{NN*}=2{m_{N*} - m_N \over m_{N*}+m_N }  \ee
 the $N_f$ dependence of these ratios is interesting: in the interval $N_f=0..3$  these chiral splittings
 are ``large" near the experimental values  (so to say, at $N_f\approx 2.5$)
 \be  \Delta^{exp}_{\rho A_1} \approx  0.45  \hspace{1cm}  \Delta^{exp}_{NN^*}\approx   0.55\ee
which are well reproduced  on the lattice.  
Yet calculated for $N_f=4$ \cite{hep-lat/0001032} and $8$ \cite{arXiv:0910.3216} theories one consistently finds about twice smaller
values, and at $N_f=12$ these splittings were not observed at all \cite{arXiv:0910.3216,Deuzeman:2010gb}. In view of the trend just discussed, as well as because of the two transition observed in \cite{Cheng:2011ic}, we expect 
 the chiral and deconfinement lines to separate, perhaps already
 at $N_f>4$ or so. 

\subsubsection{Sensitivity to the fermionic periodicity conditions.}
Introducing an arbitrary phase in the periodicity condition of the ``valence quarks", one can switch the fermionic zero mode between the dyons: this has been demonstrated using artificial configurations for calorons, e.g. in
\cite{Bruckmann:2003ag}. 
There is significant literature covering efforts efforts to understand the difference in lattice gauge configurations below and above $T_c$.
A paper presenting interesting results on the  Dirac eigenvalue spectrum  in the $SU(3)$ quenched and unquenched
ensembles is that by Bilgici et al 
\cite{Bilgici:2009tx}. Its brief  summary:\\
(i) at $T>T_c$ the Dirac spectrum has a well-determined $gap$ $G$ (no eigenvalues inside $\lambda < |G|$), growing approximately linearly  
\be G\sim (T-T_c), T>T_c \ee in the quenched case, till at least about 2$T_c$.\\
 (ii) if arbitrary (twisted) boundary conditions are used for (valence) fermions, by a phase $\phi=2\pi z$ in a periodicity condition,
 they seem to be irrelevant below $T_c$ but change the results drastically above it.   $\avg{\bar\psi \psi}$, or th density of eigenvalues at zero,
 seem to have a simple dependence on the angle 
 \be |\avg{\bar\psi \psi}| \sim c_1(T)+c_2(T) \cos\phi \ee with only one harmonics and positive coefficients $c_1,c_2$.
 For the holonomy values shifted above $T_c$ by $\pm 2\pi/3$ the phase of the cosine is shifted accordingly.
 \\
(iii)  As a result
 antiperiodic fermions $\cos{\phi}=-1$ have a density touches touches zero at $T_c$ and develops a gap, restoring chiral symmetry.
 The periodic fermions, with $\cos{\phi}=1$, never do so and their
  $\avg{\bar\psi \psi}$ grow indefinitely above $T_c$.

\subsubsection{ Chiral restoration in different Polyakov phase sectors.}
The previous issue is strongly related to lattice observations that
subensembles of quenched configurations at $T>T_c$ with {\em different Polyakov phases}  show different spectra of the Dirac eigenvalues and chiral parameters.
In SU(3) there are two sets, one with $\avg{P}$ real and another with the phase $e^{\pm i 2\pi/3}$.
 For example, Fig 1 of \cite{Gattringer:2001yu} shows that these
 two different subensembles  not only have different
eigenvalues but also drastically different ``participation ratios" (a degree of mode localization
on the lattice). 
This phenomenon gives us the opportunity to study more than one
set of holonomies in one simulation.

\subsubsection{ 
The spectral gaps at $T>T_c$ versus the fermion periodicity conditions.}  
The gap opening is clearly observed for antiperiodic fermions but not for periodic ones, see \cite{Bilgici:2009tx}. 
An explanation based on dyon-antidyon classical correlation at high $T_c$ was offered in 
\cite{Bruckmann:2009ne}.
Similar studies based on quenched configurations have been extended to adjoint fermions in 
Ref. \cite{Bilgici:2009jy}. Like for fundamental quarks, the antiperiodic adjoint fermions show a clear
gap opening above the chiral transition, while the periodic ones do not.

\subsubsection{ Identifying the topological objects via their fermionic zero modes.}
In configurations with say exact Q=1 one can locate one exactly zero mode
and see the location of the corresponding eigenvector is space-time. 
Using such lattice configurations  Gattringer and Schaefer \cite{Gattringer:2002tg}
have observed that while the eigenvector does indeed locate a single ``topological lump,"
its position and quantum number depends on the boundary phase and jumps at certain values, resembling what happens with the different types of 
constituent dyons inside classical caloron solutions. Such techniques allow
locating all kinds of dyons and potentially study their correlations/interactions.

At $T\sim T_c$, with massless fermions and restored chiral symmetry, all configurations
with nonzero $Q$ are absent, and the Dirac eigenvalues get gapped.
What are the states ``at the gap," with the {\em lowest}
Dirac eigenvalues at $T>T_c$?
As demonstrated in \cite{Gattringer:2002gn}, those have $two$ topological lumps, confirming
the picture of the paired 
 instanton-antiinstantons \cite{Ilgenfritz:1988dh}. 
As we will argue below, in the language of dyons, these molecules  are more complicated, with $2N_c$ dyons of all kinds and certain Abelian charge distribution. Therefore, 
now one should   use  this method again, to look at Abelian-projected charges, clarifying their structure further.

More recently, Bruckmann et. al. \cite{arXiv:1105.5336} looked at the fermionic states at the chiral gap at 2.6 $T_c$ for quenched SU(2) gauge simulations.
They observe that the corresponding eigenstates are well localized and correspond to a strongly modified local value of the
Polyakov line.  They have shown that the number of such objects vastly exceeds the density of the isolated topological charges
deduced from topological susceptibility, ruling out an ideal instanton gas as their source. 
They also commented, at the end of the section on topology, that their data ``do not exclude" configurations in which the topological charge cancels, like
 the instanton--anti-instanton molecules to be discussed.

\section{Dyon Interactions}

\subsection{Classical Interaction}

As is well known, 
 classical interaction of the dyons $inside$ one of the sectors -- self-dual or anti-self-dual --
are absent, as they are protected by the Bogomolny-Prasad-Sommerfeld (BPS) bound.
Although it is clear how this works  from the explicit solutions \cite{Lee:1997vp,Kraan:1998sn},
we will still discuss it in the dilute limit, as our starting point. 

The question of Higgs topology and monopole interaction has been addressed in many referances 
(e.g. \cite{Manton:1977er,Goldberg:1978ee,Magruder:1978br}). Recall that
 the usual Higgs field has 
to go to zero at the monopole center because there is no preferred 
direction of color there.  
 However with our ''Higgsing'' by 
the Polyakov loop, which is an element of $SU(2)$, and 
its value  at infinity [which defines an unbroken $U(1)$ direction],  the Polyakov loop can be viewed
as $L\in SU(2)/U(1)= S^2$ mapping. 
 At points without definite color direction 
the Polyakov loop  takes a value of $L(\vec x_0)=\pm \bm 1$: thus, 
 two types of dyons. Indeed,  the effective ''Higgs`` field at the dyon's center
it is going to zero for $M$ type and to $A_4=\pi \hat \phi\cdot \vec\tau$ for the ``twisted" $L$ dyon, where $\hat \phi$ is a unit color vector (see below).

For a caloron --the $L-M$ pair in SU(2) -- one can see from quantum numbers that
both the electric and magnetic charges are opposites, so they should both create attraction.
Another long-distance force is Higgs mediated:  because dyons of $M$ and $L$ types have 
different value of the $A_4$ at their centers this turns out to be repulsive. Furthermore, it 
\emph{exactly cancels} the attractive Coulomb forces, as required by BPS.

Now we consider  dyon-antidyon pairs, starting with $M\bar{M}$ (which do not have a temporal twist, i.e.
dyons completely static in time). We take them at a large  distance $d>>1/v$ compared to 
 the size of their cores.
  Inside some ball around the dyon (antidyon) of radius $r_0$ such that $1/v<<r_0<<d$ 
the field strength can be written as a small deviation from self-duality due to the other dyon, 
i.e. of the order $1/d$ .
 Outside of these balls the ``Higgs'' field is given by 
$|A_4|=v-1/r_1-1/r_2$, 
where $r_{1,2}$ are distances from the dyon and antidyon, so as to conform to the expected asymptotic formulas 
for dyon and antidyon (see e.g. \cite{Diakonov:2009ln}).

For a single dyon the Higgs field can be written as 
\begin{equation}
 \vec A_4=h(\bm r) \hat \phi\;,
\end{equation}
where $\hat \phi=\vec A_4/|A_4|$. Asymptotically $h(\bm r)=v-1/r$. An ansatz
that properly describes two dyons would have to obey the condition that
asymptotically $|\vec A_4|\approx v-1/r_1-1/r_2$. However the color direction  is a gauge choice that can be chosen arbitrarily at each point. Insisting that
 the Higgs points are in one color direction at some large sphere (gauge combing),  one
then has to introduce Dirac strings, as the gauge transformation cannot be made
single valued. We do not specify this gauge choice, as we only deal with the action, which is gauge invariant.

That being said, we expect that the influence of the other dyon will change the
$h$ function by introducing an additive Coulomb term near the core of the first
dyon, i.e. if $r_1<<r_2$ we have
\begin{equation}
 H(\bm r_1,\bm r_2)\approx (h(r_1)-1/r_2)\hat \phi\;
\end{equation}
where $r_1$ and $r_2$ are distances from monopoles to the point of observation.
An analogous relation holds when $r_2<<r_1$.

Now we determine the action of the system of two dyons, writing the action as an integral over three regions
\begin{multline}
 S=\frac{\beta}{2}\left(\int_1 \Tr(F^2)d^3x+\int_2\Tr(F^2)d^3x\right.\\\left.+\int_{outside}\Tr(F^2)d^3x\right)
\end{multline}
where $1$ and $2$ denote the regions around the dyon and antidyon respectively. Inside these regions we assume
self-duality (anti--self-duality) up to some small correction of the order of $1/d$ inside the cores, i.e.
$D_iA_4=\mp\frac{1}{2}\epsilon_{ijk}F^{jk}+f_i$, where $f_i$ is the field strength deviation from self-duality 
induced by the other (anti)dyon and is of order
$f=o(1/d)$. 
\begin{align}
 S_1&=\beta\int_1 d^3x\left(\Tr(D_iA_4) \epsilon_{ijk}F^{jk})+ \frac{1}{2}\Tr f_i^2\right)\approx
\nonumber\\&\approx \beta\int_1 d^3x\;
\partial_i (A_4 B^i)+o(1/d^2)=\nonumber\\
&=4\pi\left(v-\frac{1}{r_0}-\frac{1}{d}\right)+o(1/d^2)\;,
\end{align}
where the integration is over a ball of radius $r_0$ centered around the first
dyon. 

We pause to comment on similar expression in the case when we have a single
dyon. One can integrate on the surface at infinity, which would just simply
yield $4\pi v \beta$, i.e. the usual mass of a dyon. However, it is more
instructive to divide the region of integration into a small ball of radius
$r_0$ and the rest. The small ball is a total derivative and yields the
contribution $4\pi (v-1/r_0)\beta)$ to the action. Then we can write the action
as 
\ba
 S_{single \, dyon}=4\pi(v-1/r_0)\beta \\ \nonumber
 +\beta\int_{\text{outside}} d^3x\; \frac{1}{2}(E^2+B^2);\;,
\ea
where we integrate over the volume outside the ball. However in this region the
fields are Abelian and behave in an expected way. We can  write
$E_i=\frac{q_e\hat r_i}{r^2}\hat \phi$ and $B_i=\frac{q_m \hat
r_i}{r^2}\hat\phi$ and the outside integral is $4\pi/r_0$. The sum of the region inside and
outside gives the expected result $4\pi v$.

Coming back to the case of two dyons, we include the region around the antidyon
and get
\begin{equation}
 \frac{1}{2}\Tr\int\limits_{cores}d^3x\; F^2=8\pi\left(v-\frac{1}{r_0}-\frac{1}{d}\right)
\end{equation}
showing how one dyon has been
modified by the presence of the other one.
Next we write the integral outside as the sum of electric and magnetic
parts, i.e.
\begin{align}
 & E_i=\frac{\hat{r}^i_1}{r_1^2}+\frac{\hat {r}^i_2}{r_2^2}\\
 & B_i=\frac{\hat{r}^i_1}{r_1^2}-\frac{\hat {r}^i_2}{r_2^2}
\end{align}
which, upon integration, give the expected interaction $4\pi/d$ for the electric and $-4\pi/d$ for
the magnetic, and they cancel. Also there are two self-energy terms which are
given by $4\pi/r_0$ for electric and magnetic field separately, which cancel
the $1/r_0$ contribution to the inside-the-sphere integration. 

Notice that even though the electric and magnetic fields cancel outside the
cores, the dyon-antidyon system still attracts, due to the modification of their mass by the
presence of the other (anti)dyon. Thus there exists a long-distance classical Higgs-based attraction 
for the $M\bar{M}$ pair.

One  can equally well consider $L\bar{L}$ dyons with the twist, i.e. with core time dependence. 
The only modification is that
then the contribution to the action of the core is given by $4\pi\bar v$, where
$\bar v=2\pi-v$. Also for a purely self-dual sector, the interaction of the $L$ and $M$ dyon is
seen to cancel. This result is well known from the original works \cite{Kraan:1998sn,Lee:1998bb},
but here we see that since the ``Higgs'' asymptotic looks like $v-1/r_{M}+1/r_{L}$ (see \cite{Diakonov:2009ln}), the ``Higgs'' interaction is repulsive, which exactly \emph{cancels} the attractive forces of the $E$ and $B$ field. Note also that due to this effect of ``dyon mass renormalization'' we expect that if the $L$ dyon has a fermionic zero mode (which as we will see in the next subsection,
depends on this holonomy), it is renormalized by the presence of the other $M$ dyon. This 
was observed in the original papers by van Baal et. al. and it followed from the exact zero mode expression: we just identify  its 
physical origin. 

\subsection{Fermion-induced interactions}

Fermionic interactions between dyons are central for this paper. They are induced by the presence of fermionic localized modes facilitated by the time-dependent twist in the gauge fields. The fermions introduce the fermionic determinant $\det\Dslash$ factor in the partition function.
If both the dyon and anti-dyon are in isolation (at large distances), they have zero modes,
which leads to a vanishing determinant: thus, such configurations are excluded from the
ensemble. Obviously, at finite $r$ the modes are nonzero, and therefore
 the dyon-antidyon pair is attracted due to the fermions.
 
As was done for the instantons,  one can look at the Dirac operator in the basis of the localized 
zero modes of the individual  $L$ and $\bar L$ dyons. The matrix element of the $\Dslash$ zero mode
 between two of those we denote as $T_{IJ}$, where the indices run through all dyon and antidyon zero modes. Since the Dirac operator in the chiral basis connects between the left and right fermions only, the diagonal elements are all zero, and only blocks $T_{I\bar I}$ and $T_{\bar I I}$ remain, where now $I$ runs through the dyon zero modes only, and $\bar I$ through antidyon zero modes only.

It is quite clear that in the case of a dyon-antidyon pair, since $\det T=-|T_{I\bar I}|^2$, we have that 
$V_{eff}=-\ln(T_{L\bar L}(r_{L\bar L}))$. Since the matrix element is approximately $T_{L\bar L}\sim e^{-M r/2}$, 
where $M$ is the ``holonomy mass'' of the fermion, the resulting effective potential between dyons
is linearly 
confining. 

While in this paper we will focus only on the zero modes of the fundamental quarks, we would
like to mention some important works on the adjoint fermions, which naturally appear in the
supersymmetric context.  For periodic compactification the corresponding
index theorem is discussed in \cite{Poppitz:2008hr} (see also citations therein). An
extensive discussion of the zero modes for the periodic and
antiperiodic adjoint fermions can be found in \cite{GarciaPerez:2009mg}.

\subsubsection{Fermionic zero mode for arbitrary periodicity condition}
As we mentioned in the Introduction, one should not confuse the ``particle dyons" 
and ``instanton" (or self-dual) dyons:
while mathematically similar they are associated with quite different physics. The ``particle" dyons 
are time-independent 3-d objects and their
fermionic zero mode are 3-d normalizable and time independent. For the  ``instanton dyons" we need 4-d 
normalizable zero modes. Index theorems associate the latter ones
with the topological charge $Q$ of the 4-d theory: thus  an instanton ($Q=1$) consisting of $N_c$ dyons possess only  one of those, which need to be
somehow shared between the dyons.
Another important technical difference is  induced by 
the fact that in QCD-like theories the role of the Higgs boson is played by $A_4$ rather than scalars or pseudoscalars.  As a result, the corresponding gamma matrix for Higgs is $\gamma_0$ (rather than 1 or $\gamma_5$): this makes the interaction with the Higgs
chirally symmetric.  

 Let us generalize the fermionic (-) and  bosonic (+)  boundary conditions to a general 
 ``anyonic" phase
\be  \psi(\beta)=exp(-i\phi) \psi(0) \ee
which should be satisfied by a normalizable solution of the Dirac equation
\be
\Dslash \psi=0\;,
\ee
containing the gauge field  in the hedgehog ansatz
\footnote{for explicit form of function $\mathcal H,\mathcal A$ see the Appendix}
\begin{align}
&A_i^a=\epsilon_{aij}\mathcal A\hat r^j\;,\\
&A_4^a=\mathcal H\hat r^a\;.
\end{align}

We use the gamma matrix convention $\gamma^i=\sigma^2\otimes \sigma^i$, $\gamma^4=\sigma^1\otimes \bm 1$, 
so that $\gamma_5=\gamma^1\gamma^2\gamma^3\gamma^4=\sigma^3\otimes\bm 1$, and do the calculation for the right spinor component. The Dirac equation then reads
\be\label{eq:Dirac}
(\sigma^\mu)_{\alpha\beta} (D_\mu)_{AB} (\psi_R)_\beta^B=0\;,
\ee
where we explicitly wrote the Dirac indices $\alpha,\beta$ and color indices $A,B$, and where $\sigma^\mu=(\bm 1, i\sigma^i)$. Now we ansatz \cite{Jackiw:1975fn,Shnir:2005xx}\footnote{This ansatz is inspired by the
fact that the fermions have two $SU(2)$ indices, one spin and one color. Since we consider the fundamental color representation (i.e. both 
are $\bm 2$ representations), these
indices can combine into $\bm 2\otimes\bm 2=\bm0\oplus\bm3$, which is precisely the decomposition considered in the text. See e.g. \cite{Shnir:2005xx}}
\be
\psi^{A}_\alpha=\alpha(r)\epsilon_{A\alpha}+\beta(r)[(\bm{\hat r}\cdot \bm \sigma)\epsilon]_{A\alpha}\;.
\ee

We consider the matrix 
\be
\eta_{A\alpha}=-\psi_{\beta}^A\epsilon_{\beta \alpha}
\ee
and ansatz 
\be
\eta=\alpha_1(r)\bm 1+\alpha_2(r)\hat{ \bm{r}}\cdot \bm\sigma
\ee
The rule of acting with a color and the spin sigma matrices on this object is such that we multiply 
by a color matrix $\tau$ from the left, and if we multiply by a spin matrix $\sigma$, then we multiply 
from the right, and put a minus sign, i.e.
\be
\sigma \psi=\eta\epsilon \sigma^T=-\eta \sigma \epsilon\;.
\ee
The fermion density is given by
\be
{\psi^*}_\alpha^A\psi_\alpha^A=\Tr(\eta^\dagger\eta)
\ee

We now plug the ansatz into \eqref{eq:Dirac} and obtain the following two equations
\begin{subequations}\label{eq:Dirac1}
\begin{align}
&\alpha_1'(r)+\frac{\mathcal H+2\mathcal A}{2}\alpha_1+\frac{\phi}{\beta}\alpha_2=0\;,\\
&\alpha_2'(r)+\left(\frac{\mathcal H-2\mathcal A}{2} +\frac{2}{r}\right)\alpha_2+\frac{\phi}{\beta}\alpha_1=0\;.
\end{align}
\end{subequations}
where we have assumed $\psi_R\propto e^{-i\phi t/\beta}$, i.e. that the Fermion has arbitrary periodicity 
condition in the imaginary time direction. 

\begin{figure}[htbp] 
   \centering
   \includegraphics[width=0.5\textwidth]{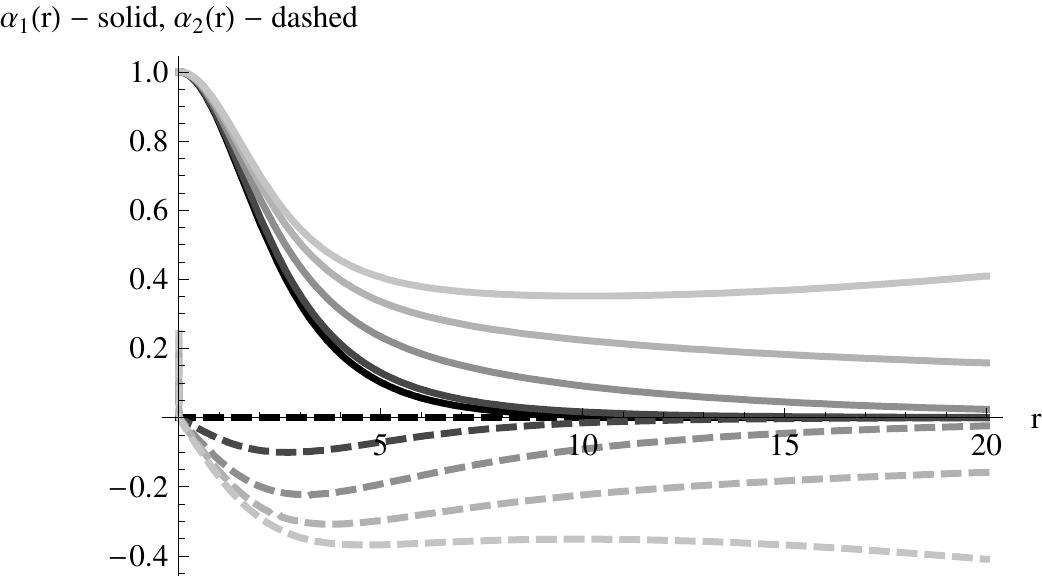} 
   \caption{ The profile of (unnormalized) zero mode components $\alpha_{1,2}$ (the solid and the dashed curves) as a function of the distance from the dyon.
   We show four 
different values of $\phi=0,0.2 v/\beta,0.4 v/\beta, 0.5 v/\beta, 0.55 v/\beta$. Note that the zero mode delocalizes 
at $\phi=0.5 v/\beta$}
   \label{fig:zero modeprof}
\end{figure}

We solve the Eq. \eqref{eq:Dirac}  in the Appendix, and the result is shown in Fig. \ref{fig:zero modeprof}. 
However here we can easily look at the asymptotic behavior of the solution, i.e. 
when $\mathcal H(r\rightarrow \infty)=v$ and $\mathcal A(r\rightarrow\infty)=0$, then
\begin{align}
&\alpha_1'(r)+\frac{v}{2}\alpha_1(r)+\frac{\phi}{\beta}\alpha_2(r)=0\;,\\
&\alpha'_2(r)+\frac{v}{2}\alpha_2(r)+\frac{\phi}{\beta}\alpha_1=0\;.
\end{align}
This equation is easily solvable by taking the substitution $\alpha_\pm=\alpha_1\pm\alpha_2$ we get
\be
\alpha_\pm=e^{-\left(\frac{v}{2}\pm \frac{\phi}{\beta}\right)r}\;.
\ee

In order for the solution to be normalizable, we must have that both $\alpha_\pm$ vanish at infinity. 
This is only possible if $|\phi|<|v|\beta/2$

\subsubsection{The zero mode hopping}

The formulas derived in the previous section explain why the zero mode ''jumps`` from one dyon to the other with the change 
in the periodicity condition of the fermions.
 However, there we assumed a static $M$-type dyon, which has  the zero mode \begin{equation}
\psi_M\sim e^{-\left(\frac{|v|}{2}-|\phi|\right)r}\;,
\end{equation}
where all dimensionful quantities are expressed in units of $\beta=1/T$. Now it is quite clear that $\phi\in (-v/2,v/2)$ will preserve the 
normalizability of the solution. But for the phase in the interval $v/2<\phi<\pi$,
one can use the equation for the zero mode on the 
$L$ dyon instead.  To do so one must first go to the static gauge in which $v\rightarrow \bar v=2\pi-v$, and then the 
 zero mode has good asymptotic behavior
\begin{equation}
\psi_M\sim e^{-\left(\frac{|\bar v|}{2}-|\phi|\right)r}\;.
\end{equation}
Furthermore, because one has to apply a time-dependent gauge 
transformation to reinstate $A_4\rightarrow v$ at infinity in the form $U=\exp(i\pi t\vec\tau\cdot\hat r)$, 
the fermions (in the fundamental representation) gain the desired antiperiodicity.
One can  replace $\phi$ by $\bar \phi=\pi-\phi$, or
\begin{equation}
\psi_M\sim e^{-\left(\frac{|\bar v|}{2}-|\bar \phi|\right)r}\;.
\end{equation}
Now we insert $\bar v=2\pi-v$ and $\bar \phi=\phi-\pi$, where $\phi$ and $v$ are the (true) holonomy and 
periodicity of fermions. We assume $\phi\in (v/2,\pi)$, so that the exponential term becomes
\begin{equation}
 \frac{|\bar{v}|}{2}-|\bar \phi|=\pi-\frac{v}{2}-\pi+\phi=\phi-\frac{v}{2}\;.
\end{equation}
and the fermion zero mode becomes normalizable for $\phi\in(v/2,\pi)$.

Finally, we explore the region $\phi\in (\pi,2\pi-v/2)$. Then we argue that the zero mode on the $L$ dyon is 
still normalizable. Indeed, the exponent now becomes
\begin{equation}
 \frac{|\bar{v}|}{2}-|\bar \phi|=2\pi-\frac{v}{2}-\phi\in (0,\pi-v/2)\;.
\end{equation}
Therefore,
\begin{align}
\psi_{M}&\sim e^{-(|v|/2-|\phi|)r} \hspace{35pt} \mbox{$\scriptstyle-|v|/2<\phi<|v|/2$}\nonumber\\
\psi_L&\sim\left\{\begin{array}{l l}
	     e^{-(|\phi|-|v|/2)r} & \mbox{ $\scriptstyle |v|/2<\phi<\pi$}\\
	      e^{-(2\pi -|v|/2 -|\phi|)r} & \mbox{ $\scriptstyle\pi<\phi<2\pi-|v|/2$}
            \end{array}
 \right. 
\end{align}

\subsection{The fermionic interaction among the clusters}
As it has already been mentioned above, on general grounds one expects the $L\bar{L}$ clusters to repel each other, 
as say atoms do, because of the Pauli principle. In this section we will show how it works using the first 
``nondiagonal" diagrams in which fermion exchange between such
clusters takes place.

The fermion determinant will be of the form
\begin{multline}
\det \Dslash=|T_{1\bar 1}|^2|T_{2\bar 2}|^2\dots |T_{N\bar N}|^2\\-T_{1\bar 2}T_{\bar 22}T_{2\bar 1}T_{\bar 11}T_{3\bar 3}\dots T_{N\bar N}\dots\;,
\end{multline}
where $T_{1\bar 1}=-T_{\bar 11}^*$. We can interpret the first term as 
a two-loop diagram, with the fermion hopping from one dyon to the antidyon and back, for each of the pairs $1\bar 1,2\bar 2$ etc. The second term is  interpreted as a one-loop process, in which the fermion is hopping from $1$ to $\bar 2$ and then from $\bar 2$ to $2$ to $\bar 1$ back to $1$. The determinant is the sum of all such terms, with the appropriate minus sign to enforce the Fermi statistics.  Note that the zero resulting from the cancellation between hose diagrams means that a dyon-antidyon pair will be repelled by another dyon-antidyon pair at certain distances. 

It is simple to see that chiral symmetry is necessarily restored if the ensemble is made of dyonic pairs. Then the determinant is dominated by the near-diagonal matrix elements $T_{I\bar I}$, where indices $I,\bar I$ go over dyons and anti-dyons respectively, which are the closest pairs, i.e. $T_{I\bar J}<<T_{I\bar I}$, for $\bar J\ne \bar I$. Then the spectrum of the Dirac operator is exactly solvable and is given by $\lambda_I=\pm|T_{I,\bar I}|$. Therefore very small eigenvalues will be given by very small matrix elements $T_{I\bar I}$ of the dyonic pairs. This matrix element is small only if the respective dyons are very far away (much further than the range of the transition element, i.e. $1/\bar v$). But since the overall configuration is anyway weighted by the determinant to the power of the number of flavors $N_f$, these configurations are strongly suppressed, and the density of such eigenvalues goes to zero at small eigenvalues, implying that, by the Banks-Casher relation, chiral symmetry is restored.

An $2N\times 2N$ matrix of the form
\be
M=\begin{pmatrix}
0& A\\
-A^\dagger & 0
\end{pmatrix}
\ee
has a determinant equal to
\be
\det M= |\det A|^2 
\ee
which is always positive. 

Let us now consider the fermionic determinant in the basis of fermionic localized modes for 2 dyons and 2 antidyons, labeled with indices $1,2$ and $\bar 1, \bar 2$ respectively. 
\be
\det \Dslash= |T_{1\bar 1}T_{2\bar 2}-T_{1\bar 2}T_{2\bar 1}|^2\;,
\ee
As an example, consider a configuration of dyons and antidyons placed on a rectangle of dimensions $a\times b$. A little thought will immediately reveal that if we put them on a square such that as we go around we have $1\bar 1 2\bar 2$, the determinant vanishes when $a=b$, or in other words clusters $1-\bar1$ and $2-\bar 2$ are mutually already infinitely repelling when $b=a$. However we can make them come closer if we orient them on the rectangle as $1\bar 1\bar 2 2$, i.e.  dyons $1-\bar 1$ and $2-\bar 2$ form independent clusters with distance $r_{1\bar 1}=r_{2\bar 2}=a$ and $r_{1\bar 2}=r_{2\bar 1}=\sqrt(a^2+b^2)$. Then the repulsion for small $b/a$ will be
\be
V_{eff}=-\ln(\det \Dslash)\sim -\log[T(a)T'(a)\frac{b^2}{a}]\;,
\ee
where $T(r_{I\bar I})=T_{I\bar I}$. Quite clearly the effective potential becomes infinite when $b\rightarrow 0$, making an effective repulsive core for two dyonic clusters.

To discuss this further we introduce the diagrammatic interpretation of the determinant  viewed as a
sum over all fermionic loops. Let us view a determinant in some basis of local fermionic states. This need not (and in
fact most certainly is not) be an eigenbasis of the Dirac operator. The basis vectors we denote as $\psi_n$, which are
localized at $\bm x_n$. Since this basis is not an eigenbasis of the Dirac operator $\Dslash$, we have that
\be
i\Dslash\psi_n=J_n\;,
\ee
where $J_n$ is a spinor resulting from the action of the Dirac operator. However we may view $J_n$ as a source of our
basis states, and interpret $\psi_n(y)=\int d^4y \Delta(x-y) J_n(y)$, where $\Delta(x-y)$ is the fermionic propagator. Then $\Dslash$ taken between two states $\psi_{m,n}$ will be
\be
(i\Dslash)_{mn}=\int d^4x d^4y J_m^\dagger(x)\Delta(x-y)J_n(y)\;.
\ee
Therefore we can view the matrix elements $T_{I\bar I}$ as being integrated propagators from one source to another.

The diagrammatic description of the determinant $A$ in the upper right quadrant is then (we assume $N_f=1$ in what follows)
\[\det A=\pair\cdot\pair\cdots\pair-\paircrs \pair\cdots\pair+\text{(all possible perm)}\]
where the black circle represents the dyon and gray the anti-dyon.

The complex conjugation can be viewed, instead of fermion going from the dyon to anti-dyon, as the opposite propagation
of an anti-fermion going from an anti-dyon to dyon.
The pictorial representation of the determinant is
\begin{multline}
\det\Dslash= |\det A|^2=\det(A)\det(A^\dagger)=\\
=\left(\pair\cdot\pair\cdots\pair
-\paircrscc \paircc\cdots\paircc+\dots\right)\\
\times \left(\paircc\cdot\paircc\cdots\paircc
-\paircrs \paircc\cdots\paircc+\dots\right)=\\
=\pairloop\pairloop\cdots\pairloop-\pairloopf\pairloop\pairloop\cdots\pairloop
\\
+\pairlooph\pair\cdots\pair+\dots
\end{multline}

\begin{figure}[!htb]
\includegraphics[width=\columnwidth]{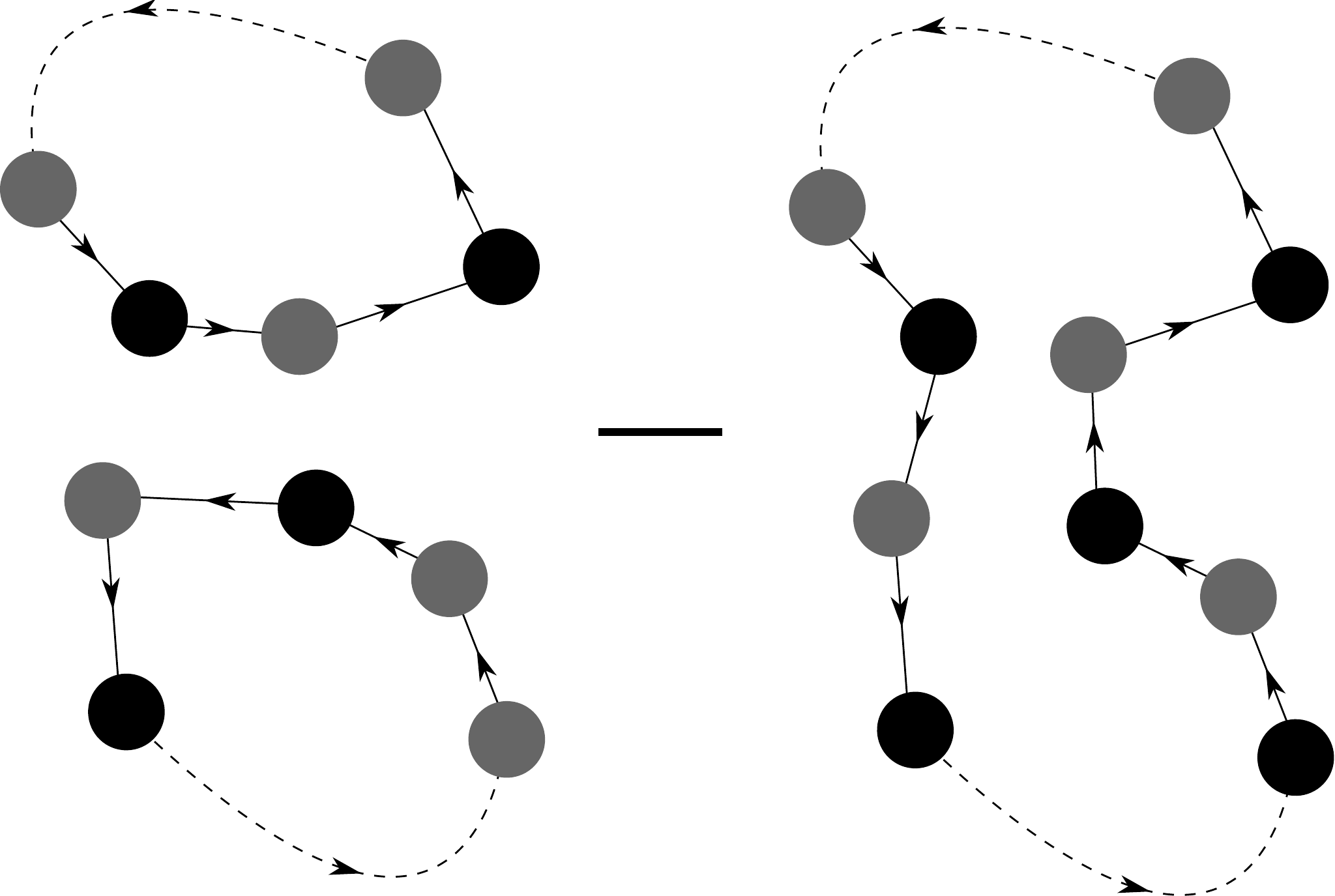}
\caption{ (Color online)A graphical interpretation of the weight in the background of the dyons. Note that the relative minus sign
will always induce a repulsion between a dyon-antidyon pair. } \label{fig:repulsion} 
\end{figure}

The interpretation of this expansion is then straightforward. The determinant can be interpreted as loop
diagrams connecting the various dyons which carry a zero mode. It is quite evident from this diagrammatic expansion
that every diagram of two loops will have a similar diagram with the opposite sign where the two loops join
via a small channel (see Fig. \ref{fig:repulsion}).

Let us now think how many pairs can we make from $N$ dyons and $N$ anti-dyons. After a bit of thought we can see that it is $N!$. All terms like that involve permuting in the above expression two positions of two (anti)dyons, and, because it requires two exchanges, the sign remains the same. The determinant will be integrated over the moduli space of the dyon-antidyon pairs, and so all of these kind of permutation can be taken to be the same. Therefore the fist term will contribute with a factor of $N!$ to the determinant.

The second term is a bit more tricky. We proceed in the following way. Let us consider $k$ 4-plets (a 2 dyon-antidyon pairs) which facilitate one loop. The number of ways we can have one 4-plet is $(N(N-1))^2$, because we can pick two dyons in $N(N-1)$ ways, and the same for anti-dyons. For $k$ such 4-plets we have the expression
\be
N_{k-4plets}=\frac{[N(N-1)(N-2)\dots (N-2k+1)]^{2}}{2^k k!}\;,
\ee
where the $k!$ factor is present to compensate for all possible interchanges of all $k$ 4-plets. The rest of dyons and antidyons can be made into pairs, and since there are $N-2k$ leftover (anti)dyons we can do this in $(N-2k)!$ ways. Combining this with the above factor we get
\be
N^k_{4-plets}=\frac{N!^2}{2^k(N-2k)! k!}\;.
\ee 

Now we consider the integrals over such matrix elements. Generally we will have that for the arrangement of pairs the integral over the moduli space (assuming flat moduli space metric -- an assumption justified only in dilute phase) will be given by
\be
N!\left(c_0\frac{V}{m^3}\right)^N\;,
\ee
where $c_0$ is a constant which depends on a particular form of the matrix element $T_{I\bar I}$. In other words we have written
the integral $\int (T(r))^2d^3r=c_0/m^3$, making it explicit that the effective volume of the integration measure is given by $1/m^3$, i.e. integrating over pairs will introduce a volume given by the range of their matrix elements $T_{L\bar L}\sim e^{-mr}$, and an overall volume corresponding to the integration over the center of mass of each pair. Notice that we also put in the factor $N!$, which is an overall degeneracy of the integral.

In the case of $k$ 4-plets and $N-2k$ pairs, we have
\be
\frac{N!^2}{2^k(N-2k)!k!} \left(c_1\frac{V}{m^9}\right)^{k} \left(c_0\frac{V}{m^3}\right)^{N-2k}
\ee
The constant $c_1$ appears in the quadrupole integral over a loop which includes 4-(anti)dyons (2 dyons and 2 antidyons). The effective volume
is now $1/m^9$, with a single, overall, volume factor.
Therefore the partition function can be approximated as
\be\label{eq:partfunctapprox}
Z\approx N! \left(c_0\frac{V}{m^3}\right)^{N}\sum_{k=0}^{N/2}\frac{(-1)^k N!}{2^k(N-2k)!k!} \left(\frac{c_1n}{c_0^2m^3 N}\right)^{k}
\ee

where $n=N/V$. Rewriting

\be
Z\approx N! \left(c_0\frac{V}{m^3}\right)^{N}\sum_{k=0}^{N/2}\frac{(-1)^kN!}{(N-2k)!k!} \left(\frac{A}{N}\right)^{k}
\ee
where the factor \be A=(c1/2c0^2)n/m^3  \label{eqn_A} \ee
 The coefficient $c1/c0^2$ is just a numerical factor, and it depends on how fast
the matrix element falls of with distance, but it does not depend on overall coefficient in front of the transition element.

The sum above is can be computed by using the identity
\[
H_N(x)=\sum_k^{N/2}(-1)^k\frac{N!}{k!(N-2k)!}(2x)^{N-2k}
\]
Then the partition function becomes
\be
Z\approx N!N^{N/2}\left(\frac{c_0\sqrt{A}}{n m^3}\right)^NH_N\left(\frac{1}{2}\sqrt\frac{N}{A}\right)
\ee

The approximation can only be valid if $A$ is small, therefore $1/A$ is large. We can employ an asymptotic form
of Hermite polynomial for large $N$ in the following form \cite{gabor:orthogonalpolynomials} 
\begin{align}
&e^{-x^2/2}H_N(x)\approx \\&\approx\frac{2^{N/2-3/4}\sqrt{N!}}{(\pi N)^{\frac{1}{4}}\sqrt{\sinh\phi}}e^{(N/2+1/4)(2\phi-\sinh 2\phi)}
\end{align}
for $x=\sqrt{2N+1}\cosh\phi$. Another asymptotic series assuming $x=\sqrt{2N+1}\cos\phi$ leads to oscillatory 
asymptotics, which is clearly a good indicator that our approximation of just including 4-plet diagrams is invalidated, 
and that higher order diagrams become important.
For such an estimate the chiral symmetry will be restored for $A<1/8$, or $n/m^3<c_0^2/2c_1$,
where $n$ is the density of one species of dyons.

\begin{figure}[htbp] 
   \centering
   \includegraphics[width=\columnwidth]{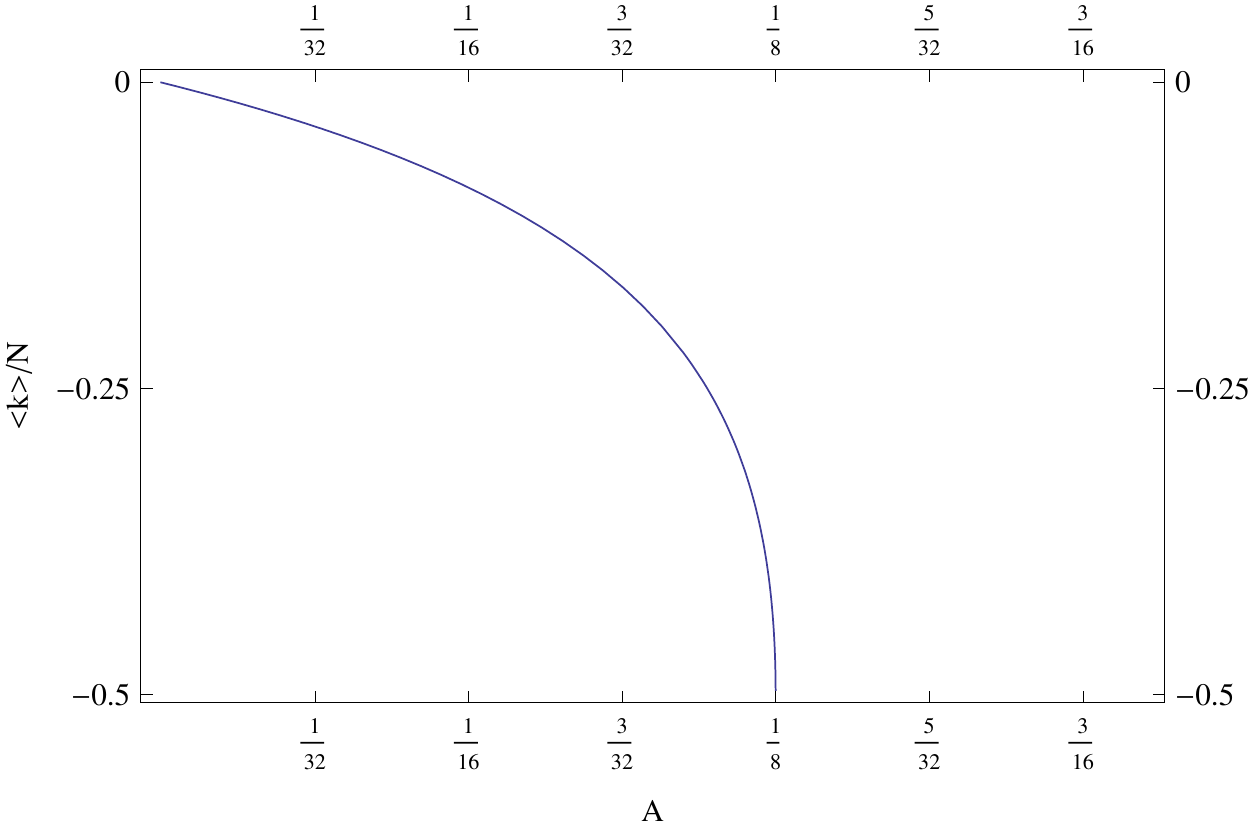} 
   \caption{$\avg k/N$ as a function of parameter $A$ defined in (\ref{eqn_A}) .}
   \label{fig:avgk}
\end{figure}

Finally going back to \eqref{eq:partfunctapprox} for a moment, we see that if we look for the quantity 
$\avg{k}=A\partial(\ln Z)/\partial A$, 
  each coefficient will have a factor of
$k$ in front. (Note that this is \emph{not} and average number of $4$-plets: each configuration has
arbitrarily many 4-plets, 6-plets, etc.)
In Fig. \ref{fig:avgk} we show $\avg{k}/N$ as a function of parameter $A$. Notice the abrupt change as we approach the critical value $A=1/8$, which we take as an indication of chiral symmetry breaking.

\subsection{Bosonic One-Loop Interactions and Electric Screening}
   The basic physics of the electric screening can be explained most simply
   following the original derivation by one of us \cite{Shuryak:1977ut} (in the Coulomb gauge). If some object
   possessing  a nonzero $A_4$ is immersed into
   a quark-gluon plasma,  those quanta from the heat bath are scattered
   on it. The simplest diagram comes from the quartic term in the gauge Lagrangian, $\sim g^2 A_m^2 A_4^2$ which couples  the heat bath gluons directly to
  square of $A_4$, but there are also other diagrams contributing to the forward  
scattering amplitude. The result was the expression for the QCD Debye mass\footnote{
In 1977, when QCD was only 4 years old, the main finding was its positivity,
which ensured \emph{screening} of a charge, as opposed to the antiscreening by the vacuum loops:
thus the ``plasma" name. } 
\be M_D^2= g^2 T^2 (N_c/3+N_f/6) \label{eqn_Debye}\ee
for massless quarks and gluons: incorporation of their
effective masses is straightforward.

The next relevant paper is by Pisarski and Yaffe (PY) \cite{Pisarski:1980md}
who calculated the one-loop action of the calorons (the finite-T instantons). Its main part
is the following correction to the instanton action
\be \delta S_{PY}= { 2\pi^2 \rho^2 \over g^2}  M_D^2 \ee
 where $\rho$ is the instanton radius.  The first factor in this expression  comes from
the (4d) dipole moment of the instanton, and the second  from the forward scattering amplitude of the thermal plasma quanta on it, for derivation see \cite{Shuryak:1982hk}. 
  This term is only present in the plasma phase,  at $T>T_c$,  as only in this
case there exist thermal quarks and gluons undergoing this scattering. 

Going forward to calorons at nonzero holonomy, a corresponding one-loop effective action
has been computed by Diakonov, Gromov, Petrov and Slizovskiy (DGPS) 
\cite{Diakonov:2004jn}. The caloron  is now a superposition of the M and L dyons,
separated by distance $r_{ML}$, and the basic expression from which the effect
comes is the following integral
\be  <A_4^2>\sim \int d^3r ({1 \over r_L}-{1 \over r_M})^2=4\pi r_{ML}+... \label{eqn_DGPS_scr}\ee
where $r_L,r_M$ are distances from the dyon centers to the observation point $\vec r$.

 (The
dots stand for corrections due to a finite dyon size: the  Coulombic $A_4$ is true only at large distance. Note also that
at large $r$ the integral converges because the integrand is $\sim r_{LM}^2/r^4$. 
This term comes again from the quartic term in the action, in which two gauge potentials are the $A_4$ 
of the instanton and two others belong
to the thermal gluons.)
 
Thus, the electric screening effectively generates the confinement of two dyons, with a potential linearly depending on the $LM$ separation:
\be V_{scr}^{LM}= r_{LM}{  2 \pi M_D^2 \over  T g^2}  \ee
At the zero holonomy this result matches the PY answer  because of the ``instanton size relation" \be \pi\rho^2 T= r_{ML}  
\label{eqn_rho_r_relation} \ee
which, so to say, relates the 4-d dipole of the instanton to the 3-d dipole of the dyon pair. The second factor is still the same
thermal integral producing $T^2$.
One obvious effect of fermions is that
 a generalization of  the DGPS result to theory with
  the generalized Debye factor as in (\ref{eqn_Debye}). 
As a result, the electric screening effect ensures $LM$ ``binding" into a finite-size 
 object with the (inverse) size 
\be <r_{ML}>^{-1} \sim T(N_c/3+N_f/6) \ee

Note also, that for $L\bar{L}$ or $M\bar{M}$  pairs with the $same$ electric charge,
there will be a plus in the integral (\ref{eqn_DGPS_scr}) above and thus the effect becomes repulsive and the integral diverges: it needs to be regulated by some opposite charges.
For molecules consisting of all 4 ($L,M,\bar{L},\bar{M}$) dyons, to be discussed shortly, the screening potential is
\be  V_{scr} = { M_D^2 \over 2 T g^2} <A_4^2>|_{LM\bar{L}\bar{M}}\ee
in which the $A_4$ now contains all 4 Coulomb contributions. This integral is of course convergent because of total the zero charge
of the molecule. 

\begin{figure}[htbp] 
   \centering
   \includegraphics[width=2in]{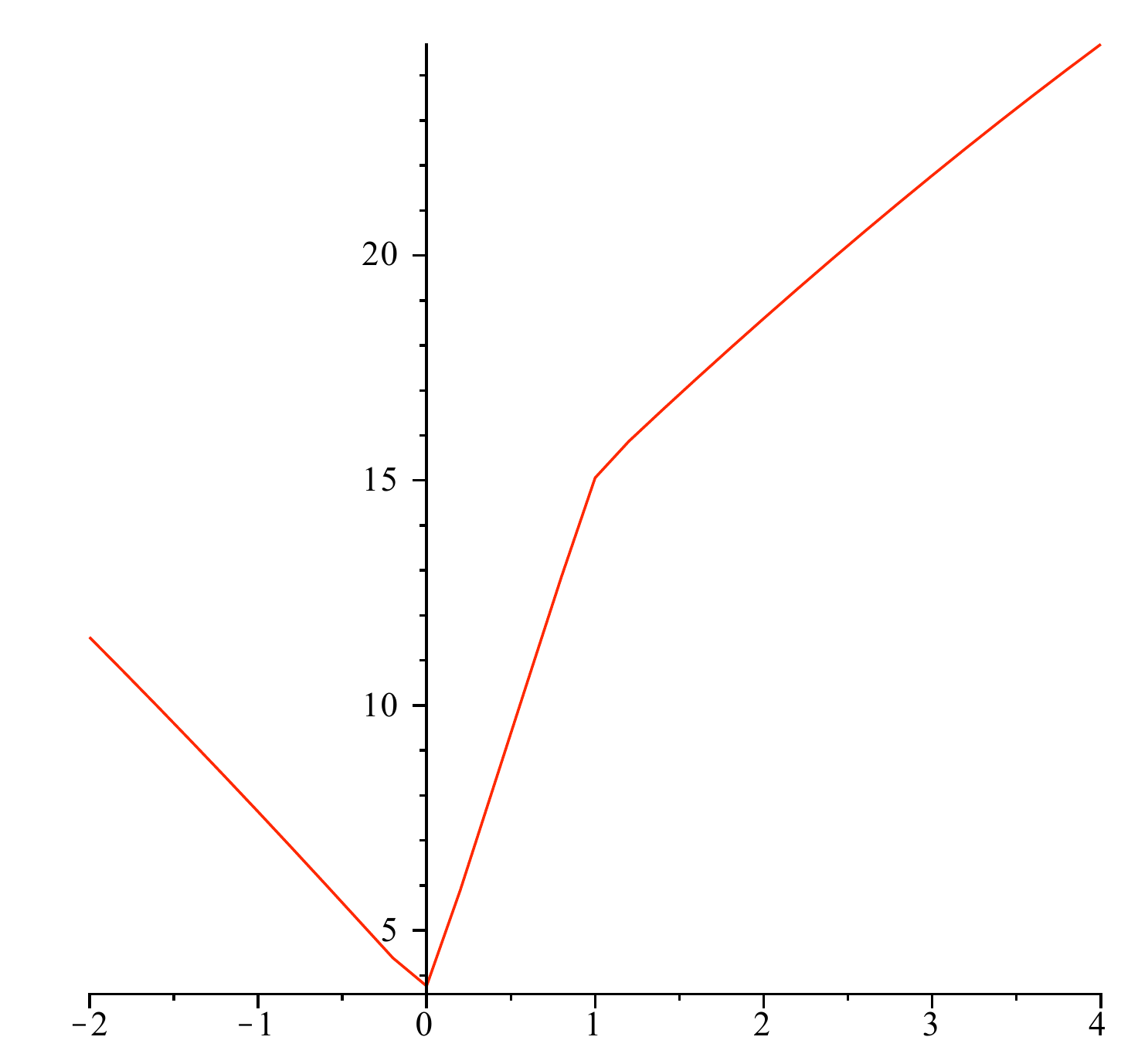} 
   \caption{ (Color online)
   The integral (\ref{eqn_4dyons}) proportional to the potential for  the $M$ (or $\bar{M})$ dyon  created by the electric screening as a function of its position.
   The charge-2 $L\bar{L}$ pair is assumed to be in the origin, and the companion dyon is placed at the point 1.}
   \label{fig_M_pot}
\end{figure}

 Let us give an example
of the electric multi-dyon screening potential created in this case. We will later see that the direct fermionic interaction  binds $L\bar{L}$ pairs stronger than the $LM$ interaction. 
Therefore, 
for simplicity one can ignore the $L\bar{L}$ cluster size and put them 
at the same point, the origin. Another simplification appears if one puts $M,\bar{M}$ and $L\bar{L}$ on one line. The integral (\ref{eqn_DGPS_scr})  changes to
\be  \int d^3r \left( {2 \over r}-{1 \over r_M} -{1 \over r_{\bar{M }} } \right)^2 
\label{eqn_4dyons}
\ee
The corresponding potential is shown in Fig.\ref{fig_M_pot}. As one can see, like for DGPS case, the potential consists of linear segments, but is now deformed away from the companion dyon. (Note,
that it is not due to their Coulomb repulsion, which is also there but will be discussed in the next section.)

\section{Statistical Mechanics of Dyons}

\subsection{Statistical Mechanics of a Single Dyonic Molecule}

  The partition function for an instanton-antiinstanton molecule can be
recovered using known elements for each of the ingredients. Let us start
with dimensional considerations, valid at high enough temperatures.    
If the fermions are all massless, the overall power of the $T$-dependence of total molecular density
can be determined  from the known power of the $\Lambda_{QCD}$
in the instanton-antiinstanton measures, namely
\be n_{mol} = {log Z_{mol} \over V}\sim T^3 \left( {T \over \Lambda} \right)^{-{11 N_c \over 3}+ {4N_f \over 3}}
\label{eqn_mol_dens}
\ee
Typically the power in the second factor is large and negative, so this density rapidly decreases with $T$.  (It is so  except near the boundary of the asymptotic freedom domain where that power is getting small: we will   not discuss this region.) 

Translation from the dyonic to instanton language at the level of the 
moduli metric and partition function has been studied for self-dual and anti--self-dual
sectors by Diakonov et al \cite{Diakonov:2004jn}. Their  expression, in the  SU(2)  
case for pure gauge theory is 
\ba dZ_{LM} & = & d^3 r_{L}  d^3r_{M} T^6 2\pi C
\left({8\pi^2 \over g^2}\right)^4 \left( {\Lambda_{PV} e^\gamma_E \over 4\pi T} \right)^{22/3}  
 \nonumber \\
&& F_{D}(r_{LM}) e^{-V_{scr}(r_{LM}}) \ea
where $r_{LM}=|\vec r_m-\vec r_M | $, numerical constant $C=1.0314...$, and the scale parameter $\Lambda_{PV}$ is for the Pauli-Villars regularization \footnote{Note that he explicit numerical factors and the definition of the $\Lambda$ parameter are not really
important since they will change
in a theory with fermions.}
 The factor
 \ba F_{D}(r)= \left(1+ {v \bar{v} r \over 2\pi T} \right) \left( 1+vr  \right)^{{4v \over 3\pi T}-1} \left( 1+\bar{v}r  \right)^{{4\bar{v} \over 3\pi T}-1}  \nonumber \ea
is the correction  appearing due to 
non-zero holonomy. If the holonomy $v=0$ or antiholonomy $\bar{v}=0$,
in the expression above $F_D=1$ and it reduces 
to the well-known caloron measure,  using
the relation  (\ref{eqn_rho_r_relation}). In the limit of large dyon separation 
one may keep only the r-terms: note that it then becomes flat and r-independent as one would expect.
The screening potential for $LM$ pair is
\ba  V_{scr}(r) &=&{2\pi r \over  \pi^2 T}  \left[ \pi T (1-{1\over \sqrt{3} } )-v  \right]  \\ \nonumber
&& \left[ -\pi T (1-{1\over \sqrt{3} } )+\bar{v}  \right]  \ea
We have excluded
one more factor in the partition function of \cite{Diakonov:2004jn}
\be exp( -V {v^2 \bar{v}^2 \over  12\pi^2 T}) \ee
which does not depend on the calorons/dyons and is just a one-loop contribution to the probability to have the holonomy $v$
in the ensemble: it certainly should not be repeated twice.

\begin{figure}[t] 
   \includegraphics[width=7cm]{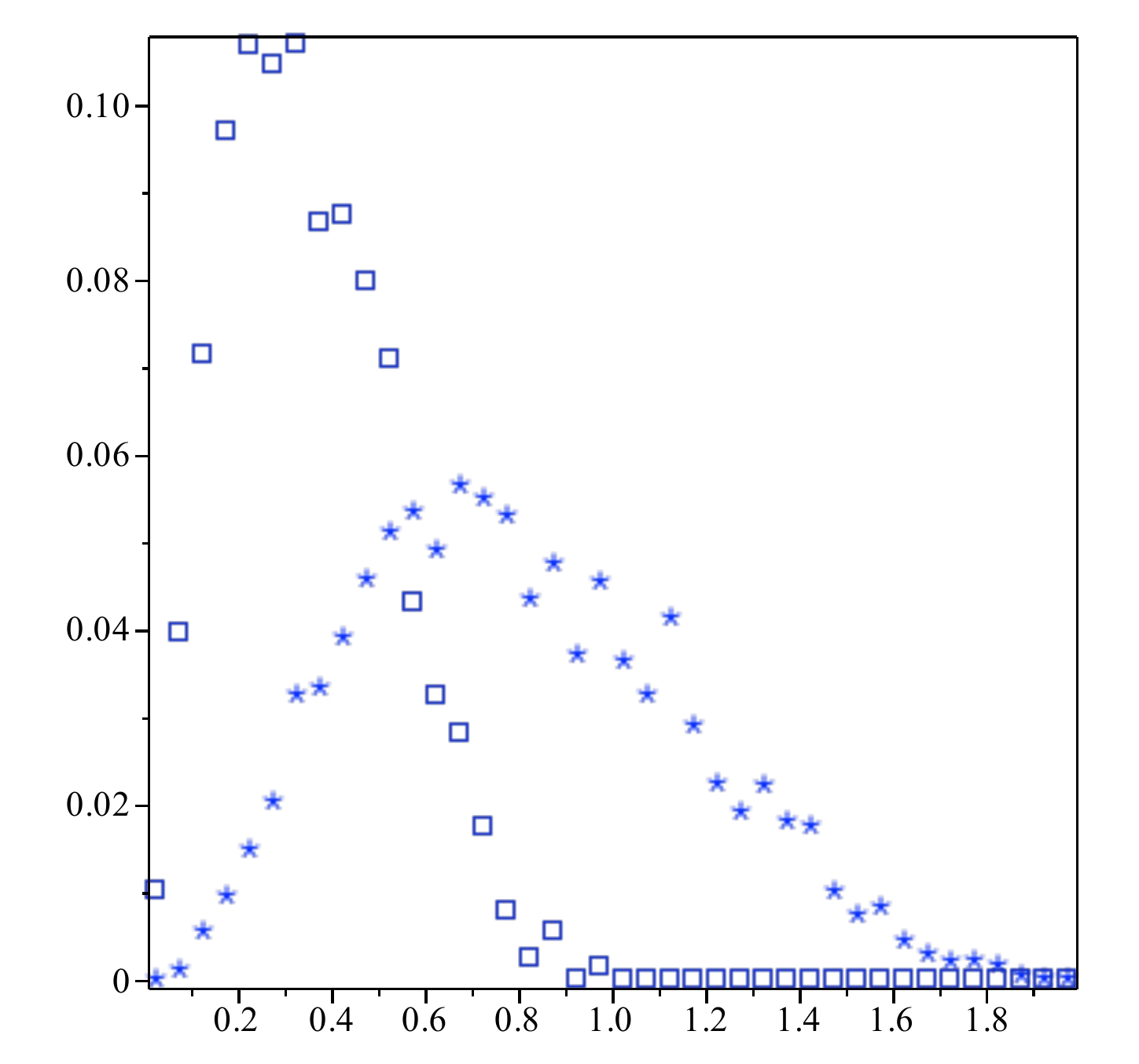}   
   \caption{ Histograms of the distributions of the distance between the $LM$ dyons (stars) and   $L\bar{L}$ (boxes)  dyons, for $N_c=2,N_f=4$ molecule, in the units of the Matsubara time $1/T$. The fermionic mass is taken to be $m_f=T$ and the holonomy $\nu=0.1$. }
   \label{fig_mol_distrib}
\end{figure}

 The same expression can be repeated for the for the $\bar{L},\bar{M}$ pair, and then the combined into the 4-particle partition function for a ``molecule".  Since, unlike 
 \cite{Diakonov:2004jn} ,  we are interested in the theories with fermions, we introduced 
extra factors which include that for zero modes as well as the nonzero mode part
\ba dZ_{mol}&=& dZ_{LM} dZ_{\bar{L}\bar{M}} \left[ {m^2 +| T_{IA}(r_{L\bar{L}})|^2  \over \Lambda^2 } \right]^{N_f} \\ \nonumber  && C(N_f)                                                     \left( \pi^2 r_{LM} r_{\bar{L}\bar{M}}\Lambda^4 \over T^2 \right)^{N_f/6}               e^{-{ V_{scr} } -V_{L\bar{L}} } \ea 
As discussed in the preceding section,  $ V_{screening}$ is defined by  the 4-particle expression for $A_4$
integrated over the volume. If one of the particle is going to large distances, the expression reduces to a dipole
and return the linear confinement result, preventing ``ionization" of a molecule.

 The bracket in the power of $N_f/6$ comes from the nonzero mode part of the fermionic determinant 
 calculated by 't Hooft. The power of $\Lambda$ in it corresponds to correct beta function of the theory with
 $N_f$ fermions, and is therefore fixed. Its dimension should be compensated by some parameters with the dimension of the
 distance, which in the case of a single instanton can only be its size $\rho$. 
 At finite temperature instantons lose 4-d spherical symmetry and another dimensional parameter -- $T$ appears, as well as a nonzero holonomy $v$.
Lacking explicit evaluation of the nonzero mode determinant,   we just used the instanton expression and
 the relation $\pi\rho^2T=r$ translating its size into the current language. At least for small-size instantons (small $r$) this should work. 
 As a result we  get factors $r^{3 N_f/2}$ in the measure, or 
  a  repulsive potential $\sim N_f log(1/r) $ trying to dissolve the molecule. Recall however that
  it only suppose to be true at small $r$, while at large $r$ the one-loop 
 electric screening effects generate an attractive potential linear in $r$ which would prevent it from happening.

We have introduced here the fermion mass $m$ in the Dirac operator, assuming it is the same for all flavors, for the normalization reasons\footnote{We also ignored small nondiagonal matrix elements of the type $m<I_1|I_2>$  resulting from non-orthogonality of the zero modes of different pseudoparticles.}. The term proportional to the masses is nothing else but a square of the independent
instanton and antiinstantons: and since their normalization have been already determined by 't Hooft, the flavor-dependent normalization constant $C(N_f)$ 
can be determined for  $\Lambda_{PV}$. 

If the fermion masses are set to zero, the fermions couple instanton to antiinstanton
via the overlap matrix element and four integrals over dyon positions produce three convergent integrals, while one 
remaining integral over the global position
 produces one factor of $V$, the box volume.

Even for the simplest case of $SU(2)$ color, when molecules contain 4 dyons, their
position space is already 12-dimensional.
 Therefore we used standard Metropolis algorithm to generate their 
statistical distributions. Fig.\ref{fig_mol_distrib} shows one typical example
of the output: in it we compare the distances between the $LM$ dyons (stars) with that of the 
 $L\bar{L}$ (boxes).  The latter are seen to be much tighter placed, forming a ``nucleus"
 of the molecule.

\subsection{Modelling the dyonic ensemble}
\subsubsection{Three  Molecular Models}

As a first step toward the understanding of the dyonic ensembles, and their role
in chiral symmetry breaking/restoration, we had formulated some simplified models.

For calculation purposes it is convenient for these models to treat the dyon density
\be n_d=n_L=n_M=n_{\bar{L}}=n_{\bar{M}} \ee
 (which is also the same as the instanton density $n_{inst}+n_{antiinstanton}$) as the basic dimensional quantity, providing the units of length $n_d^{-1/3}$. Using such length units we put $n_d=1$
 for a while, and will be expressing other dimensional quantities in these units. We will be
 working with traditional periodic boxes of some size $L\times L\times L$, with $L$ ``large",
 and thus put into such boxes  $N_d=L^3$ dyons of each kind.

For each configuration of these models we then calculate the fermionic
matrix $T_{ij}$, and calculate its eigenvalues. In this way we get part of the Dirac
spectrum built on the subspace of the dyon zero modes. 
Since antiperiodic fermionic zero
modes resign on $L,\bar{L}$ dyons only, fermionic part of the measure
ignores the $M,\bar{M}$ dyons. The matrix is thus
of the size $2N_d\times 2N_d$. For reasons of opposite chirality, two quarters of
the matrix, when both $i,j=L$ or $\bar{L}$ are zero, so fermionic hopping occurs only
from a dyon and anti-dyon. 

We assume the matrix element $T_{ij}$, where is given by
\be\label{eq:T}
T_{ij}=c \frac{e^{-M r_{ij}}}{\sqrt{1+M r_{ij}}}
\ee
where one of the indices counts $L$ dyons, and the other counts $\bar L$ dyons. The constant $c$ will be left undetermined, whereas the ``mass'' $M$ is given by
\be
M=\bar v/2=(2\pi-v)/2\;.
\ee

The form of the matrix element is not derived, but it is postulated as expected form for the matrix element of the Dirac operator in the zero mode basis. We introduce a regulator at $r=0$, so as to make deal with smooth distributions. However there will be an natural cutoff for how close the dyon and antidyon can get before they are part of a perturbative vacuum, which is roughly the size of their cores $1/M\sim1/\bar v$. Strictly speaking the definition of the distance at which the dyon-antidyon pair is irrelevant is defined as the distance at which it no longer supports a localized fermionic mode.

We proceed by three models:
\begin{enumerate}
\item The Random Gas Model
\item The Random Molecular Model
\item The Reweighted Molecular Model
\end{enumerate}

The simplest model-I is that of the ``Random Gas Model", in which all correlations between the dyons
are ignored and they are placed randomly.  The only parameter of the model is the fermion mass $M$ entering the matrix $T_{ij}$, to be
expressed in units of $n_d^{1/3}$. (In reality, both the dyon density and the holonomy,
defining $M$, will be function of the temperature $T$, but we prefer to study our models
in their parameter space before mapping some of the results  to lattice data, see below.)

In Figs. \ref{random_eigen}  and \ref{random_eigen2}  we show the results of such calculation. We use the box
of the size $6^3$ and thus 216 dyons of each kind,
and a range of fermion masses as indicated in the figure caption.
 The characteristic feature
of the  ``Random Gas Model" is a large peak near eigenvalues $\lambda \approx 0$.
Since the density of quasizero eigenvalues is proportional to the quark condensate
(Casher-Banks theorem), we conclude that this model provides large or ``enhanced"
chiral symmetry breaking. 
\begin{figure}[htb]
  \includegraphics[width=\columnwidth]{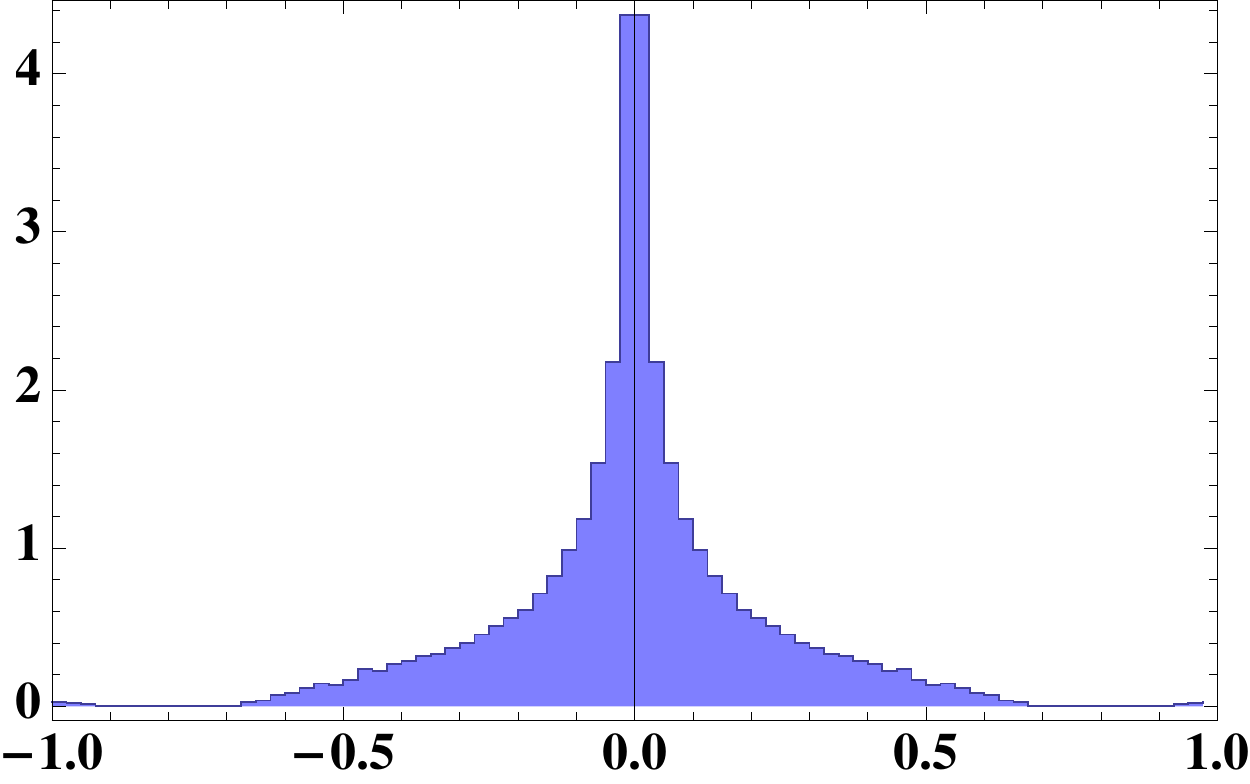}
\caption{ (Color online) The Dirac eigenvalue spectrum for the random dyon gas}
  \label{random_eigen}
\end{figure}

\begin{figure*}[htb]
  \includegraphics[width=\columnwidth]{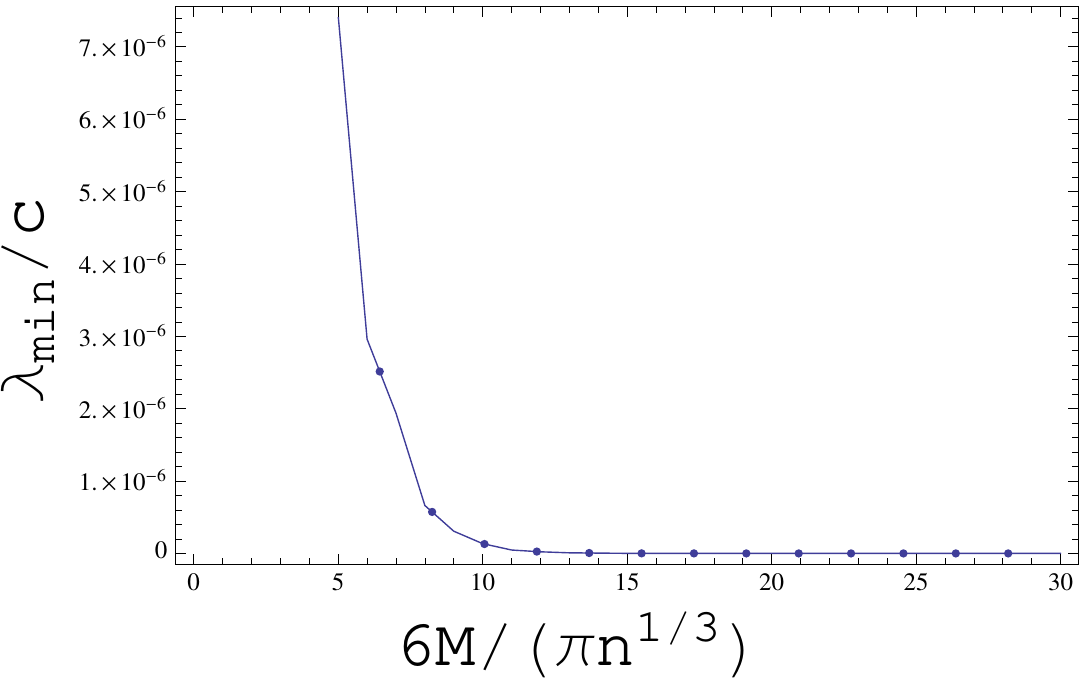} \;\includegraphics[width=\columnwidth]{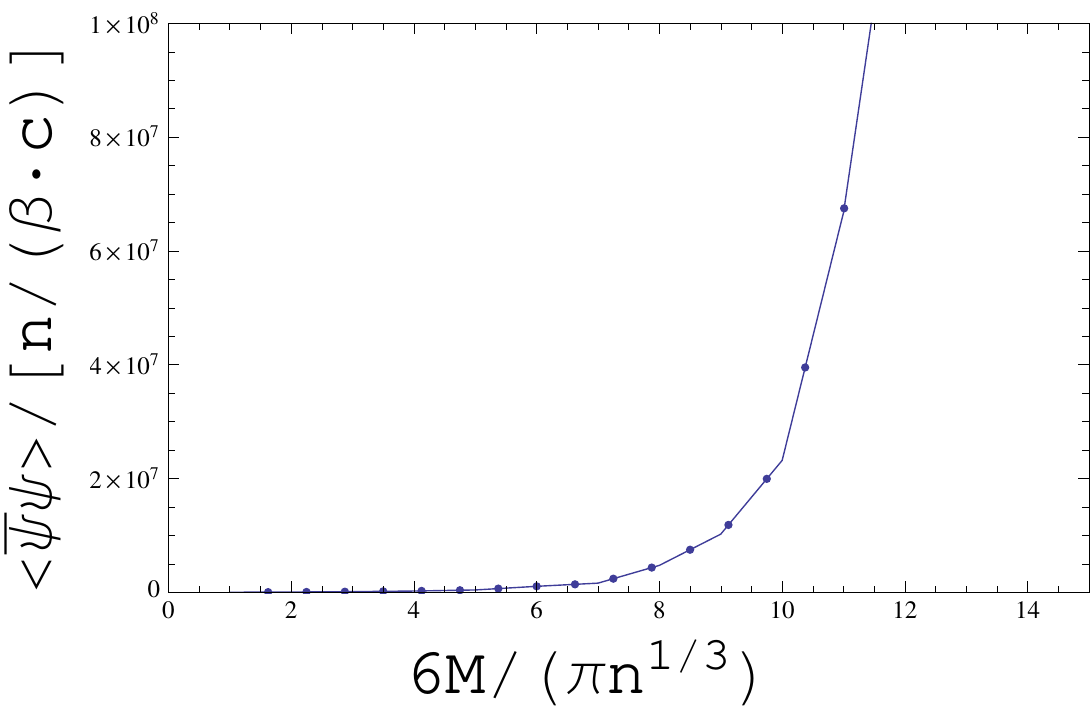}
\caption{ (Color online) The eigenvalue gap and the quark condensate for the random gas model, as a function of  the ``holonomy mass'' M.}
  \label{random_eigen2}
\end{figure*}

Our 2$^{nd}$ model is the ``Random Molecules Model", in which we include pair
correlations between   $L,\bar{L}$ dyons. 
 As we discussed above,
we expect significant attraction between those, of classical (Higgs-related) origin, as well as the fermion-induced dion-antydyon confinement. We focus here on the fermionic interaction. As the number of fermionic zero modes
grows, proportionally to the number of flavors $N_f$, we expect that at large enough
$N_f$ the molecule mean size $R_m$ decreases as $\sim 1/N_f$. We model the vacuum
as being composed out of random molecules. The following distribution of the size of molecules is used:
\be
D_{mol}(r)=N r^2\left(\frac{e^{-M r}}{\sqrt{1+Mr}}\right)^{2N_f}
\ee
where $r^2$ is due to the measure of the dyon-antidyon coordinates, and $N$ is the normalization constant. The above form is inspired by a weight $(\det T_{ij})^{N_f}$, for a dilute molecule ensemble. Average molecular size will roughly be given by $R_m=1/(N_f m)$. 

At this stage we have ignored any interaction between the molecules, placing them randomly with
random orientations.

\begin{figure*}[htbp] 
   \centering
   \includegraphics[width=0.3\textwidth]{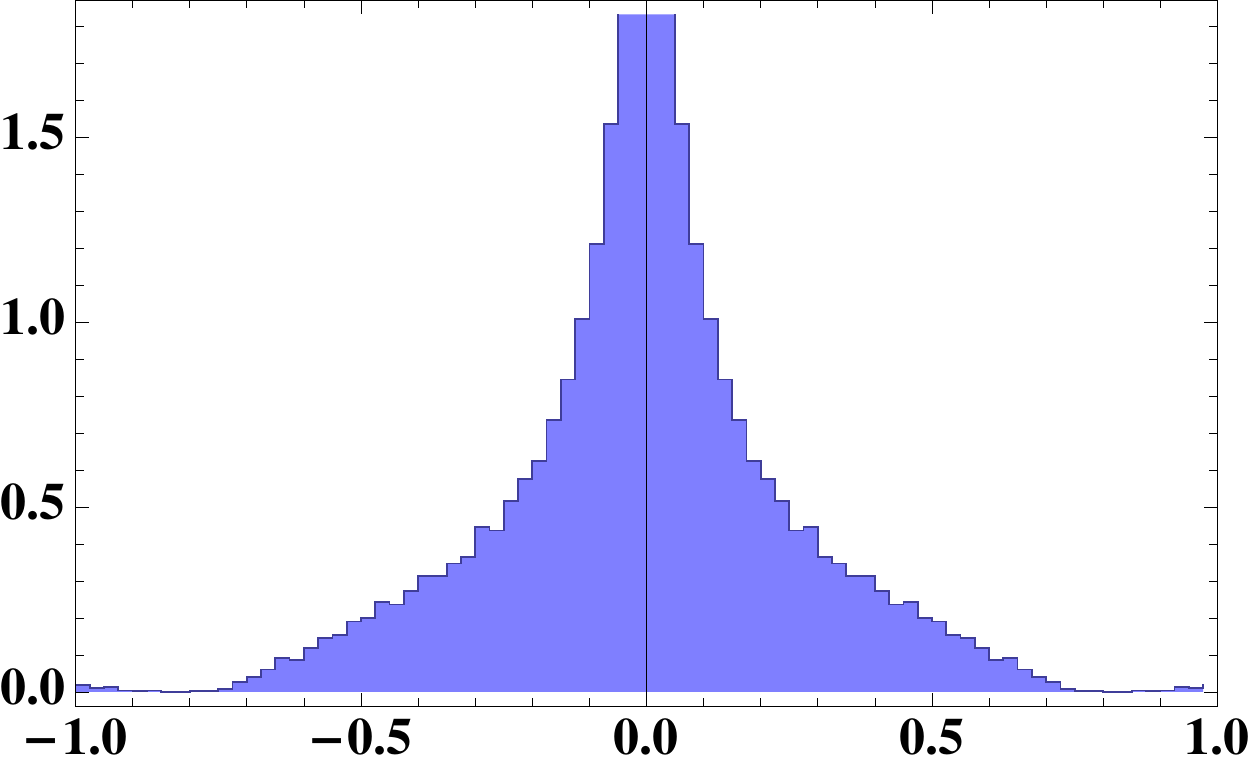}\;
   \includegraphics[width=0.3\textwidth]{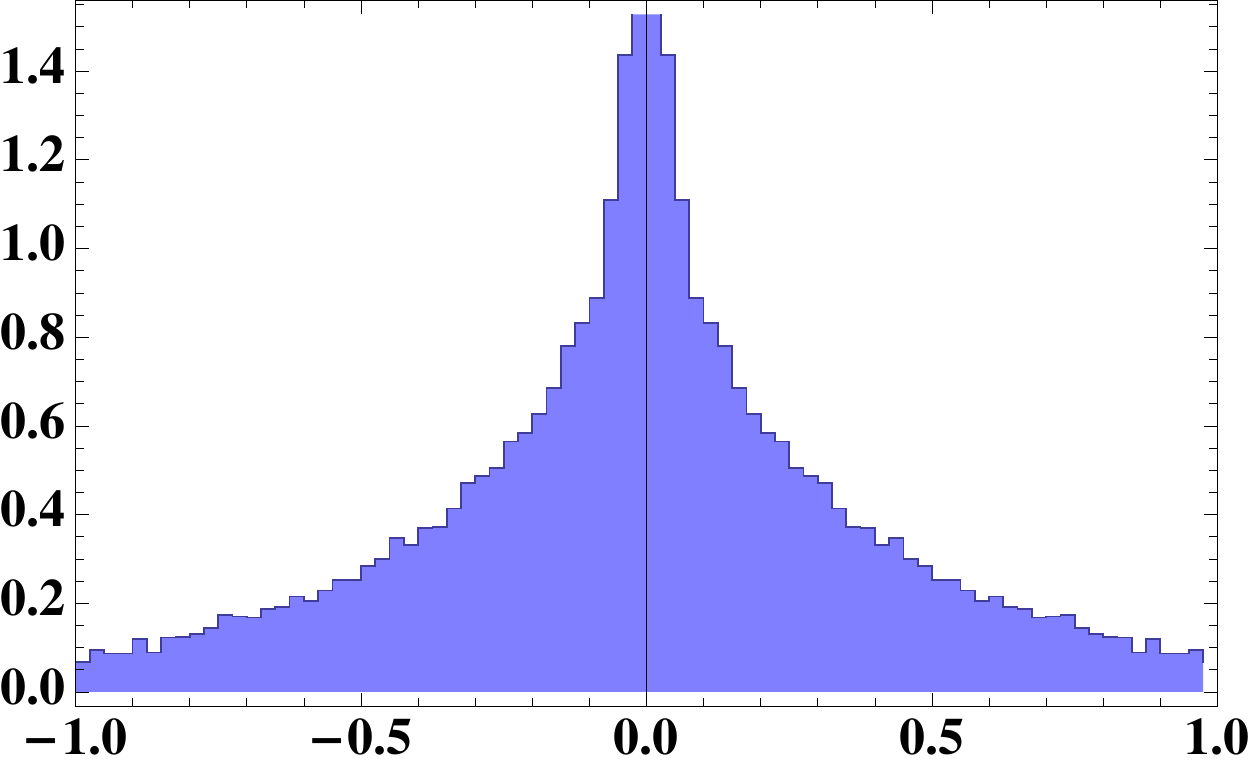}\;
   \includegraphics[width=0.3\textwidth]{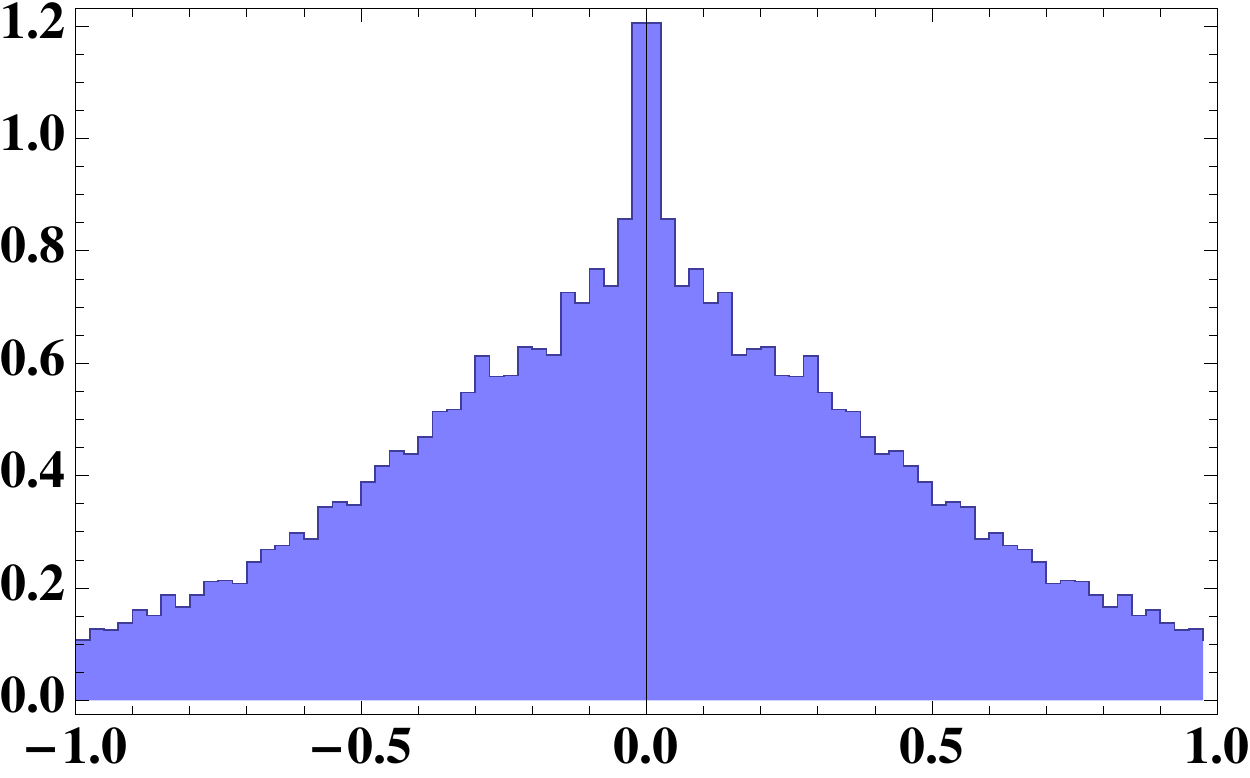}\\
   \includegraphics[width=0.3\textwidth]{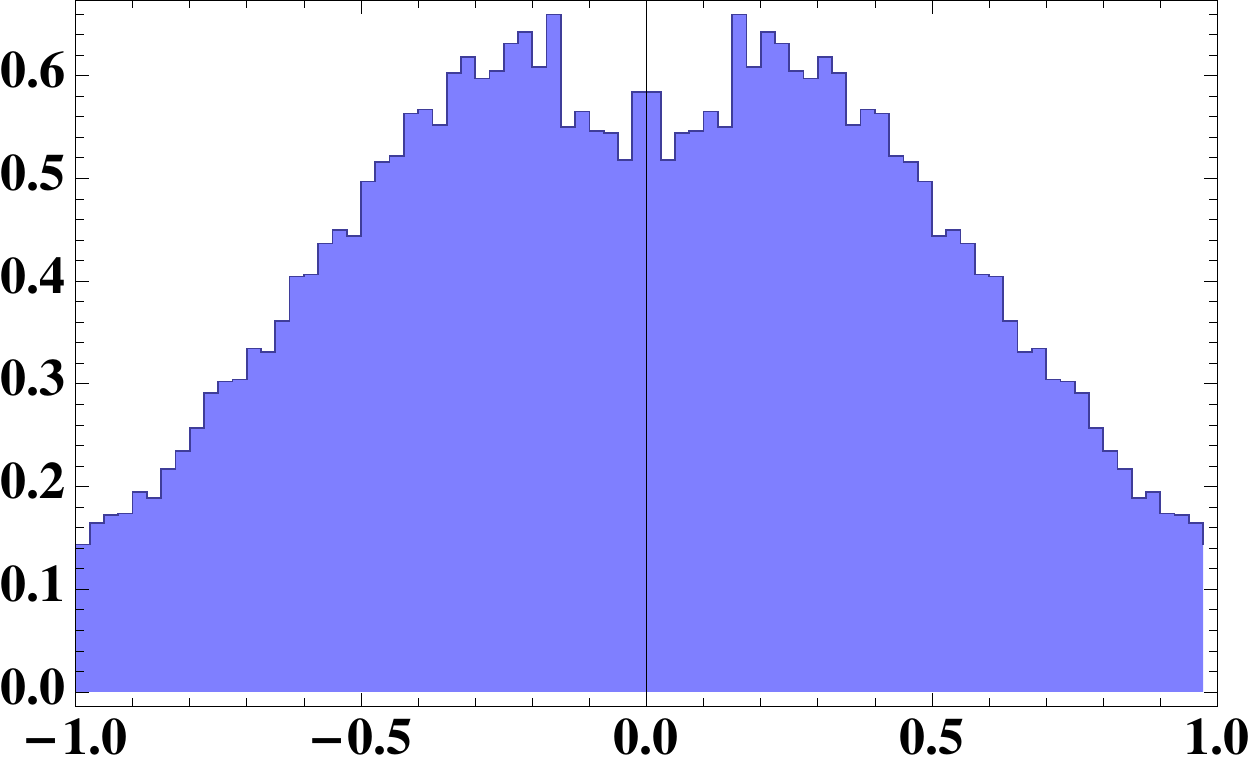}\;
   \includegraphics[width=0.3\textwidth]{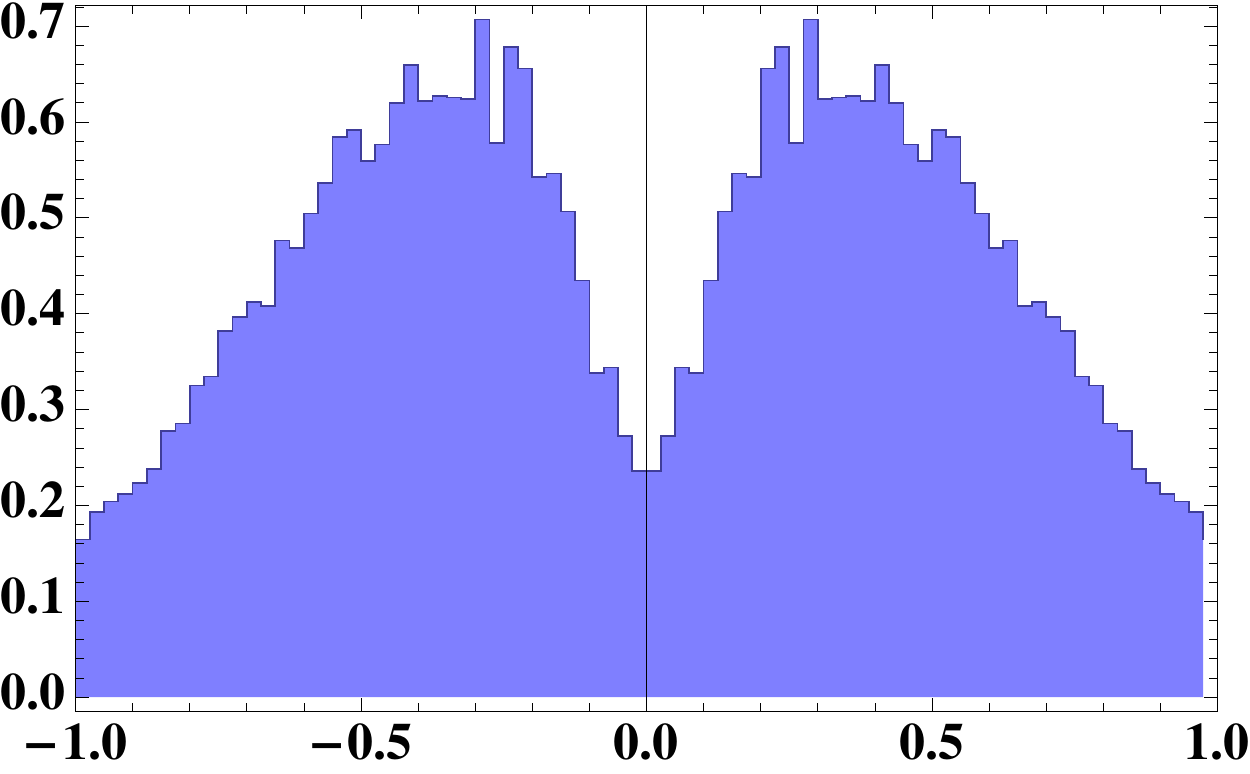}\;
   \includegraphics[width=0.3\textwidth]{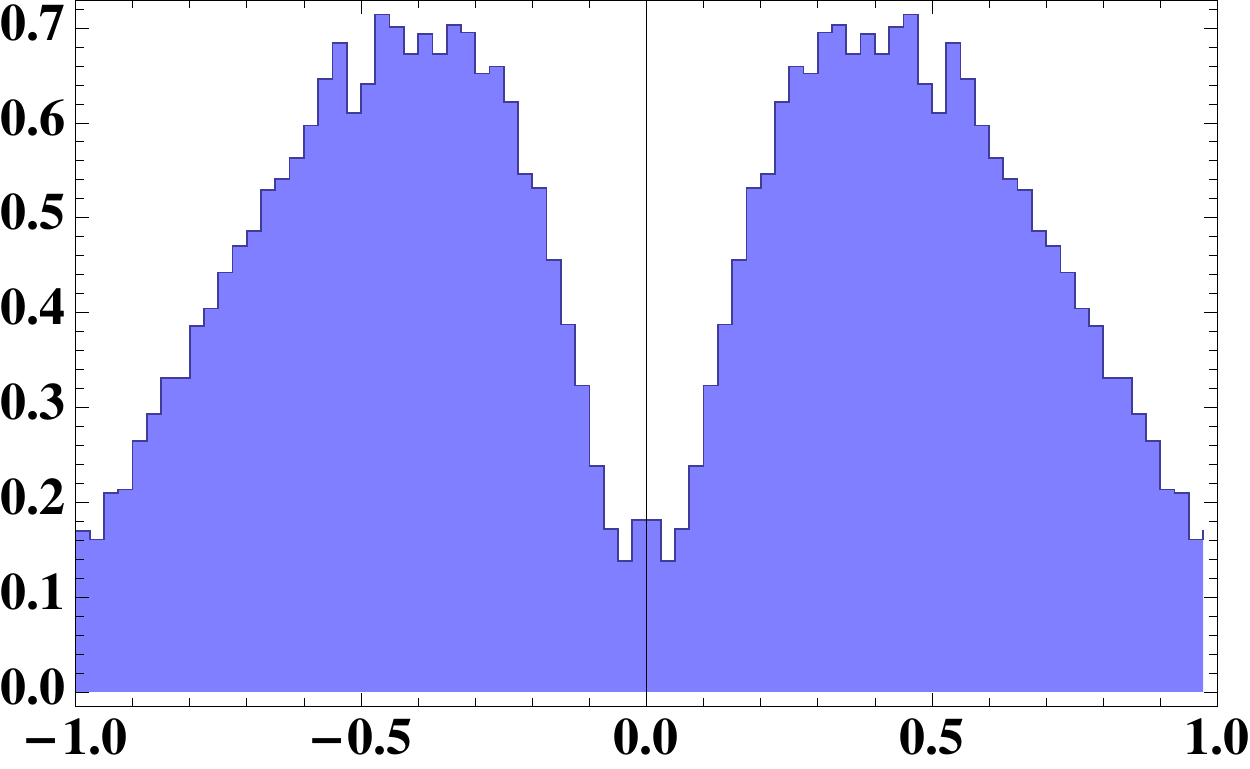}\\
   \includegraphics[width=0.3\textwidth]{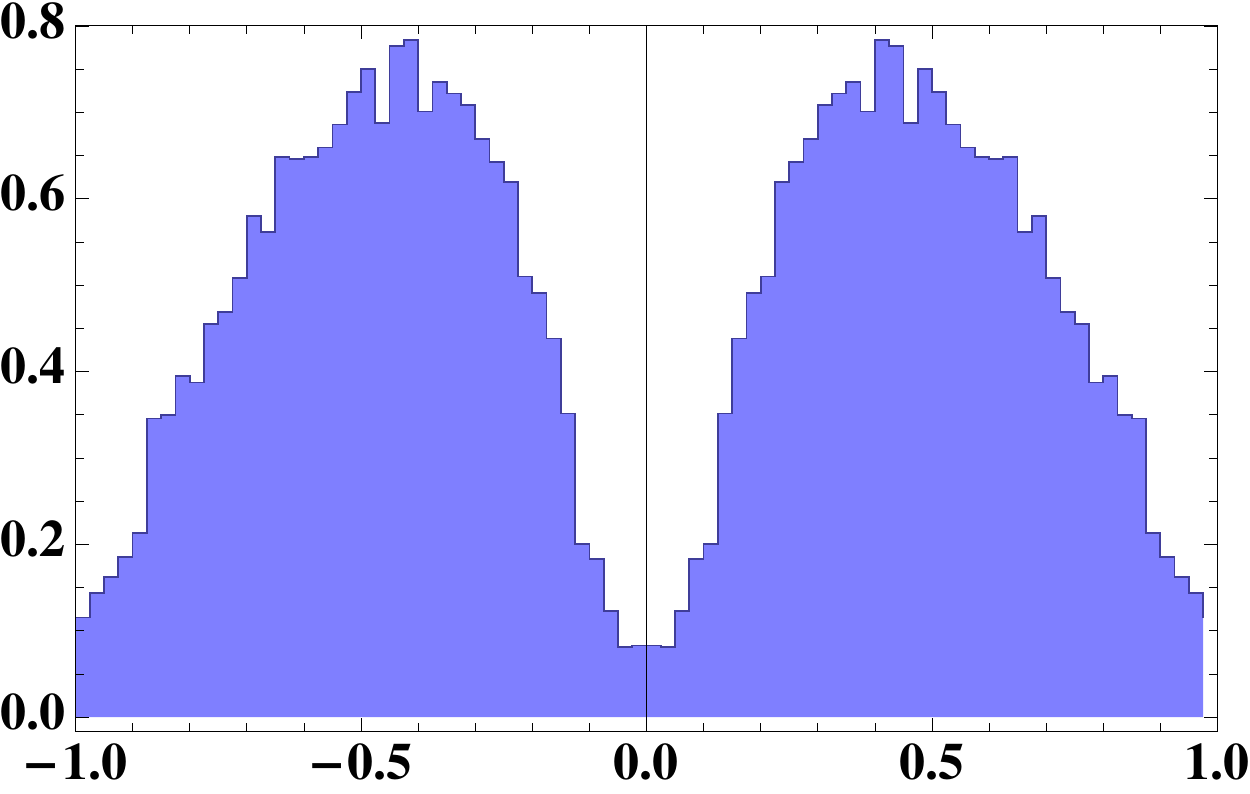}\;
   \includegraphics[width=0.3\textwidth]{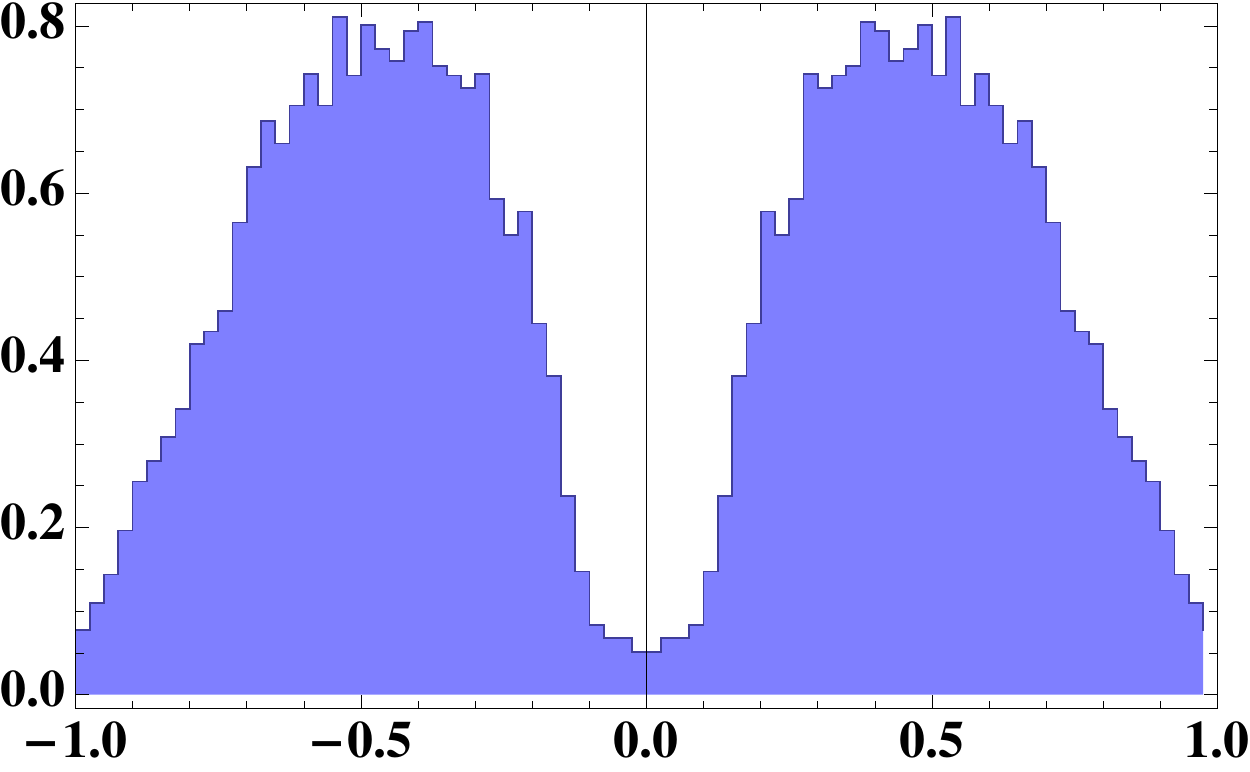}\;
   \includegraphics[width=0.3\textwidth]{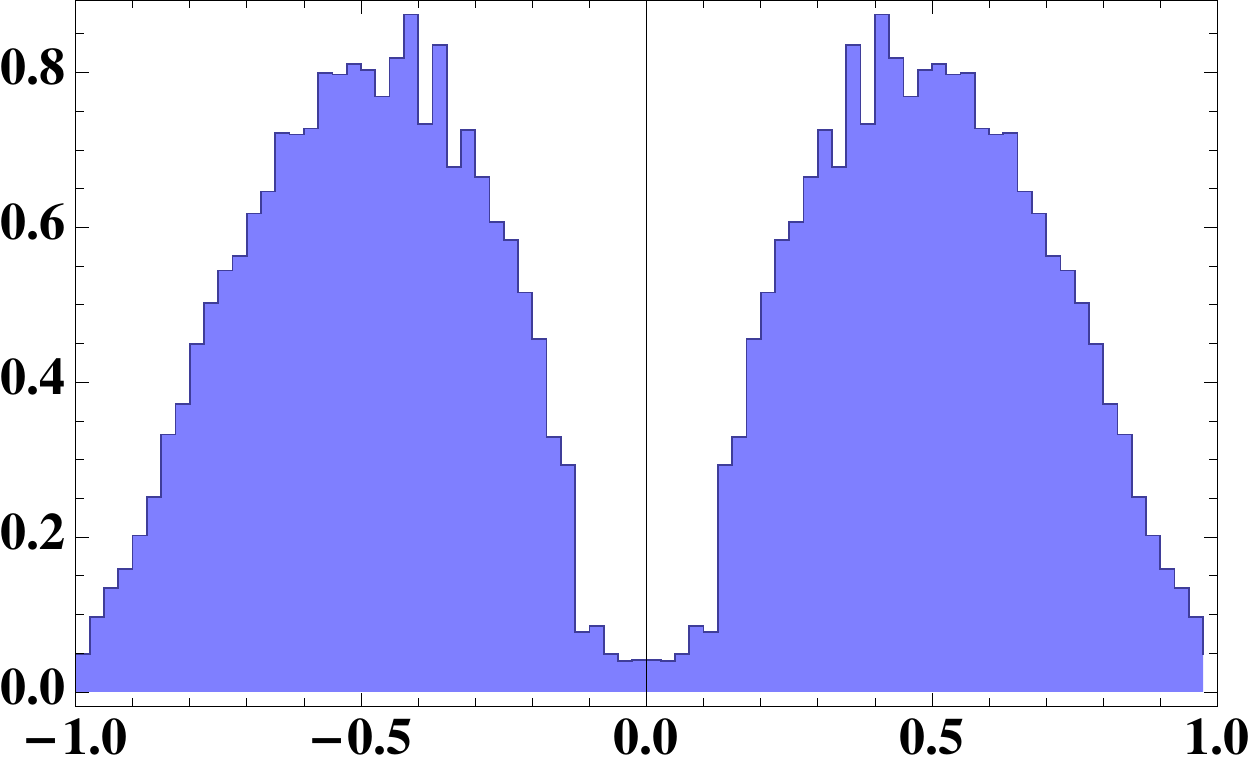}
   \caption{ (Color online) The Dirac eigenvalue spectrum. The horizontal axis is the eigenvalue of the Dirac operator $\lambda$ expressed in units of $c$ defined in the eq. (\ref{eq:T}), and vertical axis is the probability density $dP/\lambda$. The plots are for $N_f=2$ and $M=(\pi/6\dots 9)\times \pi/6 n^{1/3}$, where $n$ is the density of $L$ dyons. Note that the chiral symmetry is restored as a function of $M$ which is connected to holonomy as $M=\bar v/2=(2\pi-v)/2.$}
   \label{rand_mol}
\end{figure*}

\begin{figure*}[htbp] 
   \centering
   \includegraphics[width=2.5in]{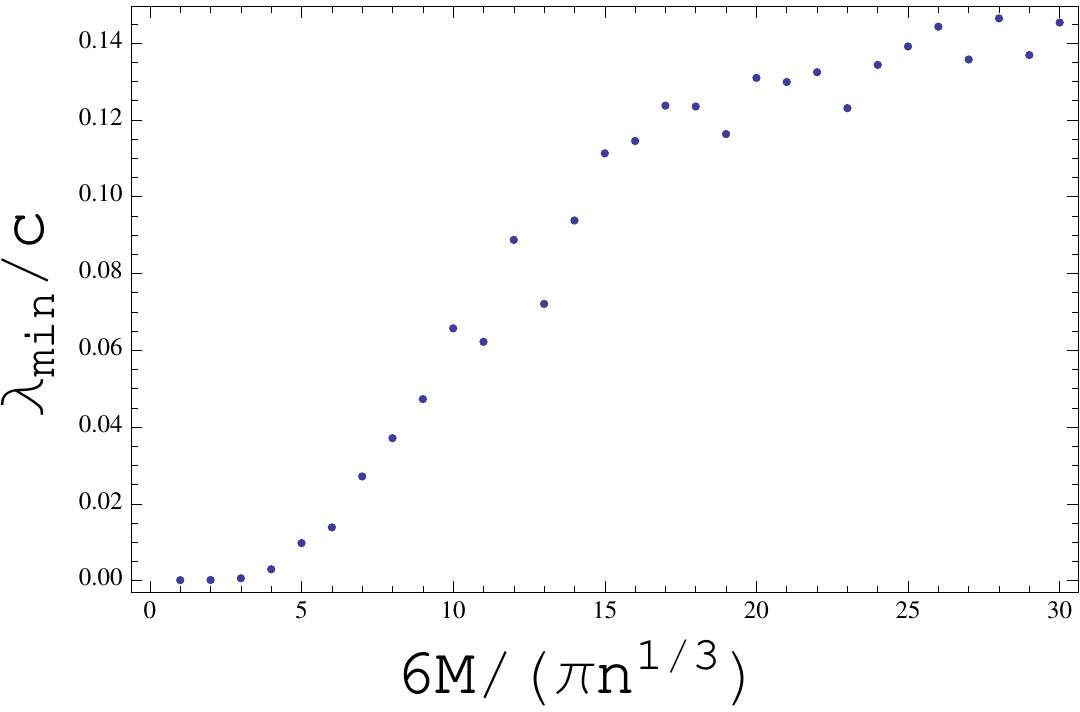}\quad\includegraphics[width=2.5in]{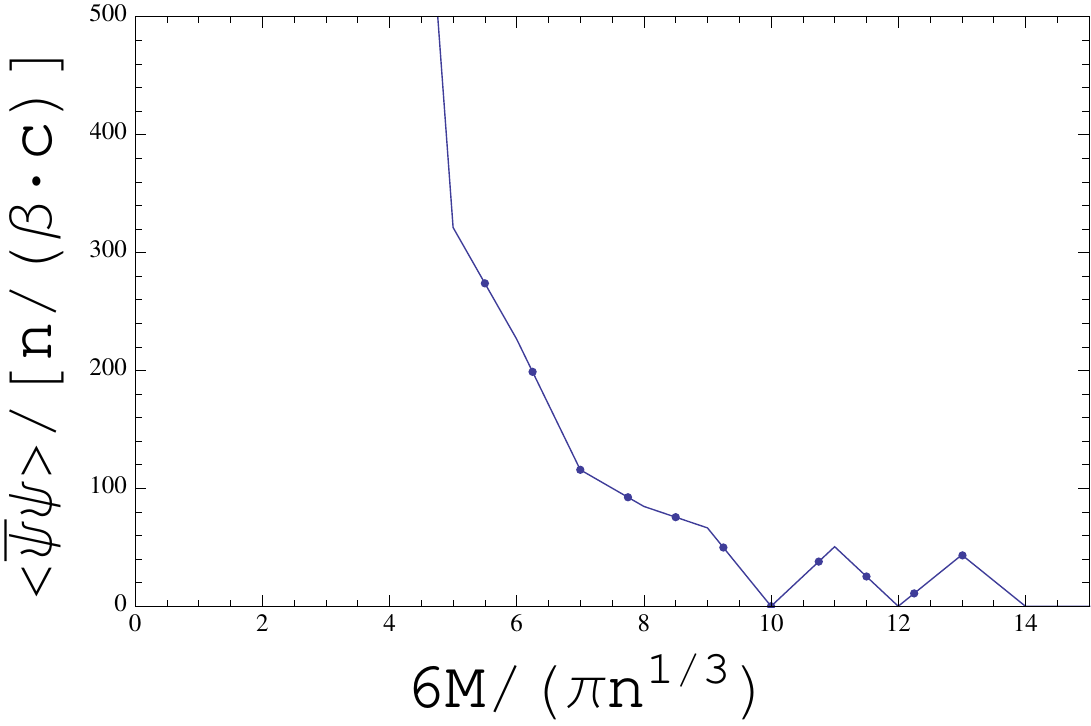} 
   \caption{ (Color online)The smallest eigenvalue and the chiral condensate for the Random Molecule model, as a function of the holonomy mass combined with the dyon density, for $N_f=2$.}
   \label{fig:rand_mol_chicond}
\end{figure*}

The model has two parameters, the holonomy ``mass''  $M$ and the number of flavors $N_f$. In Fig. \ref{rand_mol} we show the Dirac spectrum for several values of $N_f$, as a function of $M$ and in Fig. \ref{fig:rand_mol_chicond} the lowest eigenvalue and chiral condensate results. Note that here there is explicit dependence of chiral condensate on holonomy.

\begin{figure*}[htbp] 
   \centering
   \includegraphics[width=0.3\textwidth]{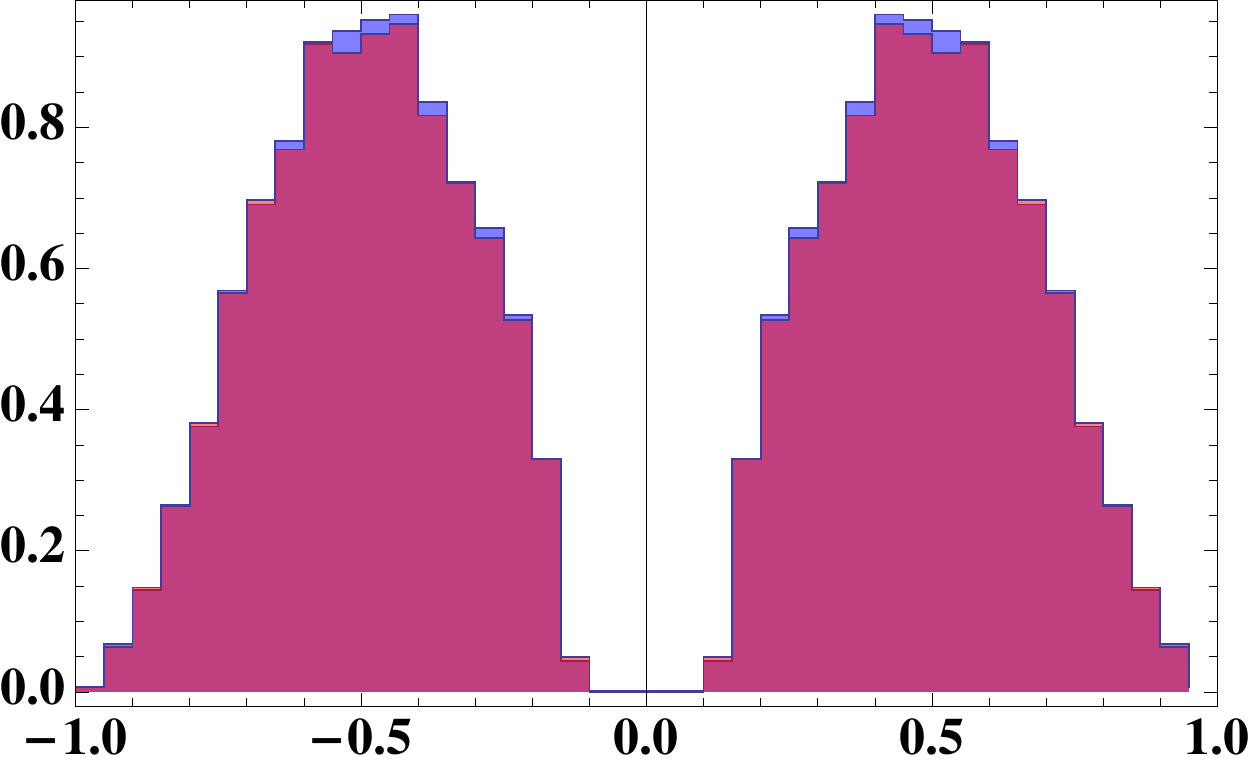}\;
   \includegraphics[width=0.3\textwidth]{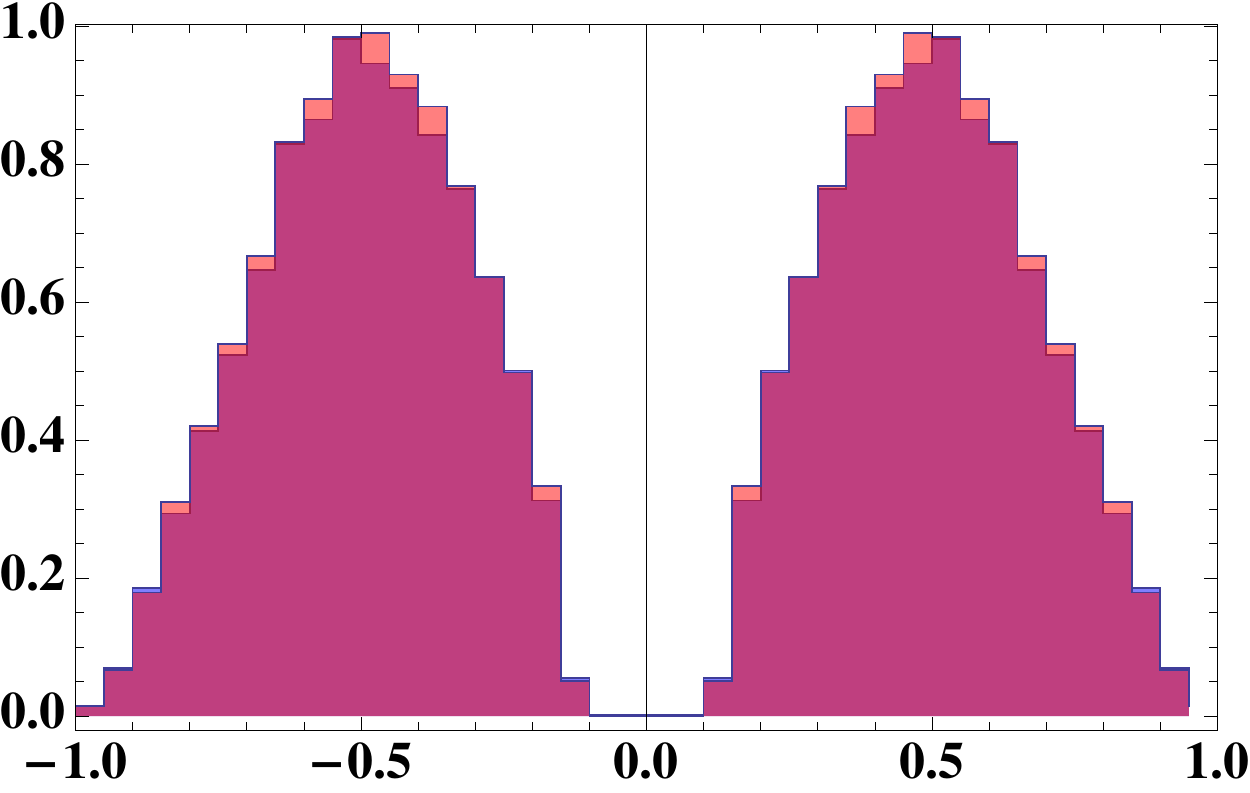}\;
   \includegraphics[width=0.3\textwidth]{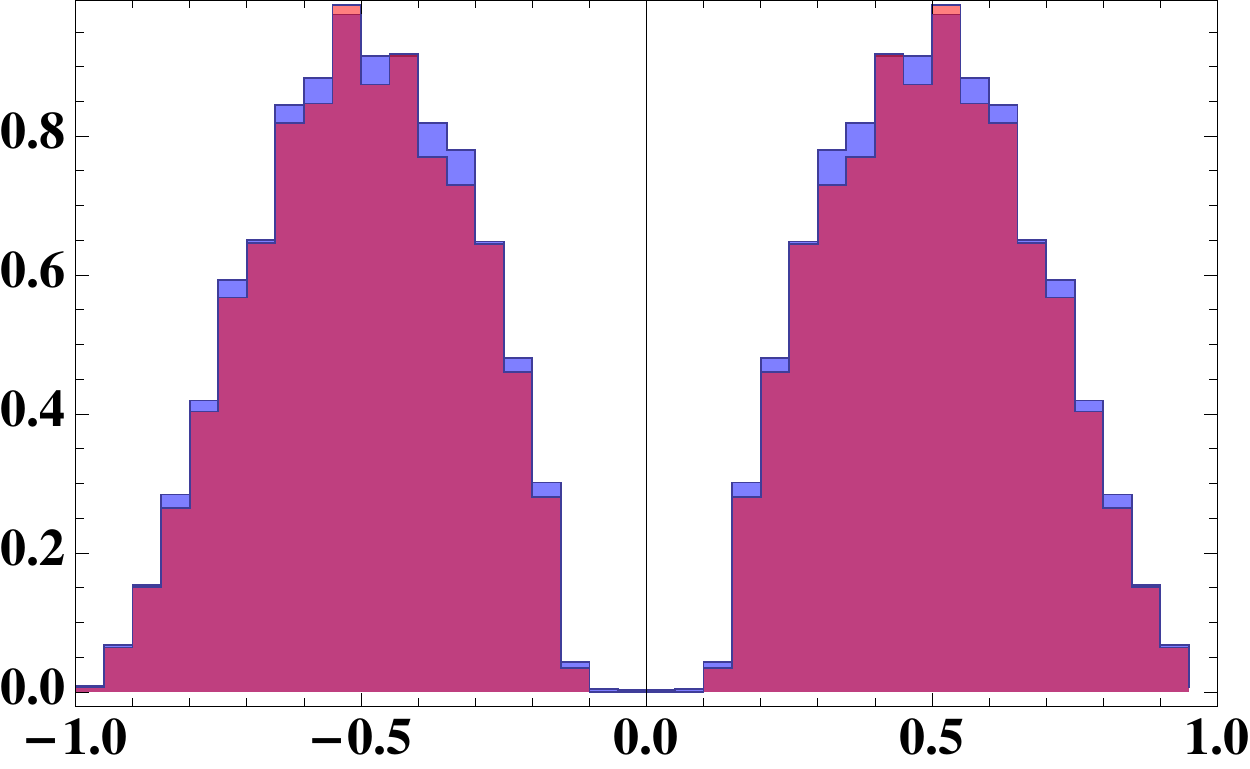}\\
   \includegraphics[width=0.3\textwidth]{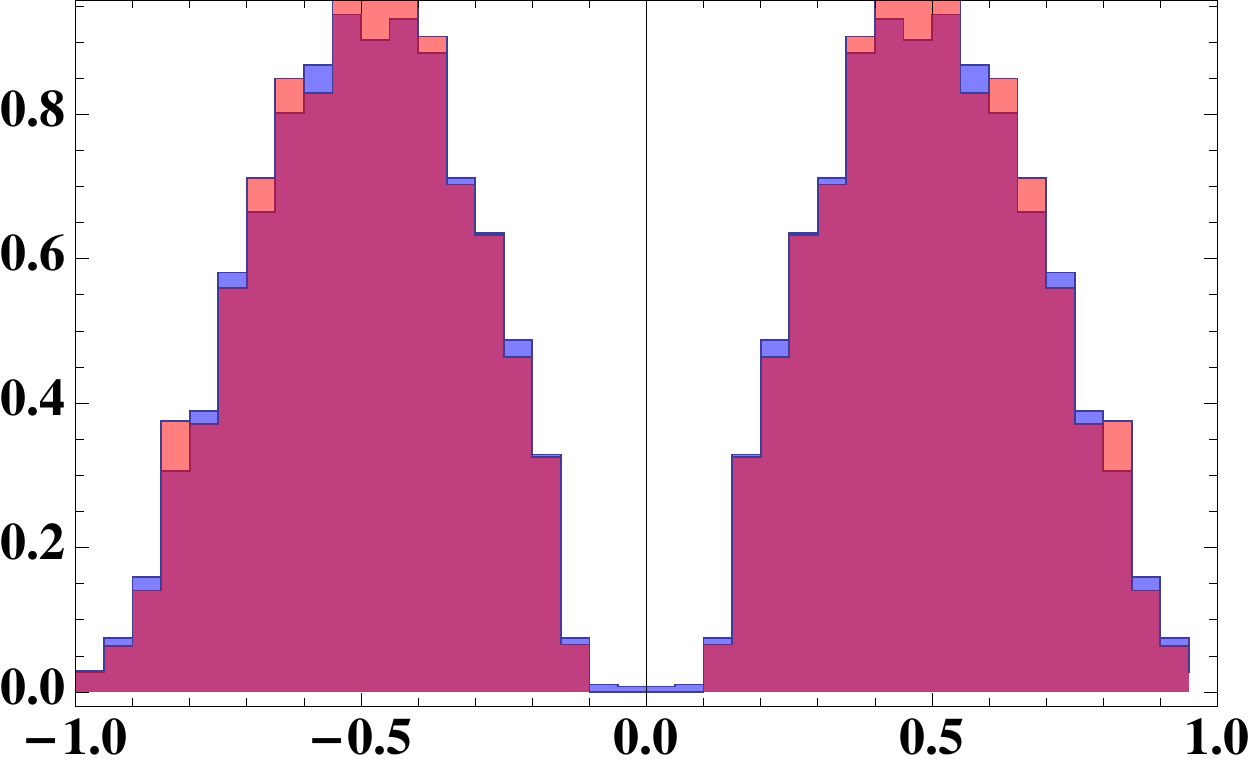}\;
   \includegraphics[width=0.3\textwidth]{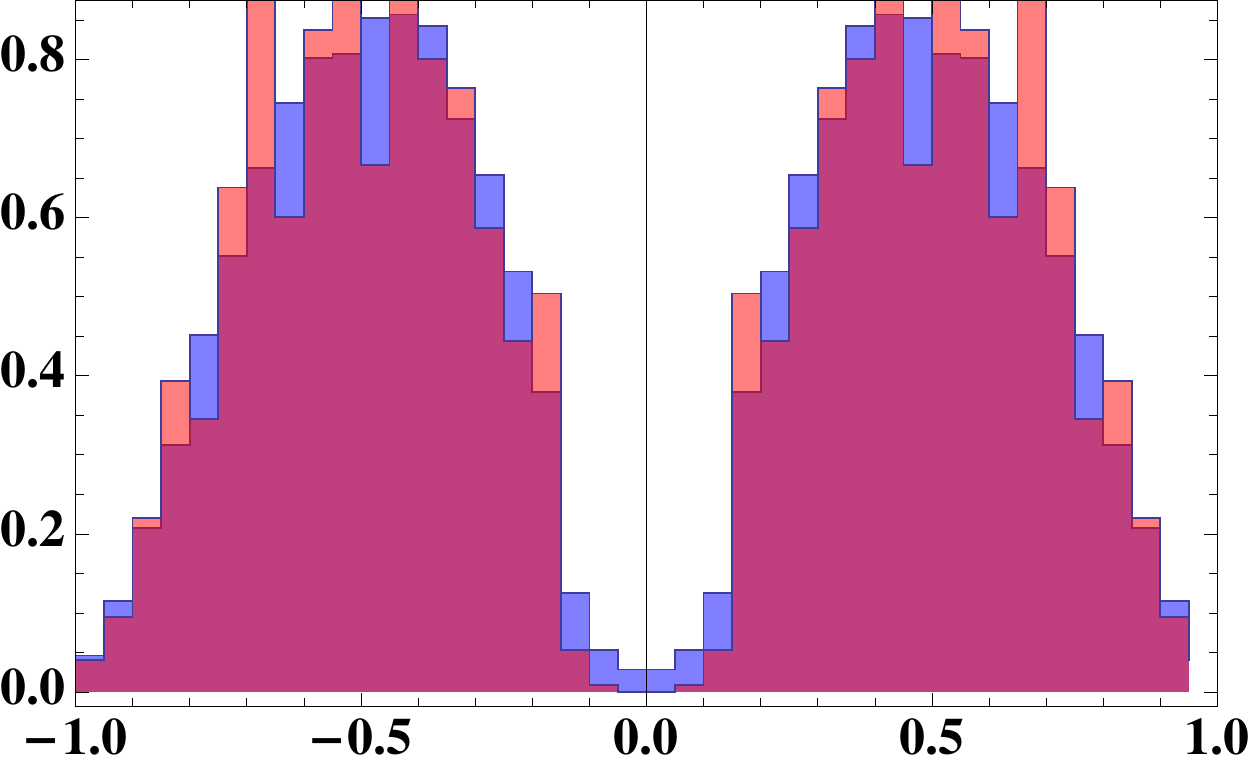}\;
   \includegraphics[width=0.3\textwidth]{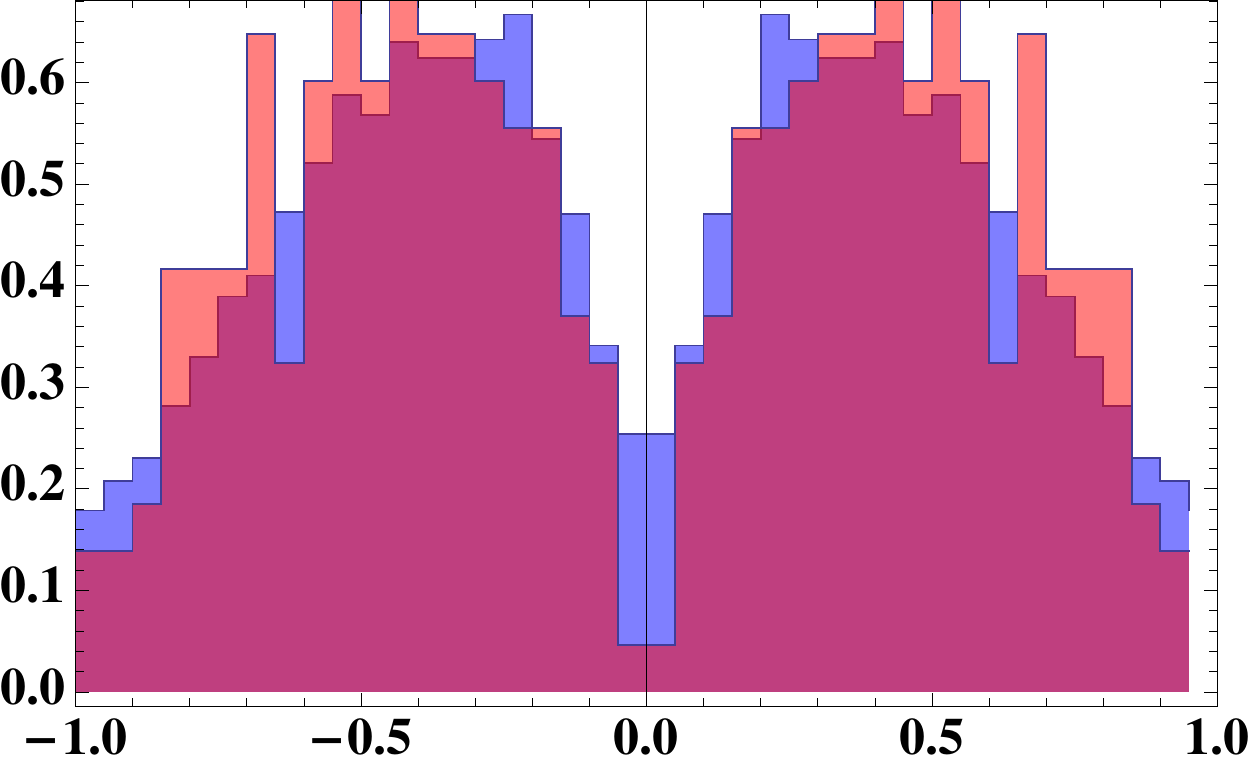}
   \caption{ (Color online) The Dirac eigenvalue spectrum for a Random Molecule Ensemble -- light (blue)  and Reweighted Random Molecule Ensemble -- dark (red). The horizontal axis is the eigenvalue of the Dirac operator $\lambda$ expressed in units of $c$ defined in the eq. (\ref{eq:T}), and vertical axis is the probability density $dP/\lambda$. The plots are for $N_f=2$ and $M=(30,25,20,15,10,5)\times \pi/6 n^{1/3}$, where $n$ is the density of $L$ dyons. The reweighting becomes unreliable in the last plot, and only one configuration dominates.}
   \label{reweight_rand_mol}
\end{figure*}

\begin{figure*}[htbp] 
   \centering
   \includegraphics[width=3in]{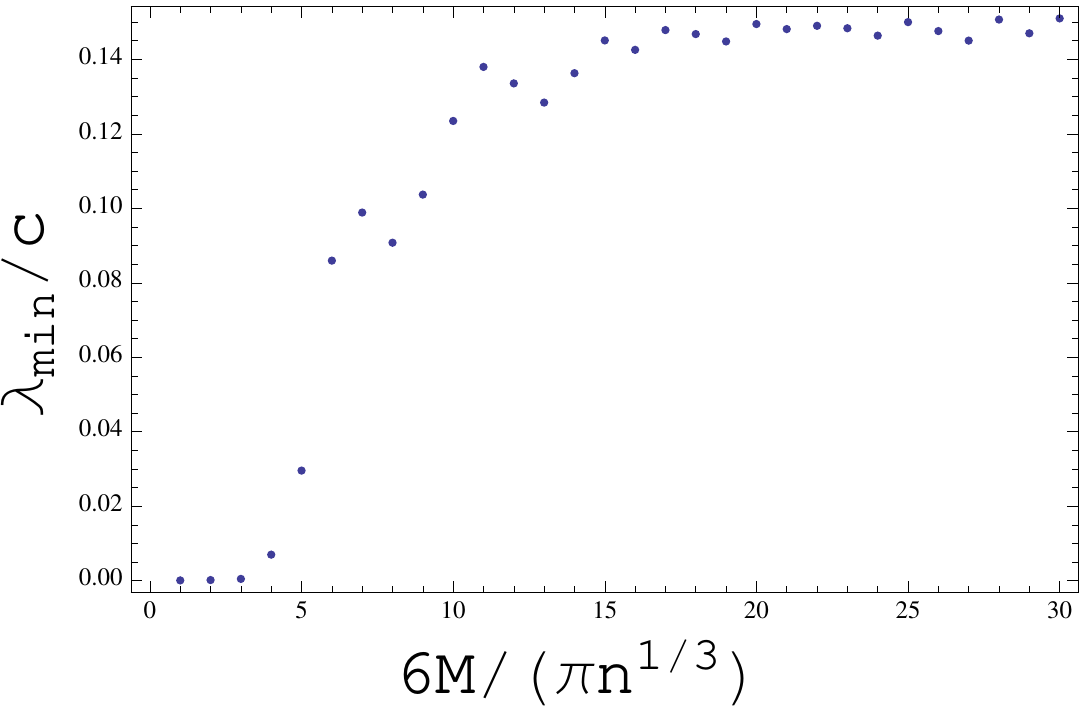}\includegraphics[width=3in]{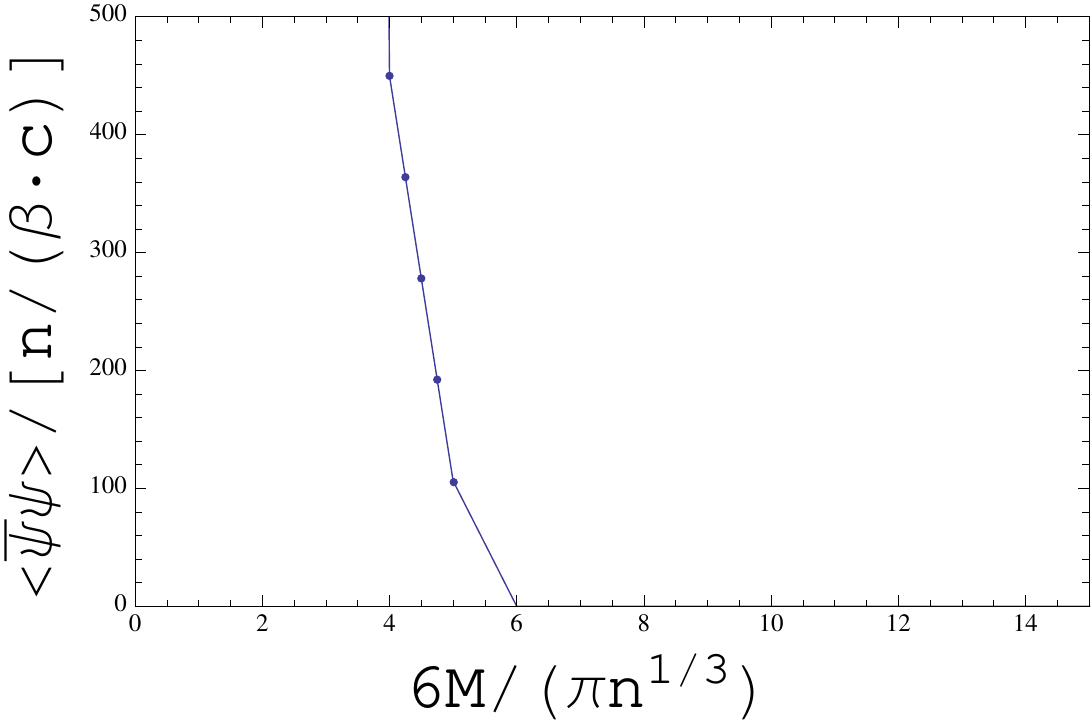} 
   \caption{ (Color online)The smallest eigenvalue and the chiral condensate for the Reweighted Random Molecule model, as a function of holonomy for $N_f=2$ and $M=\pi/6\dots 30\times \pi/6$. Note that reweighting becomes unreliable at around $M=15\times \pi/6$, and that we show this result only to demonstrate the trend that chiral symmetry persists to lower values of the holonomy mass $M$ than for the nonreweighted case.}
   \label{rw_rand_mol_lmin}
\end{figure*}

The 3$^{rd}$ model is a modification of the second by reweighting the configurations with the determinant $(\det T_{ij})^{2N_f}/\left(\prod_{i} D_{mol}(r_ii)\right)$, where $r_{ii}$ is the distance between closest neighbors. The result is shown in Fig. \ref{reweight_rand_mol} and in Fig. \ref{rw_rand_mol_lmin}

\subsubsection{Mapping the models to finite $T$ QCD}
Our three models were defined in such a way,
that each step has introduced one new parameter: with the dyon density it brings the total
number of parameters to four. 
Yet a QCD-like theories with massless fermions have
  only two parameters, the temperature $T$ and $\Lambda$.
 Thus only a 2-parameter subspace of our (up-to) four parameter model space
can be compared to reality.

Now is the time to map those parameters. 
Like lattice practitioners do, it is thus natural to measure all dimensional quantities in units of $T$. 
The dimensionless dyon density $n_d/T^3$ is one of the key parameters.
It has not yet been measured on the lattice, but it can be.
While it is the same as the density of instantons, it is $not$ given by the topological susceptibility $\chi(T)$, as neutral molecules contribute to the former but not the latter. 
Semiclassical theory tells us the large-$T$ asymptotical dependence on $T$, expression (\ref{eqn_mol_dens}).
For qualitative estimates one may use it normalized to
its value at $T_c$ . The factor in front of the power of $T_c/T$ depends on the
coupling, in a particular definition used by 'tHooft,  and the fermionic factors its value is $O( 1)$ for physical QCD or
$N_f=2$,  which one can use for absolute normalization. 

The fermion mass $M/T\sim 1/N_f$ and thus, keeping the coupling fixed while increasing $N_f$, one finds that 
the cluster  size is reducing and thus 
we are going into a regime of more dilute  gas.
However, if one wants to follow the line of ``fixed eigenvalue spectrum" and/or 
fixed $<\bar\psi \psi>$, one needs to keep
 the same diluteness
of the molecular model, or keep constant
 \be R_m^3 n_d = const \label{eqn_cond}\;, \ee where $R_{m}\sim 1/(N_{f}M)$. Thus the dyon density should grow as $N_f^3$, e.g. from $N_f=2$ to $12$ increase by a factor 216.

The only way it can be achieved is by a shift into the stronger coupling! 
A crude estimate ignoring preexponent gives a shift of
\be  {8\pi^2 \over g^2(T_c(N_f))} - {8\pi^2 \over g^2(T_c(N_f=2))} = -3 \ln (N_f/2) \ee
This  qualitatively explains why chiral restoration line $<\bar\psi \psi>=0$ derived in lattice studies (see Fig.\ref{fig:romanplot}) dramatically shifts into stronger coupling.  

Unfortunately our attempts to do it quantitatively  failed, for the following reason. The coupling $g$ in the semiclassical expressions and on the lattice (such as  Fig.\ref{fig:romanplot}) are defined in different schemes, with the scales $\Lambda_{\bar{MS}}$ and $\Lambda_{lat}$.
Perturbatively they only differ by a calculable factor, but as their ratio happen to be large, this relation is not very useful
in practice. 
For example, for QCD or $N_f=3$ theory,  the former is about 300 MeV and the latter about 5 MeV. The instanton density includes huge factor \be \exp\left(-{8\pi^2\over g^2}\right)\sim (\Lambda_{\bar{MS}}/\Lambda_{lat})^{11Nc/3-2N_f/3} \sim 60^9 \ee
Obviously, in view of such a huge factor, any small deviation from the two-loop beta function used would result in huge uncertainties which make any numerical comparison of the semiclassical expressions and lattice bare coupling meaningless. 
  
   The tests of this explanation  however can still be made using lattice data. The most straightforward one would be to measure the dyon molecule density and size
   and see if the relation (\ref{eqn_cond}) holds. To do so one can e.g. use Dirac
   eigenstates in certain interval of $\lambda$ which can identify a dyon-antidyon cluster.

%


\subsubsection{Structure of strongly coupled ``cluster liquid''}

With large number of fermions $L\bar{L}$  are strongly coupled into a charge -2 well localized
objects, compensated by the negatively charged $M$ and $\bar{M}$ dyons which are
more homogeneously distributed.

The  $L\bar{L}$ ``nuclei", with the fermions attracted to them, form mutually repulsive ``atoms".
The question is what arrangements those should have to get the lowest energy. One obvious idea is 
those with the best packing in 3d, namely the face-center-cubic ($fcc$) or the hexagonal-close-packing  ($hcp$).
 Selecting between those
one may follow the guidance  given by
ordinary atoms which are neutral and spherically symmetric by themselves. High-density solid $He^4$
is of the $hcp$ structure, and perhaps that would be the approximate local symmetry of our strongly correlated
 $L\bar{L}$ liquid. If so, each of them has 12 nearest  neighbors, organized in two  hexagons
 at two planes above and below the cluster.
 
 While large number of correlated neighbors reinforces the correlations, simple estimates show
  that  the parameter $\Gamma$ (the average interaction potential divided by $T$) does $not$
reach the critical value needed for the ensemble to get solidified. Thus, with the parameter range at hand we expect  a strongly coupled liquid, with smaller
 number of well-correlated neighbors but with their locations still correspond to those  in the crystal.
Explicit statistical simulations of it are possible, but are deferred for further studies.

 \subsection{Statistical mechanics driven by bosonic moduli space metric effects  \label{sec_salt} }
 
 Let us now discuss the opposite limit of $zero$  $N_f=0$, in which there are no $L\bar{L}$ clustering.
 Let us further imagine that for some reason one can decouple the two sectors, dyons and antidyons,
 and discuss what kind of a system would be created by ``Diakonov's determinantal forces".
Assuming that the antiselfdual sector do not exist, let us thus focus on the 
$M,L$ sector. Since the electric and magnetic charges in it have the same sign,
one may in this section simply call them $+$ and $-$ dyons.

  At large distances 
the forces between them are Coulombic, and one may think that the local crystal correlations those generate
is a simple cubic crystal of alternating charges, like e.g. the usual salt $NaCl$.
Any charge is thus strongly correlated with 6 nearest neighbors.


Classical Coulombic systems, are well-known to be unstable
against charges falling on each other. (Of course for real ions electron repulsion solves
this issue, stabilizing the salts.) We thus studied the following question:  can the  Diakonov's
determinantal forces  stabilize a cubic crystal?

\begin{figure}[htb]

\includegraphics[width=0.45\textwidth]{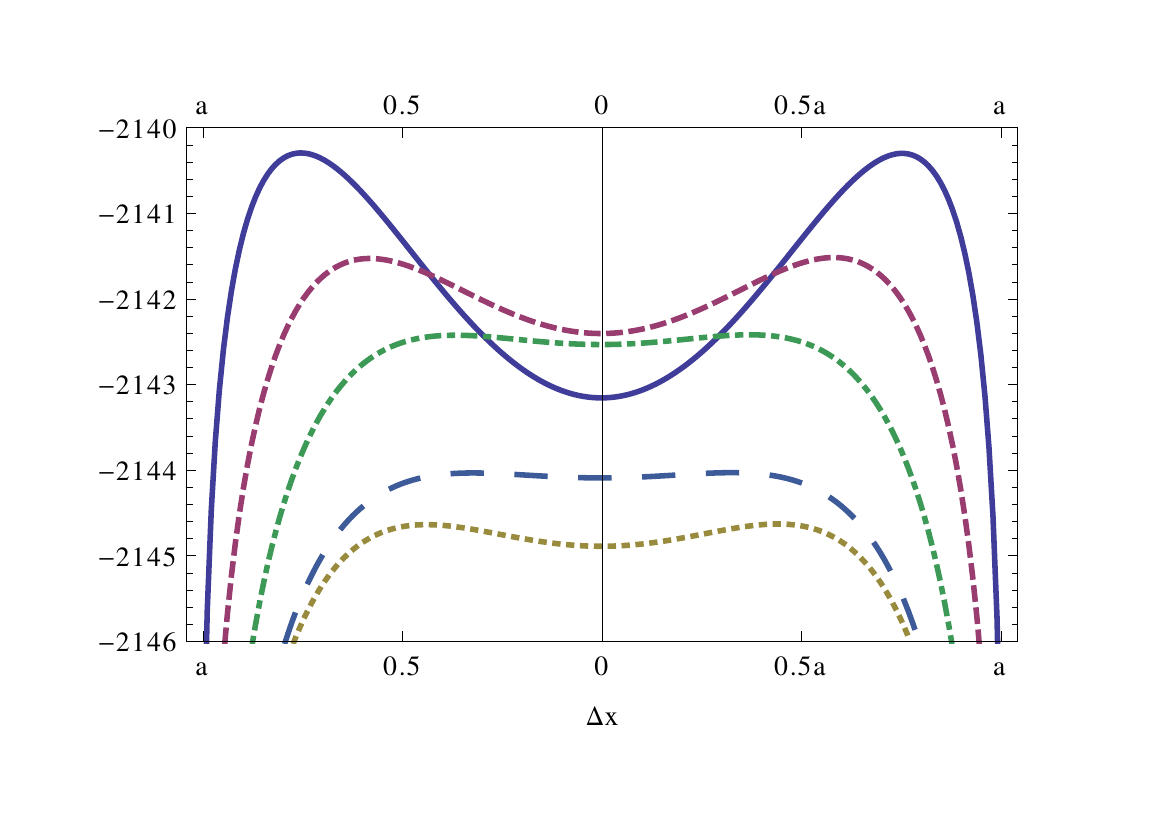}
\caption{(color online) The effective potential $V_d=-\ln(\det G)$, where $G$ is the Diakonov's determinant, 
as a function of the displacement $\Delta x$ of 
a single dyon in the 
center of a cubic $6\times 6\times 6$ crystal, in the direction of the adjacent dyon. The graphs have been rescaled for better comparison
as follows:
(blue) solid 684$V_d, a=0.1$; (red) dash 818 $V_d,a=0.25$; (brown) short dash $917 V_d,a=.5$; (green) dash-dotted $967V_d,a=0.75$; 
(blue) long-dash $1000V_d,a=1$. The units of $\Delta x,a$ is the Matsubara time. Note that as dyonic density increases by a factor
$10^3$, this one-loop  bosonic interaction creates a significant minimum at $\Delta X=0$, stabilizing the cubic structure.
}
\label{fig:diak_stab2}
\end{figure}

We use the moduli space metric $G_{ij}$ for the selfdual sector as suggested in \cite{Diakonov:2009ln} to calculate the effective potential
of a crystal configuration of $L$ and $M$ dyons, with a lattice spacing $a$, for a displacement of a single dyon
somewhere in a center. Effective potential is the log of the measure, $V_d=-\ln(\det G)$. For a purely Coulombic crystal, the crystal potential  as a function of a displacement $\Delta x$ has infinite Coulombic dips as displacement
$\Delta x$ approaches $\pm a$, which means that  for purely Coulombic interactions the alternating charges will fall on each other.
However the effective potential $V_{d}$ contains repulsion and, as shown in 
  \ref{fig:diak_stab2},  this leads to a pronounced minimum at $x=0$ for
 sufficiently small lattice spacings $a$ (high density). On closer inspection, however, there is always a small, but
clear minimum.
As expected,   the  Coulombic dips at $\pm a$  still persist.  This divergence --corresponding to small-size instantons and the factor $1/\rho^5$ in the measure - is known to be removed  by 
quantum
fluctuations, which produce a stronger factor $\rho^{11Nc/3-2N_f/3}$. Thus outside of classical approximation the problem is well defined.


On the other hand, as discussed in \cite{Diakonov:2009ln} it seems that from the point of view of the far field metric, the antiselfdual sector behaves similarly
to the selfdual one, and the interaction between $L$ and $\bar L$ is similar to the interaction between $L$ and $L$ (i.e.
repulsive), while that of $L$ and $\bar M$ is attractive. Therefore we have three possible structures depicted in Fig 
\ref{fig:salts}:

\begin{figure*}
 \includegraphics[width=\textwidth]{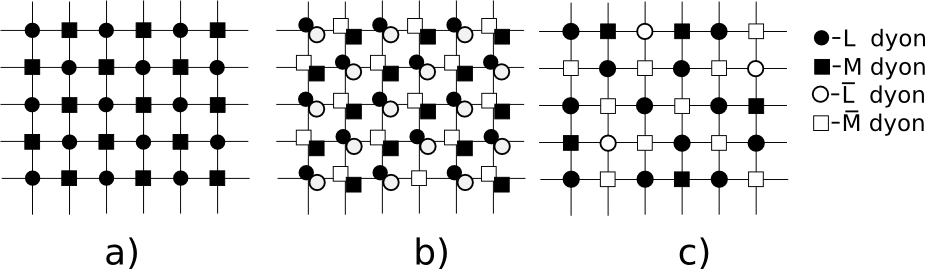}
\caption{Three possible crystal structures discussed in the text.}
\label{fig:salts}
\end{figure*}

 For nonzero 
 number of quark flavors  $N_f$ fermions correlate the $L$ and $\bar{L}$, then they will repel the other pairs, therefore 
making the hexagonal crystal, as we discussed above. For zero $N_f$ 
  they can be either strongly correlated (b)
 or form an alternating crystal (c). 
 Lattice practitioners  often  introduce the so called ``valence" 
quarks, which are not in the measure but are only used for a diagnostic purposes of the ``quenched" ($N_f=0$) theory. 
Dirac eigenvalue spectrum and chiral properties 
revealed by such studies can be computed and compared to the lattice data.
 
 In a standard way we model the Dirac matrix by a ``hopping" matrix 
 (\ref{eqn_tij1},\ref{def_TIA}),
with the matrix elements being some function of a distance between dyon-antidyon $f(r)$.
We expect these functions to be exponential in distance $r=|\vec r|$ at large $r$ and some constant we call $f(0)$ at small $r$.  

 In the first approximation we can consider only next-neighbor matrix elements, and use the cubic structure (b).
Then we can write the upper-right (or the lower-left) part of the Dirac operator matrix, in the triple-index notations with $n,m,l$
being positions in units of $a$ along the three spatial coordinates in the cubic lattice
\ba
D_{nml}^{n'm'l'}= f(0)\delta_n^{n'}\delta_m^{m'}\delta_l^{l'} \nonumber \\
+f(a)(\delta_n^{n'+1}\delta_m^{m'}\delta_l^{l'} +\delta_n^{n'-1}\delta_m^{m'}\delta_l^{l'} \nonumber \\
+\delta_n^{n'}\delta_m^{m'+1}\delta_l^{l'} +\delta_n^{n'}\delta_m^{m'-1} \delta_l^{l'} \nonumber \\
+\delta_n^{n'}\delta_m^{m'}\delta_l^{l'+1}+\delta_n^{n'}\delta_m^{m'}\delta_l^{l'-1})
\ea

\begin{figure}[htbp] 
    \centering
    \includegraphics[width=\columnwidth]{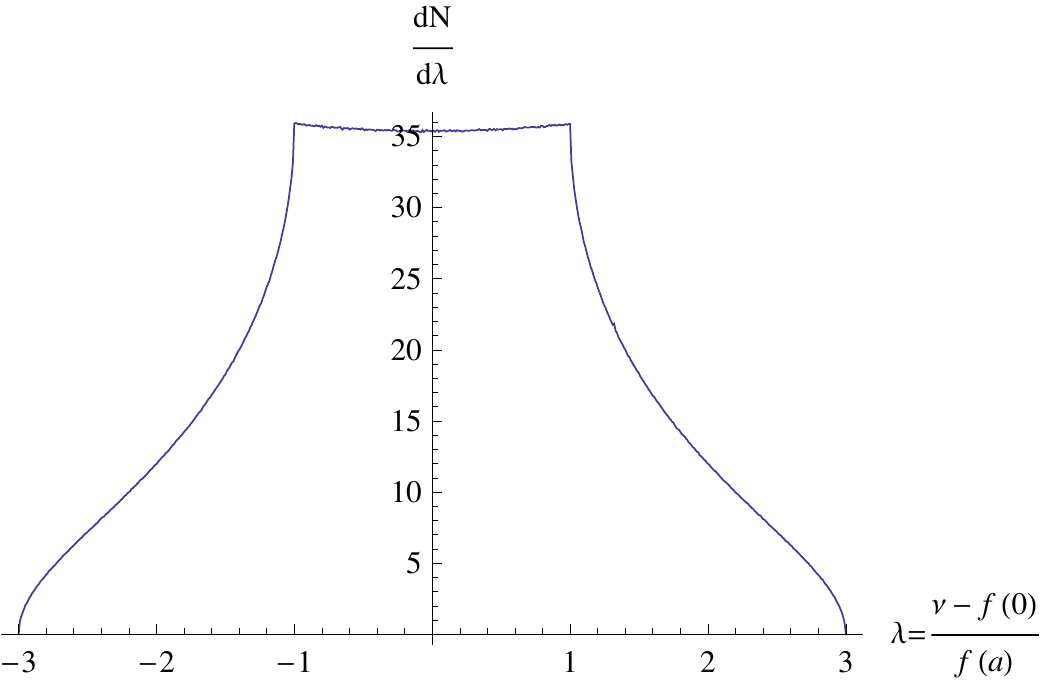} 
    \caption{(Color online)The plot of the density of states of the  cubic  lattice with the next-neighbor interactions.}
    \label{fig:crystaldensityofstates}
 \end{figure}

 Upon standard diagonalization by transformation to the dual lattice momentum states
 \be
 \ket{\bm k}=\frac{1}{\sqrt{L^3}}\sum_{\bm I=(n,m,n)} e^{\frac{2\pi i \bm I\cdot \bm k}{L}}\;,
 \ee
 we obtain that the spectrum is given by
 \be
 \nu(\vec k)=f(0)+f(a)\left(\cos{k_1}+\cos(k_2)+\cos(k_3)\right)\;,
 \ee
 where $k_{1,2,3}$ go from $(0,2\pi)$, and the elementary number of states is given by standard
 $dN=Vd^3k/(2\pi)^3$.
 The density of states is 
 \be
 \frac{dN}{d\lambda}=V\int d^3 k \;\delta(\lambda-\cos k_1-\cos k_2-\cos k_3)\;,
 \ee
 where we have used the $shifted$ eigenvalues \be \lambda=(\nu-f(0))/f(a) \ee The spectrum can be integrated to yield
 \be
 \frac{dN}{d\lambda}=\int dk_1 dk_2\; \frac{1}{|\sin k_3(k_1,k_2)|}\;,
 \ee
 where $k_3=\arccos(\lambda-\cos k_1-\cos k_2)$, and the region of integration is such that $\left|\nu-\cos k_1-\cos k_2\right|\le 1$. 
Numerical integration yield the curve shown in Fig. \ref{fig:crystaldensityofstates}

 We see that the density of states form a band with a sharp boundary,
 it goes to zero at $|\lambda | > 3$.  For scenario (b) this shape will appear in the spectrum centered around $\pm |f(0)|$, and, for vanishing $f(a)$, will be delta function-like.  Each of  those morphs into a shape of Fig.\ref{fig:crystaldensityofstates} in a type-(b) crystal as $f(a)$ increases.
 If they are separated by more than the width of the peak, the
 chiral symmetry is $not$ broken. The condition for chiral symmetry breaking is therefore
 \be \left|{f(a) \over f(0)}\right| >{1\over 3} \ee 

Alternative structure (c) can be motivated as follows. As mentioned
earlier, the long range interactions between $L$s and $M$s, become the same regardless of them being dyons or antidyons.
This means that the cubic crystal will have $L$s and $\bar L$s located at positions of ''+`` ion and $M$ and $\bar M$ at 
positions of ''-`` ion. In this case, the spectrum of the Dirac operator is
\be
\nu=2f(a\sqrt{2})(\cos k_1+\cos k_2+\cos k_3+\cos k_4)\;,
\ee
where we have approximated that, on average, each $L$ has 4 nearest $\bar L$s. Similarly as before we get that
\be
\frac{dN}{d\nu}(\nu=0)=2f(a\sqrt{2})\int \frac{dk_1dk_2dk_3}{(2\pi)^4}\frac{1}{|\sin k_4|}\;,
\ee
where $k_4=\arccos(\cos k_1+\cos k_2+\cos k_3)$, and $\cos k_1+\cos k_2+\cos k_3\le 1$

\subsection{Note on confinement of the cubic crystal}

Here we show that the Polyakov loop of a crystal configuration is indeed zero. We have that
\be
\tr L(\bm x)=\tr e^{i\left(\frac{v}{2}+V(\bm x)\right)\tau_3}=2\cos\left(\frac{v}{2}+V(\bm x)\right)\;,
\ee
where we used the gauge-combed gauge, and where $V(x)$ is some potential which goes to $-v/2$ at the position of $M$ and
$\bar M$ dyons and to $-v/2+\pi$ at the position of $L$ and $\bar L$ dyons. Using the identity that $\cos(\alpha+\beta)=
\cos\alpha\cos\beta-\sin\alpha\sin\beta$, we have that
\be
\tr L(\bm x)=2\cos\frac{v}{2}\cos(V(\bm x))+2\sin\frac{v}{2}\sin(V(\bm x))\;.
\ee
The above expression has to be integrated over all possible crystal orientations and positions. This is equivalent, 
though, to integrating over $\bm x$ in the region of one crystal unit, i.e.
\ba
\avg{\tr L(\bm x)}&=&2\frac{1}{a^3}\int_{a^3} d^3x\; [2\cos\frac{v}{2}\cos(V(\bm x)) \nonumber \\
&& +2\sin\frac{v}{2}\sin(V(\bm x)) ]
\ea

However, setting $v=\pi$ (maximally nontrivial holonomy), we see that the first term vanishes because $\cos(\pi/2)=0$, and
the second term vanishes because $V(\bm x)$ is alternating between $L$ and $M$ dyons in sign, and, since $\sin$ is an odd
function, this term too vanishes. Therefore the Polyakov loop averaged over crystal configurations vanishes, and it depends
explicitly on the holonomy being nontrivial. 

Finally let us note that such ``confinement-in-average"  phenomenon has its predecessors,
resembling a finite-density holographic model
of densely packed baryons  \cite{Kim:2007vd}. In it baryons are modeled
by instantons, which also undergo transition into a ``dyon phase" in which they restore chiral symmetry in average.
\section{Summary and discussion}
In  this work we have done qualitative study of  the interactions  of (anti)selfdual dyons, as well as some study of 
the statistical ensembles following from them. 
We emphasized in particularly the importance of the {\em dyon-antidyon} interaction, both classical
(bosonic) and fermion-induced ones, instead of the interactions
inside the selfdual and antiselfdual separate sectors.   We specifically were
focusing on the SU(2) theory and the temperatures above and near the chiral restoration phase transition.

We already summarized the overall picture
resulting from this study at the very beginning of the paper, so let us just
outline the elements of the picture which we believe explain certain lattice observations. We will also emphasize
what lattice practitioners can do to test our predictions further.

 \subsubsection{  Existence of  the ``topological molecules" }
At high $T$ the instanton/dyon density is small due to electric screening, 
the topological object must make ``molecules", neutral  in terms of all three charges involved:  topological, electric and magnetic.  
In the SU(2) theory they contain all 4 types of the dyons.
We predict  a peculiar distribution, with ``nucleus" of $L\bar{L}$ cluster at the center, with $M$ and $\bar{M}$
at the periphery, see Fig.\ref{fig:mol}. This effect is predicted to get more pronounced with the increasing
number of quark flavors $N_f$. It is important to check it on the lattice, perhaps by identifying the fermionic sates
``at the gap" in ensembles with varying $N_f$. Its generalization to any number of colors is straightforward.

 \subsubsection{ The critical line versus the fermion numbers $N_f,N_a$:}
Model simulations with such clusters, random or interacting with each other, 
predict certain distinct shapes of the Dirac eigenspectrum at small $\lambda$. We have 
in particularly found at which diluteness of the ``cluster gas" one gets particular
value of the chiral condensate, the chiral symmetry
restoration or certain size of the ``gaps".  These eigenvalue spectra and 
``lines of constant condensate"  can be compared
with the lattice ones, in order to see if chiral breaking does
happen in the ``molecular gas" regime, or at a denser regime.

The increasing number of fermions leads to stronger fermion-induced interactions,
binding the $L\bar{L}$ pairs into tighter clusters. In terms of our molecular gas model, it gets much
more dilute, unless the overall density of the dyons is significantly increased. This can only
be achieved by {\em going into stronger coupling domain}, which reduces the dyon masses and interactions. That is why 
the critical lines in Fig.\ref{fig:romanplot} go downward with increasing flavors.

We have also explained why there is a qualitative difference between the fundamental
and adjoint fermions. While the former have zero modes only with one (heavier) dyons
$L\bar{L}$, the adjoint have zero modes for all dyons, including lighter $M,\bar{M}$ (for SU(2)).
The latter are much less correlated, thus their chiral restoration temperature is much higher.

 \subsubsection{ Chiral splittings of hadronic masses versus $N_f$} 
 As this parameter is now becoming measurable on the lattice, with the progress in computer/lattice technology,
it is perhaps time to map it more consistently, and also think again about the physics it reveals.

Rapid decrease of the chiral condensate scale around $N_f\sim 4$ has been in fact predicted by the instanton liquid
simulations \cite{Schafer:1996wv} long ago. The reason for that has been a ``dip" in the eigenvalue spectrum
developed due to  the ``molecule" formation\footnote{This dip should not  be confused with that due to the finite size of the system, which is well described by the random matrix theory.}. This very phenomenon is in fact central in our current study. What has not been predicted in 1990's  was a significant shift
to the stronger coupling and drastic increase of the overall instanton/dyon density (or, in alternative language, a rapid decrease
of the hadronic scale), which makes a very small quark condensate relevant.  
It would be important to study  transition to ``molecular" topological structure
in lattice simulation.

  The dependence of the quark condensate on
 the density of ``molecules" we found in our calculations  is interesting.
  As seen  in Fig.\ref{rand_mol},  at low density there is a minimum between two ``molecular peaks", but at some
  diluteness there is a sudden appearance of a small
peak inside this  minimum.  This implies a sudden jump in the quark condensate value,
in a small interval of parameters. It is more pronounced than many crossover phase transitions: and thus we
may call this phenomenon a phase transition in the dyonic ensemble, from ``atomic" to a ``molecular" state, at $N_f>3$. 
Using quark masses as interpolating parameters between $N_f=3,4,5$, lattice practitioners can see if this change of Dirac
spectrum is also occurring in lattice simulations as well.

 \subsubsection{ Dependence of the Dirac eigenvalues on  the holonomy value and phase}
The non-zero holonomy provides Higgsing, the breaking of the color group, and thus it  naturally explains  different fermionic masses which appear in the
$T_{ij}$ hopping amplitudes.

 Since the rest of the paper has been for SU(2) gauge group, let us discuss in a bit more detail the two SU(3) options shown in
Fig 1 of \cite{Gattringer:2001yu}, namely
 the real $ <P>$ sector as well as the one with the phase of $P$ being $2\pi/3$.
 Generic holonomy in SU(3) is described by 3 parameters $\mu_1,\mu_2,\mu_3$ subject to 
 one condition $(\mu_1+\mu_2+\mu_3=0 (mod(1))$. If one imposes one more condition, such as
  the fixed phase of $<P>$,
 there is only one free parameter left. 
 If the phase is zero,  $<P>$ is real, the one-parameter family of possible holonomies  is
 \be  \mu_1=0; \mu_2= 1/2-\delta; \mu_3= 1/2-\delta \ee  
 If the phase is  $2\pi/3$  this solution is simply rotated additively to
  \be  \mu_1=1/3; \mu_2=1/3+ 1/2-\delta; \mu_3=1/3+ 1/2-\delta \ee  
 The masses of  monopoles are determined by the differences, which are the same in both cases
 \be \nu_1=\nu_3=1/2+\delta, \nu_2=2\delta \ee 
 since for gluonic observable the two sectors are identical by $Z_{N_c}$ symmetry.
  
  However the fundamental fermions notice the difference, as their masses are given by $\mu_i$, not $\nu_i$.
  Furthermore, which dyon gets the zero mode depends on the phase parameter $z$ in fermionic periodicity condition:
  the rule is that it is the one in which $\nu$ sector $z$ resides on the circle.  The antiperiodic fermions $(z=1/2)$
   picks up the second type of dyons $\nu_2=2\delta$ in the real sector and
   the first one $\nu_1=1/2+\delta$ in the one with the phase $2\pi/3$.  
  The lowest mass of the fermion is   $2\pi T \delta$ in the former case, while in the complex ones 
  it is  $(2\pi T)min(1/6,1/3-\delta) $.  
  
  Of course, in lattice subsector  with the fixed phase, the modulus still has some average and the distribution,
  determined by the effective potential of $<P>$ at a give $T$, which is known if the lattice simulation is made. 
We however dont know  the values: let us say take some generic value between 0 and 1 as a guess:  a half
\be |<P>|=1/2 = (1-2cos(2\pi \delta))/3 \rightarrow \delta\approx 0.29 \ee
If so, the fermion masses for the two sectors are $m/(2\pi T)= 0.29$ and $  0.04$, respectively.
 Such large mass difference explains
   why the participation ratios (roughly, the fraction of the box volume occupied by a mode)
  are so different: while in the real sector the Dirac modes occupy only about 1 percent of the box, 
 in the complex sectors  nearly the whole box is occupied.) 

 \subsubsection{ Dirac eigenstates ``at the gap". } The objects found via this method by Bruckmann et al \cite{arXiv:1105.5336}
are consistent with being made of $L$ and $\bar{L}$ dyons. Apparently thy must be neutral under the 
topological charge because they do not contribute to topological susceptibility. The clusters we propose in this work 
include both $L$ and $\bar{L}$ dyons, with zero total topology, yet still able to support the fermionic
localized states. Further lattice studies of the abelian projected electric and magnetic fields correlated
with those objects would further clarify their origins.

 \subsubsection{Dependence on fermionic periodicity conditions}

In several works \cite{Bilgici:2009tx,Bilgici:2009jy} effects of the temporal boundary conditions on the chiral 
condensate on quenched ensembles was explored. In addition to the restoration of the chiral symmetry for the physical,
$antiperiodic$, boundary conditions, an increase in the chiral condensate for the periodic condition above $T_c$ was observed. 

In the case of the adjoint fermions the drop of chiral condensate happens for both the periodic and antiperiodic sectors.
There is however a qualitative difference: the drop  is slower for the periodic case,  differs in shape, and was not traced to reach zero. A qualitative difference is certainly expected on the basis of Ref. \cite{GarciaPerez:2009mg} : the zero modes
for periodic and antiperiodic boundary conditions are drastically different. For periodic case, all zero modes are democratically distributed, whereas for the antiperiodic boundary conditions they all fall
onto the heavies one (the $L$ dyon). The case of antiperiodic boundary condition then restore chiral symmetry in the 
way similar to fundamental quarks, by condensation into $L\bar L$ clusters. The case of periodic 
  adjoint fermions is quite different, it has ``democratic" distribution of zero modes over 
all dyons of the instanton. This makes restoration of the chiral symmetry much more difficult, as now correlating
the ``heaviest" $L\bar L$ is insufficient and also ``light"  $M\bar M$ pairs should form. 
A more quantitative analysis of the
adjoint fermion case is postponed for future studies.

We finally comment that the case of dynamical, periodic, adjoint fermions was treated analytically in \cite{Unsal:2007vu}, 
where it was shown that the chiral symmetry is indeed restored.
  \subsubsection{ Short-range correlations  of dyons in quenched and non-quenched ensembles}

At high temperature, the
vacuum is mostly dominated by perturbative fluctuations: the coupling
is simply to small to allow any kind of large quantum effects. As we
lower the temperature, the formation of topological objects starts to
be possible. Although still suppressed, the vacuum is able to polarize
into topological objects, which can support localized fermionic modes
of small eigenvalue, but still not small enough to break chiral
symmetry. Also, the "mass" M of these zero modes, which interpolates
from $2\pi T$ to $\pi T$ i.e. reduces by a half in units of temperature as
temperature is decreased, makes it harder to break chiral symmetry at
high temperature, as the tail of zero modes, being the inverse of this
mass, does not extend very far, and, thus, at small density of
topological charge, the off-diagonal matrix elements are simply too
small to matter, and the ensemble of dyons is the ensemble of $L\bar L$ pairs
and neutral random clouds of light $M,\bar M$ dyons.

However things change drastically as temperature is decreased. The
gluon dynamics facilitate the increase of topological density, due to
a suppression of the action of any field configuration at lower
temperature by the coupling $1/g^2$. At one point the moduli space
metric of topological objects becomes the sole dictator of the
distribution of topological (dyonic) field configuration of dyons, 
the rest simply being fluctuations which is also not
suppressed at all, but does not change the important topological properties of the 
background configuration. The dynamics of
dyons becomes important solely through the geometry of the moduli
space metric.

The interaction of L and anti-L is very similar to the interaction between
L and L, i.e. they repel (Coulombic-like), and the L and anti-M attract from the point of view 
of the metric (at large distances),
just like L and M. The vacuum then needs to undergo a transition in the structure, 
from pairs of $L\bar L$ and $M\bar M$, to the crystal of alternating dyons and antidyons. 

Such an abrupt change in the vacuum structure of the quenched ensemble is absent upon
introduction of dynamical fermions. In this case the situation
changes drastically. The increase of topological density is
suppressed by the presence of fermions, as increasing density means
making pairs come closer together. By arguments in the article, such a scenario
will make the fermionic determinant be smaller and smaller, eventually
going to zero if the molecules overlap. That means that a the same
topological density is harder to develop with fermions. Increasing the
flavors makes it even harder, as the smallness of the fermionic
measure is enhanced by $N_f$. However, if $N_{f}$ is not too large chiral
symmetry may still be broken, but the nature of this transition is now
vastly different. The pairs, instead of abruptly changing their structure, are acting like slippery objects, trying to keep their distance as large as possible to all
other pairs. 

  \subsubsection{ Small comment about deconfiement} Recent work by Bruckmann et al \cite{arXiv:1111.3158} 
 have argued that uncorrelated dyons of all kinds create linear confinement. An ensemble of correlated neutral
 ``molecules" we propose would not do that and generate short-range correlations only. This potentially
 links both deconfinement and chiral restoration with molecule formation.

{\bf Note added:} After our paper was completed  we learned about important study \cite{Bornyakov:2008im} on the lattice, in which some of the tests proposed above were successfully performed.
For $periodic$ fermions their near-zero eigenmodes with small eigenvalues are indeed identified
with the type-$M,\bar{M}$ dyons, while for $antiperiodic$ fermions those are indeed related with 
the type-$L,\bar{L}$ dyons. 
This conclusion was reached by correlating the topology with the sign of the Polyakov line at the dyon center.
The difference in inverse participation ratios between the two cases
is indeed naturally explained by large difference in $M$ and $L$ actions. 
For the ``heavier" $L$ dyons, even the shape of the fermionic eigenmodes was shown to agree with
the corresponding semiclassical predictions. 
 

\vskip .25cm \textbf{Acknowledgments.} \vskip .2cm 
Strong motivation for this study during the last decade came from multiple
 discussions with Pierre van Baal. We also acknowledge more recent discussions
with D. Diakonov and F. Bruckmann.
This  work was supported in parts by the US-DOE grant DE-FG-88ER40388.

\begin{appendix}
\section{Exact solution of the zero modes}

Although zero modes of a caloron were found in all generality elsewhere \cite{Kraan:1998sn,Lee:1998bb}, 
we here use an approach which is more illuminating. One of us would like to thank A. G. Abanov
for useful discussions on this issue.

Here we solve equations \eqref{eq:Dirac1}, where
\begin{align}\label{eq:dyonsol}
&\mathcal H=\pm \frac{1-v r\coth(vr)}{r}\;,\\
&\mathcal A=\frac{1-vr/\sinh(vr)}{r}\;,
\end{align}
We will take the lower sign (antiselfdual solution).
To do this we separate the matrix $M(r)$ as
\be
M(r)=M_0(r)+M_1(r)\;,
\ee
where
\begin{align}
M_0(r)&=\left(\frac{\mathcal H}{2}+\frac{1}{r}\right)\bm 1\nonumber\\&=\frac{1+r v \coth(rv)}{2r}\;,\\
M_1(r)&=\frac{z}{\beta}\sigma_1+\left(\mathcal A-\frac{1}{r}\right)\sigma_3\nonumber\\&=\frac{z}{\beta}\sigma_1-
\frac{v}{\sinh(vr)}\sigma_3\;.
\end{align}

The solution can then be written as $\bm \alpha=\exp(-\int_{0}^rM_0(r)dr)\bm\chi$, or
\be\label{eq:chisol}
\bm\alpha=\frac{1}{\sqrt{r\sinh{r v}}}\bm\chi
\ee
Note that if we took the upper sign in eq. \eqref{eq:dyonsol}, we would get a factor of $\sqrt{\sinh(vr)}$ in front of the solution. This is clearly non-normalizable, as it should be by the index theorem.

The differential equation for $\bm\chi$ is
\be
\frac{d}{dr}\bm\chi=-M_1(r)\bm\chi\;,
\ee
i.e.
\begin{align}
&\chi_1'(r)=\frac{v}{\sinh(vr)}\chi_1(r)-\frac{\phi}{\beta}\chi_2(r)\;,\\
&\chi_2'(r)=-\frac{v}{\sinh(vr)}\chi_2(r)-\frac{\phi}{\beta}\chi_1(r)\;,
\end{align}
we may take a change of variables $\xi=r v$. Then the equations read
\begin{align}
&\chi_1'(\xi)=\frac{1}{\sinh(\xi)}\chi_1(r)-\varsigma\chi_2(r)\;,\\
&\chi_2'(\xi)=-\frac{1}{\sinh(\xi)}\chi_2(r)-\varsigma\chi_1(r)\;,
\end{align}
where we labeled $\varsigma=\phi/(v\beta)$
We now eliminate $\xi_2$, and obtain the second order differential equation
\be
-\frac{d^2}{d\xi^2}\chi_1-\frac{1}{2\cosh^2\frac{\xi}{2}}
\chi_1=-\varsigma^2\chi_1\;.
\ee

This equation has  a general solution 
\begin{equation}
\chi_1(\xi)=c_1\left(-2\varsigma+\tanh\frac{\xi}{2}\right)e^{\varsigma
\xi}+c_2(2\varsigma+\tanh\frac{\xi}{2})e^{-\varsigma \xi}
\end{equation}

with arbitrary constants $c_{1,2}$. Using the first order equations we can write $\chi_2$ as
\begin{equation}
\chi_2(\xi)=c_1\left(2\varsigma-\coth\frac{\xi}{2}\right)e^{\varsigma
\xi}+c_2(2\varsigma+\coth\frac{\xi}{2})e^{-\varsigma \xi}
\end{equation}
The function $\chi_2(\xi)$ is divergent when $\xi\rightarrow 0$, except if $c_1=c_2$, in which case $\xi_2(0)=0$. Therefore $c_2=c_1$. 
The constant $c_1$ can be determined by overall normalization. The solution then becomes
\begin{subequations}\label{eq:chisol1}
\begin{align}
&\chi_1(\xi)=2c_1\left(- 2\varsigma \sinh(\xi\varsigma)+\tanh\frac{\xi}{2}\cosh(\xi\varsigma)\right)\\
&\chi_2(\xi)=2c_1\left(2\varsigma \cosh(\xi\varsigma)-\coth\frac{\xi}{2}\sinh(\xi\varsigma)\right)
\end{align}
\end{subequations}
Finally, combining with \eqref{eq:chisol} we obtain
\be
\alpha_{1,2}=\frac{\sqrt{v}}{\sqrt{\xi\sinh\xi}}\chi_{1,2}\;.
\ee
$\sqrt{v}$ can be absorbed into constant $c_1$, and our final expression is
\be
\alpha_{1,2}=\frac{\chi_{1,2}}{\sqrt{\xi\sinh\xi}}\;.
\ee
with functions $\chi_{1,2}$ given by \eqref{eq:chisol}, $\xi=v r$, $\varsigma=\phi/(v\beta)$.
Note that the value of $\alpha_{1}(\xi\rightarrow 0)$ is given by
\be
c_1(1-4\varsigma^2)\;,
\ee
and the solution is completely regular at $r=0$.
Remarkably it turnes out that for $c_1=1/2$, the solution is already normalized (in the sense of $\int d\xi \xi^2$ integration).
\end{appendix}

\end{document}